\RequirePackage{fix-cm}
\documentclass{svjour3}                     % onecolumn (standard format)
\smartqed  % flush right qed marks, e.g. at end of proof
\usepackage{graphicx}
\usepackage{hyperref}
\usepackage{amssymb,amsmath}
\usepackage{siunitx}
\usepackage{natbib}
\usepackage{textgreek}
\usepackage{color}
\usepackage[normalem]{ulem}
\usepackage{multicol}
\usepackage{caption}

\usepackage{rotating}
\usepackage{hyperref}
\usepackage{lineno}

\usepackage{longtable,adjustbox,lscape, afterpage, csvsimple, datatool, booktabs, tabulary, geometry, graphicx, multirow, caption, verbatim}
\usepackage{newtxtext,newtxmath,amsmath}
\usepackage{color}

\defcitealias{protection1991icrp}{ICRP 60 (1991)}
\defcitealias{icru1980}{ICRU 33 (1980)}
\defcitealias{valentin2007icrp}{ICRP 103 (2007)}
\defcitealias{dietze2013icrp}{ICRP 123 (2013)}
\defcitealias{berger1984report}{ICRP 37 (1984)}

\newcommand{\la}{\;\raise0.3ex\hbox{$<$\kern-0.75em\raise-1.1ex\hbox{$\sim$}}\;}
\newcommand{\ga}{\;\raise0.3ex\hbox{$>$\kern-0.75em\raise-1.1ex\hbox{$\sim$}}\;}
\newcommand{\sla}{\;\raise0.55ex\hbox{\scriptsize$<$\kern-0.75em\raise-1.1ex\hbox{$\sim$}}\;}
\newcommand{\sga}{\;\raise0.55ex\hbox{\scriptsize$>$\kern-0.75em\raise-1.1ex\hbox{$\sim$}}\;}
\newcommand{\ssim}{\;\raise0.3ex\hbox{\tiny$\sim$}\,}
\newcommand{\sapprox}{\;\raise0.3ex\hbox{\tiny$\approx$}\,}
\newcommand{\PWN}{\raise-0.4ex\hbox{\scalebox{0.8}{\scriptsize$P$\kern-0.05em$W$\kern-0.2em$N$}}}
\newcommand{\s}{\raise-0.1ex\hbox{\scalebox{1.2}{\scriptsize$s$}}}
\definecolor{grey}{rgb}{0.52, 0.52, 0.51}
\begin{document}

%\title{Crewed Missions to Mars: Modeling the Impact of Astrophysical Charged Particles on Astronauts and Assessing Health Effects}

\title{Crewed Missions to Mars: Modeling the Impact of Astrophysical Charged Particles on Astronauts and Their Health}


\titlerunning{Crewed Missions to Mars}        % if too long for running head

\author{\underline{Dimitra Atri} \and Caitlin MacArthur$^\ast$ \and Sriram Devata$^\ast$ \and Konstantin Herbst \and Dionysios Gakis \and Shireen Mathur \and Maria Villarreal-Gomez \and Giulia Carla Bassani \and Roberto Parisi \and Azza Al Bakr \and Tammy Witzens }

\authorrunning{Atri et al.} % if too long for running head

\institute{Dimitra Atri \at
              Center for Space Science, New York University Abu Dhabi, Saadiyat Island, PO Box 129188, Abu Dhabi, UAE \\
              Blue Marble Space Institute of Science, 600 1st Avenue, Seattle, WA 98104, USA\\
              \email{atri@nyu.edu}           
           \and
              Caitlin MacArthur \at
              Pathology and Molecular Medicine, University of Otago, Wellington, 23A Mein Street, Newtown, Wellington 6242, New Zealand\\
              $^\ast$ equal contribution author
            \and 
              Sriram Devata \at
              International Institute of Information Technology, Gachibowli, Hyderabad, Telangana 500032, India\\
              $^\ast$ equal contribution author
            \and
              Konstantin Herbst \at
              Institut f\"ur Experimentelle und Angewandte Physik, Christian-Albrechts Universit\"at zu Kiel, Leibnizstra\ss e 11, D-24118 Kiel, Germany
             \and 
              Dionysios Gakis \at
              Department of Physics, University of Patras, Patras, 26504, Greece
             \and
              Shireen Mathur \at
              Blue Marble Space Institute of Science, Seattle, WA 98154, USA
              \and
              Maria Villarreal-Gomez \at
              Universidad Industrial de Santander, Bucaramanga, Santander 680002, Colombia
              \and
              Giulia Carla Bassani \at
              Polytechnic University of Turin, 10129 Torino (TO), Italy
              \and
              Roberto Parisi \at
              Department of Medicine and Surgery, Università degli Studi di Salerno, Fisciano, Salerno 84084, Italy
              \and
              Azza Al Bakr \at
              National Space Science and Technology Centre, Al Ain, UAE
              \and
              Tammy Witzens \at
              Ira A. Fulton Schools of Engineering, Arizona State University, PO Box 879309, Temepe, AZ 85287, USA
}

\date{Received: date / Accepted: date}

\maketitle
%\tableofcontents
%\vspace{0.5cm}
%\linenumbers

\begin{abstract}
\small{The impact of exposure to astrophysical ionizing radiation on astronaut health is one of the main concerns in planning crewed missions to Mars. Astronauts will be exposed to energetic charged particles from Galactic and Solar origin for a prolonged period with little protection from a thin spacecraft shield in transit and from the rarefied Martian atmosphere when on the surface. Adverse impacts on astronaut health include, for example, Acute Radiation Syndrome, damage to the nervous system, and increased cancer risk. We rely on medical studies to assess the impact of enhanced radiation dose levels on various physiological systems and the overall health of astronauts. Using a combination of radiation measurements and numerical modeling with the GEANT4 package, we calculate the distribution of radiation dose in various human body organs for various expected scenarios simulated with a model human phantom. We suggest mitigation strategies, such as improved ways of shielding and dietary supplements, and make recommendations for the safety of astronauts in future crewed missions to Mars.}


\end{abstract}

\section{Introduction}\label{intro}

Since the first crewed spaceflight in 1961 and the first human mission to the Moon in 1969, several crewed missions and long-term visits to the International Space Station (ISS) have been carried out. With new deep space missions planned for the near future, astronauts will further explore our Solar System and thus be exposed to different environments of interplanetary space and the Martian atmosphere and surface within the coming decades. Of utmost importance thereby are environmental radiation factors like galactic and solar cosmic rays, the atmospheric secondary particle environment produced by the interaction of the cosmic rays with the atmospheric components, the reduced planetary gravity (3.72 m/s$^2$ at the Martian surface) after months of traveling in microgravity, and their impact on the human body.

Interstellar space is dominated by (E)UV, soft and hard X-rays, and energetic cosmic rays of galactic and solar origin that can adversely affect human health. In particular, solar energetic particles (SEPs) accelerated, for example, in solar flares, coronal mass ejections (CMEs), or at interplanetary shocks \citep[][]{Reames-1999} are predominantly composed of 89$\%$ protons, 10\% helium, and about 1$\%$ of electrons and can have energies between several keV and hundreds of MeV. In the case of relatively strong SEP events, known as ground-level enhancement (GLE) events, SEPs with energies up to a few GeV can be found. The strongest event ever measured is GLE05 (18 February 1956). However, cosmogenic radionuclide records showed strong increases (about 15\%) in the production rates around AD774/775 \citep[e.g.,][]{Mekhaldi-etal-2015} associated with a solar superevent about 30 to 70 times stronger than GLE05 \citep[][]{Cliver-etal-2020, Usoskin-etal-2020}. Most recently, however, \cite{Brehm-etal-2021} found two even stronger events in tree ring records around 5259 BC and 7176 BC, with an increase of about 19\%.  

While being about four orders of magnitude lower than the SEP flux in the keV range, the differential particle flux of Galactic Cosmic Rays (GCRs), accelerated at supernova remnants \citep[][]{Hillas-2005, Buesching-etal-2005}, extends over more than 15 orders of magnitudes, up to more than 10$^{15}$ MeV. GCRs are known to consist mainly of hadrons, consisting of approximately 87\% protons, 12\% helium nuclei (He), and a small fraction of 1\% heavier elements up to iron \citep[e.g.,][]{Wefel-1991}. Because the solar magnetic field is frozen into the solar wind and carried outwards into the interplanetary medium, the heliospheric magnetic field is formed and extended into interplanetary space. Therefore, the low-energetic part of the GCR spectrum is modulated by changing solar activity, resulting in an 11-year solar and 22-year magnetic activity cycle variation. 

While being exposed to photons in interplanetary space can easily be prevented by adding a thin shield, charged particles can penetrate deeper into spacecraft. They may produce secondary particles capable of penetrating through even thick shielding, limited to weight restrictions on spaceflights. In particular, strong solar events and potential superevents can cause tremendous radiation hazards and risk for crewed space missions and can cause direct damage by ionization or produce harmful radicals in the body leading to secondary effects.

Besides, once charged particles enter planetary atmospheres, they are the primary drivers of planetary altitude-dependent atmospheric ionization. While the low-energy cosmic rays (CRs) lose most of their energy due to elastic collisions with atmospheric neutrals, CRs with energies above $\sim$ 1 GeV can induce extensive secondary particle cascades. Thereby, secondary mesons ($\pi^{\pm}$ and $\kappa^{\pm}$), nucleons, gamma particles, and nuclear fragments are created, which may interact further. Thus, the cosmic ray cascade evolves with increasing atmospheric depth, and the secondary particle flux continues to rise.

An astronaut's health is the number one priority for any space mission. When sending humans into space, they are constantly bombarded with ionizing radiation that could lead to severe health problems or permanent biological mutations. Before identifying the right shielding thickness and material to protect astronauts, a radiation dose estimate needs to be determined with the corresponding biological effect. To make these dose estimates, we utilize the model spectra of GCRs and historical SEPs to estimate the radiation environment in deep space and on the surface of Mars. We then use the GEANT4 numerical model to calculate the propagation of charged particles in various scenarios, namely (1) deep space, (2) the surface of Mars, (3) deep space with shielding, and (4) the surface of Mars with shielding. A standard human phantom was used to calculate radiation absorbed by various human body organs. GEANT4, an open-source code, is well-known in the community and is validated with various experiments around the globe. Our approach is, therefore, advantageous over other studies using proprietary software which cannot be validated by the community.   

In this paper, our purpose is to assess the impact of astrophysical radiation during the long-duration mission to Mars on the health of astronauts. At first, we evaluate shortly the radiation environments astronauts will encounter on their way to Mars (Section  \ref{sec:sec2}). A thorough review of the effects of ionizing radiation on the physiological systems, the genetic material, and the cancer risks is provided in Section \ref{sec:sec5}. We describe our numerical model to estimate the radiation exposure and present our results after compiling data from past and current operating missions in Section \ref{sec:sec3}. Section \ref{sec:sec6} outlines the main radiation mitigation strategies. We conclude with a summary of this work in Section \ref{sec:sec7}.

\section{Astrophysical Ionizing Radiation Environments}
\label{sec:sec2}

\subsection{Measurements in LEO and from Lunar missions}
There are three main sources of primary ionizing radiation found in low Earth orbit (LEO), namely: Galactic cosmic rays (GCRs), Earth’s radiation belts (ERBs), and Solar particle events (SPEs). GCRs are charged particles originating from outer space and range from several tens to 10$^{12}$ MeV and peak around 1 GeV. The GCR component consists of 85\% hydrogen nuclei, 14\% helium nuclei, and 1\% heavy ions called HZE (high charge and energy). On the surface of Earth, the atmosphere and the magnetic field protect us from harmful GCR radiation, while GCRs are a dominant source in space and can be extremely hazardous \citep{benton2001space, atri2014cosmic}. 
The ERBs, also known as the Van Allen radiation belts, consist of high-energy electrons and ions, primarily protons, trapped in the geomagnetic field, forming two belts in the inner magnetosphere of the Earth, where the geomagnetic field resembles a magnetic dipole \citep[e.g.][]{Koskinen-Kilpua-2022}. While the inner belt extends from 1,000 to 12,000 km and is composed of electrons of less than 5 MeV, in the outer belt (13,000 to 60,000 km), electrons with energies around 7 MeV are trapped \citep{benton2001space}. Launching an interplanetary mission, the spacecraft will pass the radiation belts for a short amount of time. Hence, it does not affect the crew and can be stopped by the spacecraft’s shielding.

Another major radiation source is the Sun, emitting high-energy particles in the form of X-rays, gamma rays, and the solar wind, a constant stream of protons and electrons. During times of higher solar activity, particles can be accelerated in solar flares and at the shock fronts of CME, detectable in interplanetary space as SPEs.  Solar flares last for a short amount of time, in the order of hours, and are restricted to a 30 – 45 degrees angle in solar longitude,  while CMEs can last for days and extend on a larger angle of 60 to 180 degrees \citep{benton2001space}. Each of the three primary sources varies depending on altitude, the inclination of the spacecraft’s orbit, and the phase of the solar cycle. 
 
The International Space Station (ISS), orbiting the Earth in LEO at around 400 km, is an active space laboratory with much research being conducted. In one study, astronaut Scott Kelly was sent to the ISS for a year to study the impact of long-duration spaceflight on the human body. Results showed that changes were no more significant than that observed in his twin brother Mark Kelly who stayed on Earth \citep{garrett2019nasa}. Also, 91.3\% of Scott’s gene expression levels returned to normal a few months after returning to Earth, suggesting that health can be sustained during this time in space. Although some biological mutations were observed in Scott, they were not fully understood since Scott was the only astronaut that went through this experiment. Thus, more research needs to be done on a more significant number of astronauts. 

The Apollo program is one of the most complex and most extensive scientific exploration programs in the history of humankind and the farthest point traveled by humans beyond low Earth orbit. Although Moon does not belong in the current plans for the human transfer to Mars, collective data from the Apollo missions are presented in this section to understand better the radiation dose and its effect on astronauts, which is crucial for developing and planning for future deep-space missions. Six out of seven missions planned for lunar landing were successful. This includes Apollo 11, 12, 14, 15, 16, and 17, putting twelve astronauts on the Moon's surface. Table~1 presents the dosimetry data from Apollo missions, retrieved from \citet{english1973apollo}.

\begin{table}[t]
    \centering
    \begin{tabular}{c|c|c|c}
    \hline
     Mission & Time (days) & Absorbed dose (mGy) & Mean dose rate ($\mu$Gy/day)\\
     \hline
     Apollo 11 & 8 & 1.8 & 220 \\
     \hline
    Apollo 12 & 10 & 5.8 & 570\\
     \hline
    Apollo 14 & 9 & 11.4 & 1270\\
     \hline
    Apollo 15 & 10 & 3 & 240\\
     \hline
    Apollo 16 & 11 & 5.1 & 460\\
 \hline
Apollo 17 & 12 & 5.5 & 440\\
 \hline
    \end{tabular}
    \caption{Dosimetry data from Apollo missions \citet{english1973apollo}}\label{tab:1}
\end{table}

%Table~\ref{tab:1}
Table~1 shows the mean dose rate per day as a function of the mission duration. The dose rate varies depending  on the lunar surface activity, intra-vehicular Command and Lunar Module activity, and the changes due to cosmic ray modulation and solar activity. According to a NASA report, the doses received by the Apollo crew were small and had no harmful impact on their health \citep{Johnston1975}. However, this is not surprising since, luckily, no major SPE occurred throughout the missions. However, some significant biological observations were reported in \citet{english1973apollo}, including vestibular disturbances, inflight cardiac arrhythmia, reduced postflight orthostatic tolerance, reduced postflight exercise tolerance, postflight dehydration, and weight loss. Several causes have been suspected. Although the flight diet was adequate, the food consumption was suboptimal; in the end, it decreased the red cell mass and the plasma volume, and resulted in a negative inflight balance trend for nitrogen, calcium, and other electrolytes. In addition, increased inflight adrenal hormone secretion has been reported.

\subsection{Measurements from deep space and the Martian Surface}

One of the main challenges of sending humans to Mars is the exposure to high ionizing radiation over a long period. The journey to Mars takes up to around six to eight months. While coming back to Earth will need another launch window, spending around 500 days on the Martian surface, and another six to eight months journey back to Earth, making the total duration of a roundtrip around two-three years. However, since the martian atmosphere is much thinner than that of the Earth, the radiation exposure on Mars is much higher. 

In 2012, NASA successfully landed the Curiosity rover on the Martian surface in Gale Crater. Since then, Curiosity’s Radiation Assessment Detector (RAD) has been measuring the energetic particle radiation environment, e.g., in the form of charged particle and neutral particle radiation, giving the total dose rate and particle spectra from 10 to $>$100 MeV/u \citep{Hassler:2012,Ehresmann:2014,Kohler:2014}. 

\begin{figure}[!t]
\centering
\includegraphics[width=\textwidth]{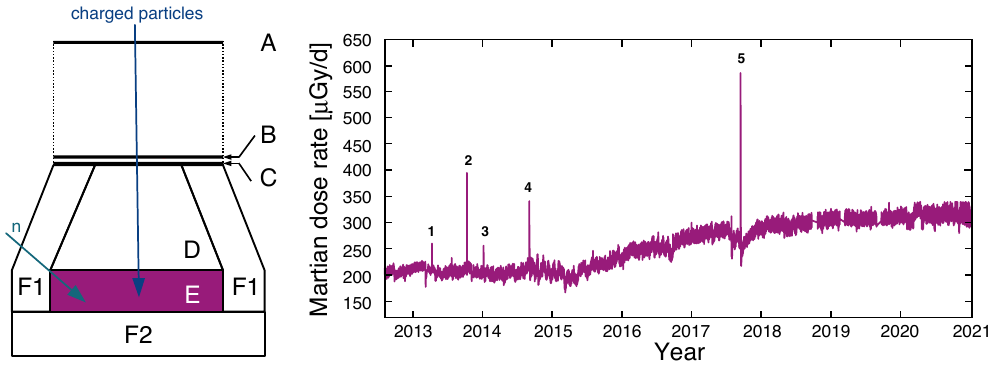}
\caption{Left panel: Schematic view of the RAD sensor head with the three silicon detectors (A, B, and C), detector D (a caesium iodite scintilator), and the plastic scintilator E. Right panel: Martian dose rate detected in E between landing and December 2020. Numbers 1 to 5 mark the SPEs detected at the surface of Mars during that time. Figures after \citet{Guo-etal-2021} (under a Creative Commons Attribution 4.0 International License).}
\label{fig:RAD}
\end{figure}

Figure~\ref{fig:RAD} shows a schematic view of the RAD instrument (left panel) and the measured Martian dose rates detected since Curiosity’s landing (right panel). In 2013, the radiation dose rate varied between 180 and 225 $\mu$Gy/day while in 2020 it varied roughly between 300 and 350 $\mu$Gy/day. In addition, throughout the mission so far five SPEs have been detected on the Martian surface, with the most prominent one on 10 September, 2017. The latter also has been detected as Ground Level Enhancement at the Earth's surface; thus, it is the first SPE observed on two planetary surfaces. 

It was also found that the average GCR dose rate at Gale Crater varied between 0.210 $\pm$ 0.040 mGy/day, while the Mars Science Laboratory (MSL) spacecraft carrying Curiosity measured a dose rate of 0.48 $\pm$ 0.08 mGy/day during the cruise phase \citep{hassler2014mars}. Accordingly, the quality factor Q characterizing the severeness of a radiation environment was found to be 3.05 $\pm$ 0.3 on the Martian surface, much lower than the value of 3.82 $\pm$ 0.3 in deep space. This difference is also driven by the heliospheric conditions, which reduced the dose rate by a factor of two compared to the cruise phase. 

Two scenarios are examined for the human journeys to Mars: the conjunction-class and the opposition-class mission \citep{Saganti:2004}. The first type of mission involves a more extended stay on Mars than the second \citep{Cucinotta:2002} and seems the preferable choice to maximize the stay on Mars because of the lower doses there. Based on the measurements and calculations for the conjunction-class mission, the total mission dose for an entire round trip has been estimated to be around 1.01 Sv, which may differ according to the specific mission characteristics and the solar conditions. 

Evidently, any decisions concerning the shielding strategies on the surface of Mars must be taken in alignment with the type of journey chosen, and vice versa \citep{Wilson:1997}. For example, additional shielding might be unnecessary during very short stays as exposure levels would be much lower than the acceptable limits.  Likewise, if adequate shielding is ensured on the planet's surface, we could increase the duration of the human presence there. 

%

%\textcolor{red}{\textbf{Ionizing Radiation: Theoretical Models of Solar Proton Events + Galactic Cosmic Rays}
%To estimate the Ionizing dose, we use The European Space Agency (ESA) Space Environment Information System (SPENIVIS) to model the space radiation environment. We look at three different target materials (tissue, bone and water). Figure 3 shows that the dose absorbed by the target material decreases and we increase the shielding thickness.
%(Figure)
%Fig. 3. Dose as a function of shielding thickness in tissue, bone and water respectively}

\section{Impact of Ionizing Radiation on Astronaut Health}\label{sec:sec5}

Humankind is well-protected from ionizing radiation by our planet’s atmosphere and magnetosphere. The International Space Station sits in low-Earth orbit, where the magnetosphere protects it. Until today, the only human beings that traveled outside the magnetosphere were the Apollo astronauts. Our knowledge of the harmful impacts of radiation exposure on the human body comes mainly from studying radiation therapy. However, radiation therapy does not employ neutron exposure; therefore, 1 Sv is approximately equivalent to 1 Gy in patients receiving radiotherapy treatment. Similarly, as most experimental studies utilize low photon energies, 1 Sv is approximately equivalent to 1 Gy in such publications \citep{Dainiak2003473}. Some knowledge further comes from historical radiation disasters like the atomic bombings of Hiroshima and Nagasaki in 1945 and the Chernobyl disaster of 1986. 

Acquiring knowledge of how radiation exposure affects the human body is crucial for the future of space travel (such as to Mars), where exposure to galactic cosmic radiation and solar particle events poses a threat to astronaut health. The following literature review material focuses on the effects of radiation on various human physiological systems and genetic material. 

\subsection{The Impact of Ionizing Radiation on the Physiological Systems}
\subsubsection{Acute Radiation Syndrome}
In August 1972, a series of solar flares led to severe solar storms, solar particle events, and geomagnetic storms. These events occurred only five months after Apollo 16 had been launched and five months before Apollo 17 was launched in December 1972. Initially, the launch of both Apollo missions was set for August 1972. %August 1972 was considered a potential launch date for either Apollo 16 or 17 during the initial planning stages. 
If Apollo 16 or 17 had been launched during the dates of the solar storms, and specifically during the extreme solar flare of August 4th, any in-flight astronauts Would have been exposed to an absorbed radiation dose of around 0.5 Gy \citep{Jones202039}. The Centers for Disease Control and Prevention (CDC) defines 0.7 Gy as the threshold of radiation exposure at which acute radiation syndrome (ARS) can occur and reports that mild symptoms can be observed at exposures as low as 0.3 Gy \citep{CDC2018ARS}. Thus, ARS is a health concern for astronauts during SPEs. NASA has commented that there is a risk of ARS for astronauts due to the known occurrence of SPEs and the poor ability to predict SPEs \citep{NASA2019}.  

ARS is divided into three medically distinct presentations: hematopoietic, gastrointestinal, and neurovascular. This is not an exhaustive list of the tissues that can be affected by ARS. In medical terms, ‘acute’ means that the illness presents severely in a short amount of time. The three divisions of the syndrome present at different thresholds of radiation exposure: The hematopoietic syndrome at 0.3-5 Gy, the gastrointestinal syndrome at 5-20 Gy, and the neurovascular syndrome at above 20 Gy. Although the gastrointestinal and neurovascular syndromes are unlikely to be a risk for astronauts at spaceflight-relevant radiation doses, the hematopoietic syndrome occurring at doses as low as 0.3 Gy, might be an actual threat+ since it includes low numbers of white and red blood cells, which can lead to infection and anemia. The higher the radiation dose an individual is exposed to, the lower the survival rate. If bone marrow is sufficiently damaged during the hematopoietic presentation, it can lead to hemorrhage and impaired ability of the immune system to fight off infection, both of which can lead to death \citep{CDC2018ARS}.
\subsubsection{The Nervous System}
Radiotherapy for patients with brain tumors produces a wide range of acute and  delayed effects on the nervous system. This is valuable information for considering the potential effects radiation exposure may have on the nervous system of astronauts in long-term space missions. For patients who have been recently diagnosed with brain cancer and treated with radiotherapy, observed side effects include mood changes (anxiety and depression) and difficulty with reading, writing, and expression of language. Late side effects of long-term survivors of radiotherapy-treated brain tumors can also include physical and mental fatigue, anger and frustration, and difficulty concentrating.  External beam radiation therapy is the most common radiation therapy used to treat brain cancers, utilizing X-rays and gamma rays. Proton therapy is a novel technique gaining momentum and is used more often for children. Radiation doses in brain tumor radiotherapy can be around 20 Gy in adults and 24 Gy in children, delivered in fractions of 1.8 to 2.0 Gy directly to the tumor. These dose rates can cause delayed cognitive impairment.

A common side effect of brain tumor radiotherapy and one of the particular concerns for astronauts is fatigue. Fatigue is an acute side effect usually observable within several weeks of the patient’s first radiotherapy treatment and usually persists for one to three months after treatment has finished. A defining characteristic of radiotherapy-induced fatigue is that it is not typically alleviated by resting \citep{Butler2006517}. This, in particular, concerns astronauts who need to be alert when partaking in dangerous interplanetary space travel and exploration. Fatigue can be accompanied by depression and cognitive dysfunction, but some symptoms can be relieved by medication such as methylphenidate. Hair loss and skin erythema (redness of the skin) are other acute side effects that often occur alongside fatigue due to radiotherapy \citep{Butler2006517}. It is important to note that the radiation dose employed in cancer radiotherapy is much higher than what astronauts are expected to be subjected to during interplanetary space travel. However, the risk of symptoms like impaired cognitive performance \citep{Craven1994873} as well as skin erythema and hair loss are well-documented \citep{Letaw1989285}.

\cite{Butler2006517} discuss that brain irradiation can lead to the demyelination of white matter as a delayed reaction. Demyelination involves damage to the myelin of the neurons. Myelin is a fatty sheath that wraps around neurons and allows signals to propagate through the nervous system at a higher rate. Thus, the demyelination of neurons in the brain's white matter would considerably slow the rate at which action potentials (signals) are propagated, accounting for some of the cognitive dysfunction and mood changes that can be observed in patients following radiotherapy. The cognitive dysfunction that is observed as a delayed reaction to radiotherapy involves slowing of cognitive functions, difficulty concentrating and multitasking, and decreased memory \citep{Butler2006517}. An impeded function of neurons is characteristic of the pathogenesis of several neurodegenerative diseases, such as Parkinson’s Disease, where there is an observed decrease in the number of dopamine-producing neurons in the substantia nigra. The substantia nigra is the part of the brain that is important for movement, thus producing the movement disorders associated with Parkinson’s Disease \citep{Lotharius2002932}. Also, damaged neurons can result in decreased neurotransmitter production. Low levels of neurotransmitters like dopamine and serotonin can cause anxiety and depression. Thereby, the ability of radiation to damage neurons is potentially hazardous for astronauts. 

A significant component of deep space radiation is high-energy protons. Proton radiotherapy is a novel development of high interest in medical fields as it allows for the treatment of tumors without such a high degree of damage to the surrounding healthy tissue due to an ability to control the depth of the proton beam inside the tissue. Proton radiotherapy is a commonly used treatment for rhabdomyosarcoma (RMS), a highly aggressive form of cancer that derives from muscle cells with an impaired ability to differentiate. Orbital rhabdomyosarcoma is a head and neck subtype that accounts for 10\% of pediatric RMS cases and commonly affects the orbit (the skull socket where the eye and its appendages sit). Standard photon radiotherapy as a treatment of orbital RMS can lead to impaired vision in the irradiated eye in most patients, and 10\% of patients receiving this treatment develop a later complication that requires enucleation (surgical removal of the eyeball from its orbit). Standard photon radiotherapy that utilizes X-rays and gamma-rays also renders children more prone to a second tumor development than adults.

\citet{Yock20051161} conducted a study that followed seven patients treated via proton radiotherapy for orbital rhabdomyosarcomas and who had also received dactinomycin/vincristine-based chemotherapy. The patients were all treated at Massachusetts General Hospital or the Harvard Cyclotron between 1995 and 2001, with a median dose of 46.6 CGE (cobalt gray equivalent) delivered in conventional fractions. The study followed up the patients at a median of 6.3 years following radiotherapy treatment. At the follow-up, it was found that all seven patients were disease-free, and only one patient had required enucleation and stereotactic radiotherapy due to the treatment not being locally controlled. The remaining six patients had good vision in the radiotherapy-treated eye. There was additionally no evidence of cataract formation or neuroendocrine effects in any of the patients. All patients displayed mild to moderate orbital bony asymmetry or enophthalmos.  Enophthalmos causes the eye to sink and be abnormally positioned in its socket due to an injury or medical condition. Despite this, overall results showed that healthy tissue surrounding the cancer mass was better spared by proton irradiation than by photon irradiation \citep{Yock20051161}. This is interesting, considering the high abundance of protons in deep space radiation, and highlights an avenue for further research into proton irradiation in radiotherapy and space medicine.

HZE particles (high-energy nuclei) are a crucial component of galactic cosmic radiation, and it is essential to consider their biological effects compared to the standard X-rays and gamma-rays commonly used in radiotherapy. A study by \citet{Craven1994873} showed calculations on the number of cell nuclei in the brain likely to be struck by HZE particles during interplanetary space travel. The study determined that 3.4\% of the neuronal nuclei at the brain's center would be crossed by an HZE particle once every three years of interplanetary space travel. As previously discussed, neuronal damage plays a role in neurodegenerative diseases and mental illnesses. Neurons of the central nervous system could experience much damage from being hit by a high-energy HZE particle. Neurons have a very low regeneration rate making repair and replacement in response to damage a prolonged process. The study also reports on iron as an abundant component of galactic cosmic radiation and an element capable of creating breaks in the DNA strands in neuronal nuclei. Due to the low regeneration rate of neurons, these radiation-induced DNA changes may not be phenotypically present until later in the astronaut’s life. The accumulation of DNA damage and direct damage to neuronal nuclei could result in premature aging for exposed astronauts.  In addition to damaging the neurons themselves, the heavy-charge radiation particles could disrupt the complex interactions (synapses) of neurons in the central nervous system, impairing the mental and physical abilities of the astronaut. Optimal cognitive and physical performance is essential for astronauts partaking in the dangerous endeavors of interplanetary space travel \citep{Craven1994873}.

\subsubsection{Behavioral Effects}
As previously discussed, ionizing radiation can impair the central nervous system (CNS). Damages occur in basic neuronal processes, such as neuronal firing and synaptic excitability, and also in the brain structure due to white matter necrosis, demyelination, and vascular changes \citep[e.g., ][]{Trivedi20122009}. \citet{Cucinotta2009191} described some side effects in behavior and cognition; according to their exposure, they can be classified as acute or chronic. Acute CNS risks include altered cognitive function, reduced motor function, and behavioral changes, which may affect performance and human health. Chronic CNS risks are possible neurological disorders such as Alzheimer’s, dementia, or premature aging \citet{Cucinotta2009191}. Radiotherapy patients manifest behavioral changes, such as chronic fatigue and depression, and neurocognitive degeneration in verbal and visual memory, attention, speed of thought, and verbal fluency \citep{Cucinotta2009191}. Children are especially susceptible to these effects. A review by \citet{Butler2006184} showed a decline in intelligence and academic achievement, including low intelligence quotient (IQ) scores, verbal abilities, and performance IQ in children after radiotherapy treatment for brain tumors. 

Besides the hazard for the physiological systems, there is a growing concern for the social and cognitive welfare of astronauts on long-term missions due to the prolonged radiation exposure and confinement they will experience \citep{Frederico2009278}. As behavioral effects are difficult to quantify, different tests evaluate changes in conduct and cognitive performance. In each assay, results always depend on the age of the individual and radiation exposure dose \citep{Cucinotta2009191}. \citet{Trivedi20122009} exposed three-month-old male mice to doses of 3 Gy, 5 Gy, and 8 Gy (Tele 60C0 unit gamma irradiation). Ten days post-irradiation, they detected anxiety-like symptoms in the three irradiation groups and a decrease in locomotor activities and execution in the novel object recognition test (a paradigm used to assess object recognition and spatial memory) for 5 Gy and 8 Gy groups \citep{Trivedi20122009}. In another study, \citet{Pecaut2002329} simulated space radiation conditions with mono-energetic protons (250 MeV) for 3 or 4 Gy doses in two-month-old female mice. Their results also evidenced deficits in locomotor activities and acoustic startle habituation. However, two weeks later, the effects of radiation were negligible \citep{Pecaut2002329}.

A recent investigation by \citet{Kiffer-etal-2019a} analyzed female mice nine-months the late effects in cognition, behavior, and neuronal morphology post-exposure to lower doses of 0.1 Gy or 0.25 of  $^{16}$O. Nine months post-exposure, they estimated animal conduct changes. 0.1 Gy and 0.25 Gy induced deficiencies in object recognition memory and deficit in social novelty only at higher doses. \citet{Cherry2012e53275} found consistent results in mice three-month; they evidenced cognitive and memory space impairment after exposure to 0.1 Gy and 1 Gy of $^{56}$Fe radiation at 1 GeV/$\mu$. These findings suggest that recognition memory is most affected after space-type radiation and that these damages may persist over time. One of the long-term effects is the possible development of the neurological disorder Alzheimer’s disease (AD), characterized by a progressive cognitive decline over several years. One of the causes of this decline is an ongoing chronic neuroinflammatory process. The accumulation of amyloid-beta (an abnormal protein) in the brain forms plaques; these are vital players in neuroinflammation and are one of the principal histopathological hallmarks. Monitoring plaque progression is a diagnostic tool for humans and allows the gauge of disease severity \citep{Cherry201212}.

Mice receiving doses as low as 0.1 Gy showed alarming amyloid accumulation, behavioral deficits, and reduction in cognitive abilities. Mice do not possess the genes capable of developing symptoms related to AD. However, there are genetically modified mouse models to express proteins associated with this disease \citep{Cherry201212}. One of them is the protein ApoE-4 which tends to produce amyloid-beta accumulation that precedes the symptoms of AD \citep{Donohue20172305,Corder1993921}. A study by \citet{Villasana2011567} with male transgenic mice expressing ApoE-4 evidenced deficiencies in spatial memory after 3 Gy radiation exposure. \citet{Poage2008167} further reported deficits in displaying the NOR test in female transgenic mice expressing ApoE-4 after radiation exposure of 2 Gy. 

\citet{Rudobeck201712} evaluated spatial learning and memory with the water maze (WM) test and amyloid-beta deposits in 3-month-old APP/PSEN1 transgenic (TG) and wild-type (WT) mice irradiated with protons (150 MeV; 0.1$\pm$1.0 Gy; whole body). After irradiation, TG mice performed significantly worse than WT mice in the WM test. Also, TG mice exhibited an increase in amyloid-beta deposits in the brain cortex. They concluded that although irradiation with protons increased A$\beta$ deposition, the complex functional and biochemical results indicate that irradiation effects are not synergistic to AD pathology \citep{Rudobeck201712}. Nevertheless, Alzheimer’s disease is another concern that should be taken into account and needs further studies for long-term space missions.

\subsubsection{The Ocular System}
\citet{Chalupecky-1897} started the first studies of the effects of ionizing radiation on the eye. Since then, subsequent investigations have tried to demonstrate the possible harmful impairments on the ocular structures as a warning to therapists \citep{Vaeth1972346}. Later  \citet{Rohrsehneider1929} listed the oculars structures according to their sensitivity to radiation, with the lens being one of the most sensitive, followed by the conjunctiva, cornea, uvea, retina, and optic nerve. The use of radiotherapy to treat different conditions in the ocular structures, such as ocular lesions, non-neoplastic tumors, eye tumors, and orbit tumors, is well-known. Radiation doses vary between 6-12 Gy for a single dose, and between 20 - 100 Gy for various doses during periods of several weeks (more than two weeks); however, the appearance of side effects after several weeks or even years of exposure is a current concern \citep{Vaeth1972346}. Possible damages are erythema, telangiectasis \citep[chronic dilation of the capillaries and other small blood vessels][]{Flores2011163}, lacrimal gland atrophy, hyperemia  \citep[increased amount of blood in the vessels of an organ or tissue in the body, see][]{Bliss19984}, thinning of the cornea, and cataracts. According to the International Commission of Radiological Protection (ICRP), cataracts have become a growing concern due to the low radiation threshold necessary to detect lens opacities. For a single acute exposure, only 0.5–2.0 Gy is required. For highly fractionated or protracted exposures, 5 Gy is required. Annual dose rates for yearly highly fractionated or protracted exposure for many years are $>$0.1 Gy/a; however, the cataract formation is a late effect \citep{Fish2011}. Even some epidemiological studies from Hiroshima and Nagasaki and children survivors from Chernobyl have demonstrated that lower doses can trigger detectable opacity \citep{Ainsbury20091}.

NASA's Lifetime Surveillance of Astronaut Health (LSAH) data provides a historical record of cataract incidence in astronauts, and the NASA Study of Cataracts in Astronauts (NASCA) aims to precisely determine the types, severity, and progression rates of lens opacification in astronauts. It was a five-year longitudinal study where the status of the lens and visual function was measured in 171 astronauts (active and retired) of two control groups: military aviators (active and retired) and people without any military aviation background \citep{Chylack200910}. The Nidek EAS 1000 Lens Imaging System \citep[an imaging system for the anterior eye segment, which offers the possibility of recording Scheimpflug and retro illumination images][]{Wegener199255} measured the status of the nuclear, cortical, and posterior subcapsular (PSC) lens of the subjects. Other characteristics measured at baseline were age, demographics, general health, nutritional intake, and solar ocular exposure \citep{Chylack200910}. The study led to the conclusion that the variability and median of cortical cataracts were significantly higher for exposed astronauts than for the control group. Within the astronaut group, PSC opacity was more significant in subjects exposed to higher space radiation doses. There was no evidence of an association between space radiation and nuclear cataracts. These results suggest increased cataract risks at smaller radiation doses than have been reported previously \citep{Chylack200910}. \citet{Cucinotta2001460} discussed their findings in another study, with 295 astronauts participating in NASA’s Longitudinal Study of Astronaut Health (LSAH) and individual occupational radiation exposure data. The results suggest an increased risk of cataracts for astronauts with higher lens exposure doses ($>$8 mGy) compared to other astronauts with lower exposures.

Another interesting phenomenon is the observation of phosphenes by crew members of the Apollo missions 11 to 17 \citep{Pinsky1974957}. Phosphenes are the visual sensation as flashes of light without light entering the eye induced by various stimuli \citep[mechanical, electrical, magnetic, ionizing radiation, and some others, see][]{Bokkon2008168}. Apollo 11 Lunar Module pilot Edwin Aldrin first described these flashlights, and subsequently, the Apollo 12 and Apollo 13 crews were briefed on the phenomena and asked to report their observations. After these reports, Apollo 14, Apollo 15, and Apollo 16 dedicated a one-hour session to observe these flashlights with a simple blindfold (designed to avoid pressure on the eyeballs). In the end, three types of flashes were reported: the “spot” or “starlike” flashes, the most common ones, the “streaks” (most likely caused by particles with trajectories approximately tangent to the retina), andbflashes referred to as “clouds” that always were seen in the eye periphery. On average, the flashed were reported 19.3 minutes after starting the darkness adaptation phase. The Apollo 16 and Apollo 17 missions added a device known as the Apollo Light Flash Moving Emulsion Detector (ALFMED) to understand the origin of this phenomenon \citep{Pinsky1974957}. \citet{Chapman1972} suggest that the flashlights observed might be due to primary cosmic rays passing the vitreous humor generating Cherenkov radiation or by direct ionizing interactions with the retina. Over 80\% of astronauts from NASA and ESA (European Space Agency) programs have perceived phosphenes, at least in some missions \citep{Sannita20062159}. It is necessary to study in-depth the possible long-term effects of this particular phenomenon.

\subsubsection{The Thyroid Gland}
The thyroid gland is a human organ positioned in front of the trachea and under the larynx, which is characterized by the presence of two lobes that are interconnected by the presence of the central isthmus. It is covered by two capsules, one given by the pretracheal layer of the deep cervical fascia and another by the peripheral gland of the glandular tissue. From a microscopic point of view, the thyroid is organized in round-shaped structures called folliculi. The cells creating these structures are called follicular cells, responsible for synthesizing mature thyroid hormones. The thyroid hormones T3 and T4, synthesized from iodine and thyroglobulin, have several effects on the human body: they increase the basal metabolic rate (an increase of oxidative processes), the heat production, the oxygen delivery to the tissues, and the intestinal absorption of folates and B12. 

Exposure to ionizing radiation leads to DNA damage and increases the rates of thyroid cancer. Thereby, the thyroid gland is one of the most radiation-sensitive organs in our body. Even doses as small as   50-100   mGy   have   been   associated   with   an   increased   risk   of   thyroid malignancy in children \citep{Sinnott2010756}. During a CT scan, the thyroid is exposed to 15.2-52 mGy, a radiation dose that would increase the cases of thyroid malignancies in up to 390 per million exposed people \citep{Mazonakis20071352}. Most of the current information derives from past nuclear accidents, as they happened in Hiroshima (1945) and Chernobyl (1986). Although the radiation doses were high in both cases, considering the currently recommended human health threshold, the radiation characteristics differed in both scenarios. In the Hiroshima bombing, people were subjected to rapid irradiation with high-energy gamma rays and neutrons, while in the Chernobyl accident, many people, especially in Belarus, were invested with beta and gamma rays and 
different iodine radioisotopes. 

A   study   conducted   on   Hiroshima   survivors   demonstrated   that   among   105,401 subjects, 371 thyroid cancers were identified between 1958 and 2005, and the excess risk was higher for people ten (or less) years old at the time of the bombing \citep{Iglesias2017180}. In the same study, it was defined that a mean dose as low as 0.05-0.1 Gy may significantly increase the risk of thyroid cancer, a correlation that continues being linear up to doses as high as 20-29 Gy. Different studies linked the Chernobyl disaster to an increased incidence of thyroid cancer in children. For example, \citet{Acar2011} analyzed about 4,000 cases of thyroid cancer in 0-18 year old participants between 1992 and 2000. About 75\% of these cases occurred in children aged 0-14 years. Ionizing   radiation (IR)  may   induce   specific   genetic   instability   in   human   thyroids.   In particular, two genetic alterations have been intensively studied, and they both belong to the same intracellular signaling pathway: 
\begin{enumerate}
    \item ET/PTC, the genetic   translocation   induced   by   IR,   whose   variant   1   is characterized  by  a centromeric  inversion   on  chromosome  10, which has  been observed in children exposed to the Hiroshima and Nagasaki bombings. Its presence is correlated to a more aggressive cancer phenotype \citep{Nagataki1994364}.
    \item BRAF/AKAP9, the paracentric inversion on chromosome 7q, which impairs the autoinhibitory structure of BRAF, a portion of the protein that also contains RBD (Ras Binding Domain), leading to a Ras-independent activation of BRAF. It has been described in two cohorts of Belarussian children and adolescents exposed to radiation after the Chernobyl accident \citep{Ciampi200594}.
\end{enumerate}

\subsubsection{The Skeletal System}
The skeletal system of the human body is vital for structure and movement, and damage to bones underlies several medical conditions, including osteomalacia (bone softening due to malnourishment). An increased risk of bone fracture is a concern for cancer patients post-radiotherapy. Following radiotherapy treatment, radiation damage to the healthy skeletal tissue near a tumor can render it more vulnerable to fractures. For example, in breast cancer patients receiving radiotherapy, post-treatment rib fracture rates can be as high as 19\%. The standard radiation therapy for breast cancer patients involves X-rays and gamma rays. Nowadays, the experimental use of proton therapy occurs more frequently. Radiotherapy can impede bone structure integrity through loss of minerals and damage to the essential spongy bone tissue (trabecular structures). A paper by \citet{Baxter20052587} investigated the relative bone fracture risk of three cancer types in women that are commonly treated via radiation therapy: cervical, rectal, and anal cancer. The study looked at women over the age of 65 and showed that women who were treated for cervical cancer by radiation therapy had a resulting increased relative risk of hip fracture of 65\%. Further, the study subsequently identified the increased relative risk of hip fractures for rectal and anal cancers to be 66\% and 214\%, respectively. 

\citet{Willey201154} have made significant advances in investigating the possible effects of space radiation on bone health. They compared the radiation exposure of astronauts during spaceflight to that of cancer therapy patients. Radiotherapy as a cancer treatment typically involves delivering fractions of 1.8-2.0 Gy directly to the tumor mass for a few minutes. The total size and number of fractions delivered depends on the cancer type. For example, gynecological cancers are treated with a total of 54 Gy delivered over six weeks in fractions of 1.8 Gy at a time. For comparison, in-flight astronauts exposed to a solar particle event lasting 8-24 hours would receive whole-body cumulative doses of 1.0-2.0 Sv or 1.0-2.0 Gy of protons \citep{Jones202039}.

The radiation doses that cancer radiotherapy patients receive are higher than what astronauts are expected to be exposed to in spaceflight. This somewhat weakens the comparison of radiotherapy health risks to space radiation health risks. However, spaceflight-relevant radiation doses have been shown to induce bone loss. At proton doses as low as 0.5 Gy, bone loss persists nine weeks after spaceflight-relevant radiation exposure. At a dose exposure of 1 Gy, bone loss persists for four months following the exposure. Thus, the 1.0-2.0 Gy of protons that astronauts could be exposed to during a solar particle event could ensue long-term adverse effects on bone health \citep{Willey201154}.

Two major effects of spaceflight-relevant radiation exposure are a decrease in osteoblast activity and an early increase in osteoclast activity. Osteoblasts are large cells containing a single nucleus that function in synthesizing bone. Bone synthesis, however, requires a number of individual osteoblasts to work as a group of connected cells. Radiation exposure has been shown to damage osteoblast precursors and osteoprogenitors (cells that develop into osteoblasts). Osteoprogenitors are specifically damaged by the oxidative stress of irradiation, which impairs the ability of the mesenchymal stem cells to undergo osteogenic differentiation. The impact of radiation-induced oxidative stress makes research on antioxidants of target interest. Antioxidants such as beta-carotene have been seen to prevent radiation-induced oxidative stress damage \cite[e.g.,][see also Section~\ref{sec:sec6} for further information]{Bairati20165805}. The process of osteoblast differentiation involves a key transcription factor, RUNX2. Levels of RUNX2 are seen to decrease in response to radiation exposure, suggesting a weakened ability of osteoblast precursors and osteoprogenitors to develop into fully functioning, bone-synthesizing osteoblasts. Irradiation results in a decline in the number of mesenchymal stem cells able to respond to osteogenic stimulation, which delays the recovery of the bone from the osteoblast damage. Osteoblast damage has other consequences on bone health, including reducing the size of the blood vessels that supply oxygenated blood to the bones, which can cause hypoxia \citep[inadequate oxygen supply; e.g.,][]{Willey201154}.

MC3T3-E1 is an osteoblast precursor cell line derived from the calvaria of Mus musculus (house mouse). These MC3T3-E1 cells can undergo differentiation to form osteoblasts. A study by \citet{Kook2015255} demonstrated the impaired capacity of MC3T3-E1 cells to differentiate into osteoblasts following radiation exposure. MC3T3-E1 cells were exposed to radiation doses ranging from 0-8 Gy (spaceflight-relevant doses). The study produced results on the specific effects of irradiation on MC3T3-E1 cells. The study reported that irradiation enhances the biochemical Nrf-2/HO-1 pathway, which interferes with the role of  transcription factor RUNX2 to drive osteoblast development. Osteoclasts are large multinucleated cells that damage bone tissue in response to, for example, a bodily demand for calcium. The early response of osteoclasts to irradiation is the opposite to that of osteoblasts, with an initial increase in osteoclast activity. The dissolution of bones by osteoclasts involves an enzyme, tartrate-resistant acid phosphatase (TRAP5b). The prevalence of TRAP5b in circulation can be measured using serum markers. In response to radiation exposure, it is observed that serum markers for TRAP5b are elevated in the first 24 hours \citep{Willey201154}. The study further identified bisphosphonate antiresorptive risedronate as a potential pharmacological agent for suppressing radiation-induced increases in osteoclast activity, thereby suppressing the resulting damage to the bone tissue.

\subsubsection{The Immune System}
There are many different types of cells in the human body, each with its defining characteristics. The regeneration rate of a particular cell type influences its sensitivity to radiation-induced damage. Immune cells (white blood cells) are one of the most radiation-sensitive cell types in the human body and have one of the highest regeneration rates. In comparison, nerve cells have one of the lowest regeneration rates and show less radiation sensitivity than immune cells. Neutrophils are the most abundant type of leukocyte (immune cell), accounting for 60-70\% of the whole leukocyte pool in the body. Neutrophils regenerate every 1-5 days. Neutrophils are a crucial part of the body’s immune system and can engulf and digest pathogenic microorganisms through phagocytosis. The neutrophil releases antimicrobial products such as hydrolytic enzymes and reactive oxygen species to kill the phagocytosed pathogen. These antimicrobial products can cause damage to the surrounding healthy human tissue if their production is unregulated. To reduce the detrimental effect on the surrounding tissue, a neutrophil undergoes apoptosis (programmed cell death) after successfully phagocytosing and killing a pathogen. This means that the pool of neutrophils needs to be continuously replenished, resulting in a high neutrophil regeneration rate. Cells that regenerate more frequently are replicating their DNA more frequently, making it more likely that harmful radiation-induced DNA mutations (see DNA damage section) will be expressed. 

The effect of space radiation on astronauts' immune systems is a big concern in space travel science. Due to the rapid regeneration rate of leukocytes, the immune system is particularly susceptible to local and systemic damage by particle irradiation. A subtype of leukocytes, lymphocytes, show increased chromosomal aberrations following long-term space missions in low Earth orbit, such as in the International Space Station \citep{FernandezGonzalo2017}. Lymphocytes are critical to the adaptive immune response, a robust immune response that allows the immune system to remember previous infections. \citet{Gridley200255} carried out Earth-based experiments on the response of the immune system in mice to proton irradiation. %, as protons make up a significant proportion of deep space radiation \citep{Gridley200255}. %It is estimated that if a solar particle event lasting 8-24 hours were to occur whilst astronauts were travelling in deep space, the exposure of the in-flight astronauts would be 1.0-2.0 Gy \citep{Jones202039}. 
In their experiments, \citep{Gridley200255} delivered total doses of 0.5, 1.5, or 3.0 Gy in fractions of either 1cGy or 80cGy per minute. It was observed that such irradiation produced a linear or dose-dependent decline in the leukocyte and lymphocyte populations. B cells were identified as the most radiation-sensitive leukocytes, followed by T cells, and natural killer cells as the least radiation-sensitive leukocytes. B and T cells are adaptive immune cells, while natural killer cells can kill virally-infected and cancerous cells. Cancer risk is increased in astronauts following radiation exposure (see increased cancer risk section), which could be significantly worsened following a decline in T cells and natural killer cells. The simulated deep space radiation not only affected leukocyte population size but also impacted the functional capabilities of the immune cells \citep{Gridley200255}. \citet{Chancellor20184} discussed the limitations of using Earth-based radiobiology studies to predict the health risks of space radiation for astronauts. Such limitations include the discrepancies between animal and human models, difficulty simulating the distribution of radiation exposure across the entire body, and the discrepancies between terrestrial and space radiation types.

Research on the impairment of leukocyte function by irradiation is compiled from various journal articles by the International Atomic Energy Agency. The compiled research shows the effects of different levels of radiation exposure. After exposure to 5 Gy for an hour, concentrates of lymphocytes in saline had a significant fall in ATP levels, indicating implications for the metabolic and energy processes of the cell. At exposure to 10 Gy, leukocyte cytoplasms show increased vacuolization, indicating increased cell death. At 20 Gy, radiation exposure is high enough to eliminate leukocyte mitotic ability, impairing the ability of leukocytes to replicate. A subtype of leukocytes called granulocytes seems less susceptible to radiation damage than other leukocytes. Granulocytes are immune cells releasing granules containing antimicrobial chemicals to kill infecting pathogenic microorganisms. At 30 Gy, lymphocytes show decreased responsiveness to phytohemagglutinin; a chemical that promotes lymphocyte activation and proliferation \citep{Castelino1997}. The inflammatory response is part of the immune system that is of concern for astronauts. \citet{Hayashi2003129} took blood samples from 453 atomic bomb survivors and assessed the plasma levels of C-reactive protein (CRP) and interleukin-6 (IL-6), both of which play key roles in the inflammatory response. The study showed that exposure to just 1 Gy produced a 28\% increase in the plasma levels of CRP (p-value of 0.0002) and a 9.8\% increase in the plasma levels of IL-6 (p-value of 0.0003). The inflammatory response is a quick response of the body to infection, characterized by increased blood flow to the infected area to allow higher numbers of immune cells to reach the infection site. Inflammation, when unregulated, can damage healthy body tissue and have pathological consequences. Even low levels of the ionizing radiation-induced inflammatory response are accepted as a risk factor for developing both cancerous and non-cancerous diseases such as cardiovascular disease. 

\subsubsection{The Cardiovascular System}
Nuclear bombs generate high-intensity fluxes of x-rays, gamma rays and neutrons, as well as other radioactive nuclei. Cardiovascular disease is the world’s leading cause of death and is of particular concern in cancer patients receiving radiotherapy to the thoracic area of the body. It is seen in breast cancer patients that a dose of $<$2 Gy can increase the risk for the later development of radiation-induced cardiovascular disease. Results from survivors of the Chernobyl accident have shown that a dose as low as 0.15 Gy can significantly increase one’s chances of developing radiation-induced cardiovascular disease \citep{Hughson2018167}. The Chernobyl disaster resulted in the release of several radioactive elements, with the radioactive isotopes of iodine-131 and caesium-137 being two of the most prominent and most harmful to those exposed \citep{OECD2002Chernobyl}.  The Life Span Study of individuals who survived the atomic bombings of Hiroshima and Nagasaki in 1945 indicates that ischaemic heart disease and hypertension (high blood pressure) are the two most common late effects seen in survivors of radiation exposure, and both are involved in cardiovascular disease \citep{Hughson2018167}. 

The International Space Station sits in the LEO where, while it is not protected from space radiation by Earth’s ozone-rich atmosphere, it is protected by the Earth’s magnetic field. The only humans to have traveled outside of the realms of the magnetic field are the Apollo astronauts. \citep{Delp201629901} investigated the effects of deep space radiation on astronauts in relation to cardiovascular disease. The study compared the cardiovascular disease mortality rates of astronauts who had never flown orbital missions in space, astronauts who had flown in low-Earth orbit (LEO), and Apollo lunar astronauts. The study uncovered that the cardiovascular disease mortality rate was 4-5 times higher in Apollo lunar astronauts than in LEO or non-orbit astronauts. The study found that the cardiovascular disease mortality rate did not differ between low-Earth orbit and non-orbit astronauts, but that the Apollo lunar astronauts showed a significantly higher cardiovascular disease mortality rate. However, the study by \citet{Delp201629901} has been opposed by other publications such as \citet{Cucinotta201653} highlighting downfalls in the methodology of the study. In addition, \cite{Elgart20188480} investigated the excess risk of cardiovascular disease mortality from space radiation exposure and found no significant correlation.

It is interesting to investigate the cardiac abnormalities observed in Apollo astronauts when considering this increased risk of radiation-induced cardiovascular disease. As is noted in Section~\ref{sec:sec2}, the Apollo astronauts displayed some homeostatic disruptions, including inflight cardiac arrhythmias. Cardiac arrhythmia is a condition whereby the heartbeat rate is irregular, either too slow or too fast. Throughout the Apollo 15 mission, the Lunar Module Pilot showed arrhythmias, whilst the Command Module Pilot did not.  After docking with the Command Module Pilot, the Lunar Module Pilot showed abnormal cardiac behavior. Firstly, at 178 hours of Ground Elapsed Time (GET), there were five premature ventricular contractions experienced by the Lunar Module Pilot in just 30 seconds. Premature ventricular contractions occur when the wrong "pacemaker\footnote{A pacemaker is a heart structure that regulates the timing of the heartbeats.}" initiates the heartbeat. Premature ventricular contractions are induced when the Purkinje fibers initiate the heartbeat instead of the sinoatrial node, which is the heart's normal pacemaker. A single premature ventricular contraction is not considered to be dangerous, but when they become frequent, they are of more serious concern and can lead to arrhythmia-induced cardiomyopathy\footnote{Cardiomyopathy is a disease that damages the heart muscles and impedes the ability of the heart to pump blood.}. Later at 179 GET, the Lunar Module Pilot showed a sudden onset of a nodal bigeminal rhythm, an abnormal or “off-beat” heart rhythm. Before the onset of the bigeminal rhythm, the Lunar Module Pilot’s heart rate peaked at 120 beats per minute. The heart rate then slowed to 95 beats per minute immediately preceding and during the arrhythmia \citep{Johnston1975}. An adult's average resting heart rate is between 60 to 100 beats per minute. 

Different types of radiation are observed to instigate different biological effects in many tissue types, including the cardiovascular system. HZE particles (abundant in space radiation) alter gene expression, which can impair cardiovascular function and induce harm. Exposure to HZE particles is also shown to impair the production of new blood vessels, the physiological process of angiogenesis. Studies have observed these effects by HZE particles to be potent at low-to-moderate doses exposures. Neutrons are shown to damage cardiovascular tissue more severely than ground-based gamma rays. High-LET particles (such as alpha rays, protons, and neutrons) are shown to induce damage to endothelial cells, which can trigger the immune system’s inflammatory response. Inflammation is an effective method of fighting off infections, but when it is not needed, it can cause damage to human tissue. Thus, this can lead to the subsequent development of radiation-induced cardiovascular disease if cardiac tissue is damaged. In the inflammatory response, immune cells commonly utilize reactive oxygen species to kill infectious microorganisms. One of the main factors in radiation-induced damage to the cardiovascular system is oxidative stress. This has led scientists to propose a diet rich in nutrients that counteract oxidative stress as a preventative- and counter-measure of radiation-induced cardiovascular disease. For example, nitrate, nitrite (both from green vegetables), and nicotinamide riboside (from milk and yeast-derived foods) facilitate the production of nitric oxide, which can counter oxidative stress, and lycopene found in tomatoes is an antioxidant nutrient with additional radioprotective properties \citep{Hughson2018167}. See more on dietary mitigation strategies in Section~\ref{sec:sec6}. 

Space science and medicine are undoubtedly intertwined. It is challenging to study the effects of space radiation on astronauts without comparison to the effect of radiotherapy on cancer patients. The Childhood Cancer Survivor Study followed a group of adults who received radiotherapy treatment for childhood cancer between 1970 and 1986 and survived at least five years after the treatment. The study reported that patients who had received radiotherapy to the chest had a significantly increased risk of developing illnesses such as cardiovascular disease and cerebral infarction (stroke). It was also reported in the study by \citet{Marmagkiolis2016} that for patients who underwent radiotherapy between 1960 and 1995, cardiovascular disease was responsible for 9-16\% of mortality. Hodgkin’s disease is a cancer of the lymphatic system and is commonly treated with radiotherapy. Specifically, Hodgkin’s disease can be treated by external beam radiation therapy methods such as conventional radiotherapy and intensity-modulated radiation therapy (utilizing X-rays and gamma rays), but proton radiotherapy is becoming more common \citep{Chera20091173}. Following patients who received radiotherapy for Hodgkin’s disease between 1960 and 1998, 10\% clinically presented with apparent coronary artery disease at a median time of 9 years following radiotherapy, and 6.2\% clinically presented with significant valvular dysfunction at a median time of 22 years following radiotherapy \citep{Marmagkiolis2016}. Both are involved in the development of cardiovascular disease. 

\subsubsection{The Pulmonary System}
Over the next few decades, there are plans to try and land the first humans on Mars. The NASA Artemis missions aim to establish sustainable crewed exploration of the Moon by the end of the 2030s in preparation for sending astronauts to Mars later. The Martian surface poses many environmental threats to human health, including its high radiation levels and the Martian soil, which contains high levels of perchlorate. Perchlorate compounds are toxic to humans, particularly to the thyroid gland. However, some bacterial lifeforms have shown developed tolerance to perchlorates \citep{Wadsworth20174662}. Also, perchlorate-reducing bacteria like \textit{Dechloromonas genus} and \textit{Wolinella succinogenes} have been discovered \citep{Shrout2002261}. The Martian surface sees irradiation by high levels of UVB and UVC, the more dangerous ultraviolet rays that the Earth’s surface is predominantly protected from by ozone. Perchlorate irradiated by the high UV levels of the Martian surface produce increasingly more dangerous chemicals, including hypochlorite and chlorite, which were shown to be lethal to the previously described perchlorate-tolerant bacterial cells \citep{Wadsworth20174662}. Martian soil is of particular concern for the pulmonary (lung) health of astronauts due to its hazardous chemical composition, fine dust particles, and interactions with UV radiation.

A study by \citet{Lam2008901} exposed laboratory mice to low and high doses of Martian soil simulant, with follow-up examinations either 7 or 90 days post-exposure. The low dose of the Martian soil simulant was 0.1 mg/mouse, while the high dose was 1 mg/mouse. The examination that followed seven days after the exposure showed observations of mild fibrosis (scarring) at various points throughout the mouse lungs, and particle-laden macrophages\footnote{Macrophages are immune cells that function in phagocytizing pathogens, and the lungs have their own macrophages called alveolar macrophages.}. The examination that took place 90 days following Martian soil simulant exposure showed significant amounts of particle-laden macrophages, mild-to-moderate inflammation of the alveoli (the site of oxygen and carbon dioxide gas exchange), as well as inflammation surrounding the pulmonary blood vessels and bronchioles, and debris within the alveoli \citep{Lam2008901}. These are all factors that significantly reduce the correct functioning of the lungs.

Radiotherapy is commonly used to treat non-small cell lung cancer (NSCLC), which makes up around 80\% of lung cancer cases. Proton therapy is shown to induce less damage to the nearby tissue, such as the heart and the esophagus, than standard photon therapy \citep{Auberger20073}. One of the main concerns of radiotherapy for lung cancer is the development of radiation-induced lung fibrosis\footnote{Fibrosis is the laying down of scar tissue and can result from a sustained inflammatory response whereby immune cells and fibroblast cells attempt to fight off an infection or heal an injury \citep{Ding20131347}.}, which clinically manifests as a dry cough, dyspnea, and severe respiratory insufficiency. Scar tissue is laid down when an area is too damaged to return to healthy tissue and impedes the ability of that organ or tissue to carry out its original function. Damaged lung tissue that is replaced by fibrotic scar tissue will not be able to provide the adequate oxygen and carbon dioxide gas exchange needed to supply the body. As discussed previously, radiation can induce oxidative stress, which can mimic some of the antimicrobial defenses characteristic of the inflammatory response. 

\citet{Oh2012343} analyzed radiation-induced lung fibrosis in 48 NSCLC patients treated with postoperative radiation therapy (PORT) and not receiving simultaneous chemotherapy. The radiation dose delivered to these patients was in typical fractionations of 1.8-2.0 Gy/day, with a total overall dose ranging from 44 Gy to 65 Gy. The study used some key parameters to analyze the appearance of radiation-induced lung fibrosis. ‘Vf’ was the fibrosis volume, and ‘V-dose’ was the volume of lung tissue that received more than the reference radiation dose. The results showed a significant correlation coefficient of 0.602-0.683 (p < 0.01) between the dosimetric parameters of the fibrosis volume and the V-dose, meaning that as the volume of lung tissue receiving the radiation dose increased, so did the level of fibrosis or scarred lung tissue. A strong correlation coefficient of 0.726 (p < 0.01) between the mean lung radiation dose and the fibrosis volume was also found. The study's main conclusion was that the fibrosis volume increased continuously with the V-dose, as the reference radiation dose increased. Despite these results, caution must be taken when using radiotherapy and historical radiation disaster studies to predict the health risks of space radiation exposure for astronauts. Lifestyle factors such as underlying health conditions, physical fitness, age, and smoking history cannot be ignored in terms of their potential influence \citep{Cucinotta201102}.

\subsubsection{The Integumentary System}
The integumentary system of the body refers to the skin, attached hairs, and nails. In other animals, it includes appendages such as feathers, fur, scales, and hooves. The integumentary system provides a protective barrier between the body and the outside environment. For example, the skin provides a barrier to pathogenic microorganisms. Because the top layer of skin (the epidermis) is composed of dead cells and sheds regularly, microbes cannot infect it. Additionally, the presence of keratin and fatty acids on the skin creates a dry and acidic environment unfavorable to many pathogenic microorganisms. The skin has many other roles, but notably, in humans, the skin is essential for allowing the body to synthesize vitamin D, which, within the skin, is produced through interactions of ultraviolet radiation with compounds such as 7-dehydrocholesterol. Vitamin D is required for healthy bones and other body tissues. While exposure to UV radiation is helpful in vitamin D synthesis, it is also a significant risk factor for skin cancers. 

Skin cancer is one of the most highly prevalent forms of cancer, with over 3.5 million cases diagnosed annually in the US. On Earth, the ozone-rich atmosphere absorbs most of the harmful UV radiation and prevents it from reaching the surface. The much greater UV radiation exposure accompanying interplanetary space travel and Mars exploration threatens astronaut health. There are three main types of skin cancer: basal cell carcinoma, squamous cell carcinoma, and melanoma. Both basal and squamous cell carcinoma arise from keratinocytes (skin cells that produce keratin) and are less invasive and aggressive than melanoma. Melanoma arises from melanocytes (skin cells that produce melanin pigment) and is the most aggressive form of skin cancer, responsible for most skin cancer deaths. Melanoma is more harmful as it has often metastasized to other parts of the body (such as the brain or bones) by the time it is clinically presented. The risk of developing any of the three forms of skin cancer increases dramatically with UV radiation exposure \citep{degruijl19992003}. An individual’s risk of developing melanoma can be tripled with just one sunburning incident every two years. 

UV radiation exposure can cause a variety of genetic mutations that contribute to skin cancer. Basal cell carcinoma can be caused by mutations of the PTCH1 and PTCH2 genes, which lead to the activation of transcription factors, as well as genes that promote cell proliferation and angiogenesis \citep[blood vessel formation, e.g.,][]{Kim201911}. Mutations of the p53 tumor suppressor gene commonly accompany squamous cell carcinoma. \citet{Brash199110124} report the evidence that such p53 mutations are induced by UV exposure, as the mutation induced involves two cytosine bases being replaced by two thymine base pairs. This is a specific mutation attributed only to the action of UV radiation . The most common mutation associated with melanoma is a mutation of the CDKN2A tumor suppressor gene. A mutation of the CDKN2A gene can lead to uncontrolled cell proliferation. \citet{Monzon1998879} report that in families with a genetic predisposition to melanoma, 20\% have CDKN2A gene mutations. 

\citet{Kim-etal-2006} discussed the risk of skin cancer for future astronauts partaking in missions to the Moon and Mars. Considering that on Earth - with an ozone-rich atmosphere to absorb a lot of the harmful UV radiation - there are such high rates of skin cancer, the risks for astronauts traveling to Mars, which has an atmosphere of less than 1\% the thickness of Earths atmosphere, needs to be assessed. The risk of UV-induced skin cancer is significant in areas of the body that are exposed tomore sunlight \citep[e.g., face and hands, see][]{MyungHee20061798}. The risk is increased in individuals with fair skin, accompanied by an inability to tan, red or light-colored hair, and freckles. This is due to an associated lack of the melanin pigment that produces the color in darker-skinned individuals \citep{Sturm2002405}. There are a variety of hardware mitigation strategies that would allow for blocking UV radiation exposure; however, this would likely impact vitamin D levels in astronauts as well and lead to another health consideration. The study by \citet{Kim-etal-2006} reports that astronauts on the Lunar surface during a solar particle event could have greatly increased risks of skin cancer development. The study also evaluated radiotherapy cases, with a study of patients with tinea capitis (fungal infection of the scalp) treated with radiation therapy who received around 0.1 to 0.5 Gy (X-rays) to the face and neck region. There was a strong correlation between ionizing radiation exposure and the development of basal cell carcinoma in these patients \citep{MyungHee20061798}. Ionizing radiation exposure has also been observed to induce hair loss and skin erythema (reddening of the skin) in cancer patients \citep{Butler2006517}. 

\citet{Wao2014890} analyzed numerous skin samples from female C57BL/6 mice after being launched in space for 13 days inside animal enclosure modules (AEMs). Several genes involved in reactive oxygen species (ROS) metabolism were upregulated (Als2, Cat, Fmo2, and Noxa1). Also, there was an alteration in the number of genes responsible for antioxidant defense, like Ehd2, Prdx5, Ptgs2, and Gsr (Glutathione reductase - see DNA damage section). Also, genes involved in extracellular matrix remodeling were analyzed, and 11 of them appeared upregulated, while four appeared downregulated. In particular, the gene coding for collagen II alpha-1 was downregulated, while collagen IV alpha-1 was upregulated. At the same time, researchers found an increase in many integrins, MMP15, and Timp3, the inhibitor of metalloproteinase 3. Also, MMP15 belongs to the membrane-type metalloproteinases (MT-MMP) family, whose upregulation is a typical trait of several human cancers.

\citet{Su2010e16} created a 3D skin model system to conduct simulations of exposure to ionizing radiation and to assess the consequent DNA damage. For the experiment, scientists used MCR5 cells (lung fibroblasts) and Human EpiDerm, a skin model with only primary keratinocytes, and EpiDermFT, which contains fibroblasts too. These tissues were irradiated with a Cs-137 radioactive source delivering a dose of 0.82 Gy/min. This is the first example of a 3D artificial skin that appeared to be good for conducting radiation experiments for astronauts. The potential damage to the skin through increased exposure to UV and other ionizing radiation is a concern for astronauts and showcases the need for further research in this area.

\subsubsection{The Reproductive System}
\citet{OglivyStuart1993109} outlines some of the key effects of irradiation on the reproductive system in humans. In biological males, aspermia (lack of semen in an ejaculation) can result from direct exposure of the testis to a radiation dose of  $>$0.35 Gy, and aspermia can become permanent if the radiation dose is $>$2 Gy. At a considerably higher dose of  $>$15 Gy, the production of testosterone by Leydig cells can become impaired. In biological females, direct irradiation of the ovaries with a dose of 4 Gy may cause a 100\% incidence of sterility in women over 40 years of age and 30\% sterility in younger women. The reproductive glands, or gonads, can be directly affected by irradiation of the abdomen, pelvis, spinal cord, or testicles in radiotherapy. Cranial irradiation, such as in the treatment of a brain tumor via radiotherapy, has the potential to damage the hypothalamic-pituitary axis, which has multiple consequences, including the unusually early occurrence of puberty (precocious puberty), unusually high levels of prolactin in the blood (hyperprolactinemia), and deficiency of gonadotropin (hormones that act on the reproductive glands). Deficiencies of the pituitary gland are also observable following the irradiation of nasopharyngeal tumors, likely as a secondary result of damage to the hypothalamus. There are many interactions between the hypothalamus and pituitary gland that are crucial to the proper functioning of the male and female reproductive systems. 

Radiotherapy of women's craniospinal, abdominal, or pelvic regions can have consequences on fertility and pregnancy.  \citet{Wo20091304} reported a dose-dependent relationship between ovarian radiotherapy and the premature onset of menopause. Those undergoing radiotherapy of the craniospinal region had consequent impairment of hormonal regulation, leading to difficulty becoming pregnant later in life. Those undergoing radiotherapy to the abdominal or pelvic regions showed the consequences of birth abnormalities such as uterine and placental abnormalities, early labor, low birth weight, and miscarriages. \cite{Norwitz2001929} reported the case of a 23-year-old woman who experienced uterine rupture at 17 weeks gestation after a history of receiving whole-body irradiation therapy for the treatment of childhood leukemia. At 17 weeks gestation, after the woman reported acute abdominal pain, a pathological examination showed an atrophic (wasting away) uterus and myometrium that had thinned to about 1-6mm, being completely absent in some areas. The study hypothesized that the prior radiation therapy the woman had experienced as a child had led to uterine injury and subsequent thinning of the myometrium. The woman also showed clinical signs of placental percreta, an abnormal condition where the placenta implants deeply into the uterine wall and is not delivered normally during childbirth, which can be life-threatening. The post-pubertal uterus is a relatively radiation-resistant organ in adult women, but the study reports that high-dose radiation exposure in childhood can cause irreversible injury to the uterine musculature and vasculature. 

\subsubsection{The Digestive System}
The gastrointestinal (GI) system, composed of various organs and an associated microbiome, can be severely affected after exposure to ionizing radiation. The resulting adverse effects depend on both the dose and the organ involved. Doses as low as 1 Gy can diminish gastric motility (with delayed gastric emptying) and, in some occasions, induce sphincter incompetence, late suppression of gastric acid secretion, and release of neurohumoral factors and histamine, which contributes to diarrhea \citep{Jones202039}. One of the organs with moderate radiosensitivity of the GI system is the stomach, which sustains doses of up to 40 Gy without severe secondary complications. Adverse effects occur with doses greater than 45 Gy, where the risk of developing ulcerations, perforations, and obstructions increases to 25-50\% of cases. For exposures between 55 Gy and 64 Gy, this percentage can grow to almost 63\% of cases. The ulcerations appear after a month of exposure and present with symptoms such as intractable nausea, pain, vomiting, and hematemesis \citep[vomiting of blood, which may be red or have an appearance similar to coffee grounds, e.g.,][]{Wilson199085,Vaeth1972346}.

The intestine represents a particular interest as it is the portion most sensitive to radiation. Injury of the small intestine gives rise to more acute and long-term suffering and risk of death than any other abdominal viscera. Intestinal radiation toxicity is classified as acute (early) or chronic (delayed). Acute toxicity occurs during or within three months with a cumulative dose of- 45 Gy or daily doses between 1.5-2.0 Gy (5-day week schedule). It leads to epithelial injury, resulting in the breakdown of the mucosal barrier and mucosal inflammation (intestinal mucositis). The clinical presentation includes vomiting, abdominal pain, and diarrhea. It occurs in 60-80\% of patients who receive intraabdominal or pelvic radiotherapy \citep{Vaeth1972346,Hauer200723}. After treatment with high daily radiation fractions of 25-30 Gy, chronic toxicity begins. Some characteristics of the internal intestinal compartments change, generating mucosal atrophy, intestinal wall fibrosis, and vascular sclerosis manifested in malabsorption syndrome, dysmotility, transit abnormalities, vitamin B12 deficiencies, diarrhea, weight loss, and acute or intermittent small bowel obstruction [64][65]. Higher doses, like 50-60 Gy, can lead to Chronic Radiation Enteritis, with symptoms similar to 25-30 Gy doses. In both cases, symptoms can appear between six months and 20 years after radiotherapy or ionizing radiation exposure \citep{Theis201070,Hauer200723}.

After radiation exposure, ionizing radiation is also a risk factor for colorectal (CR) cancer due to a higher incidence of pre-malignant lesions, such as colonic polyps. Some NASA projects have tried to estimate the risk of space radiation-induced intestinal tumors. For example, a study by \citet{Fornace2008} employed the adenomatous polyposis coli (APC, a mouse model used to demonstrate radiation-induced intestinal cancer). They studied survival and the progression of colon cancer after full body exposure of APC mice to a simulated Solar Particle Event (SPE) with varying energies using a total dose of 2 Gy over 2 hours. Therefore, they used 2 Gy of monoenergetic proton for reference radiation exposure and X-rays at a dose rate of 20 cGy/min. As a result, the SPE stimulation induced an increase percent number of polyps\footnote{A polyp is defined as any mass protruding into the lumen of a hollow viscus \citep{Shussman20141}}, and invasive adenocarcinomas\footnote{a malignant epithelial tumor with a glandular organization \citep{NCBI132020}} compared to reference radiation exposure. These findings suggest that exposure to low dose rate SPE protons has significant biological effects that may have functional consequences on colon cancer progression \citep{Fornace2008}.

In addition, \citet{Suman-etal-2017} evaluated the intestinal and colon tumorigenesis in male and female APC mice exposed to different doses (10 or 50 cGy) of energetic heavy-ion radiation ($^{56}$Fe or $^{58}$Si) and monitored 150 days after radiation exposure. Intestinal tumorigenesis and intestinal tumor size in male and female mice were similar for both radiation types for high and low dose rates tested. Colon tumor frequency in male and female mice was also not significantly different after high and low dose rates of energetic heavy ions. In conclusion, intestinal and colonic tumor frequency and size were similar, suggesting that the carcinogenic potential of energetic heavy ions is independent of the dose rate.

The human microbiome is the collection of microbes that live in and on our bodies, and it plays an essential role in health \citep{Lorenzi2011}. As previously mentioned, the GI tract has a microbiome associated; it includes viruses, archaea, fungi, protists, and the main population of bacteria. The GI microbiome in each individual has vital functions, which include the production of vitamins, carbohydrate fermentation, absorption of nutrients, and the suppression of pathogenic microbial growth creating a more hostile environment \citep{Kennedy201410,Kumagai201810}. It also connects to other body systems, including the digestive, immune, endocrine, and, notably, nervous system, which is in the gut-brain axis. This axis influences many factors and disorders, including cognition and mental health \citep{Jones2020193}.

Dysbiosis is the disturbance and imbalance of the GI microbiome’s regular composition. It can occur after diet changes, enteric infections, antibiotic treatment, and radiation exposure \citep{Kumagai201810}. Dysbiosis triggers the release of pro-inflammatory cytokines that eventually aggravate mucosal damage. It also seems to be associated with several gastrointestinal diseases and systemic conditions such as Type I and II diabetes, autoimmune disorders, neurodegenerative diseases, obesity, and psychiatric episodes \citep{Jones2020193,Kumagai201810}. 

A study in fecal samples from 19 rhesus macaques, Macaca mulatta, exposed to 7.4 Gy cobalt-60 gamma-radiation, evidenced significant differences in the microbiome populations four days after irradiation. Ten animals demonstrated diarrhea after these days, revealing an increase in Lactobacillus reuteri, Veillonella sp., and  Dialister sp., and a decrease of Lentisphaerae and Verrucomicrobia phylum and Bacteroides genus in comparison with non-diarrhea animals. These differences show the potential association between the prevalence of microbiomes and differential susceptibility to radiation-induced diarrhea \citep{Kalkeri2020}. Studies on radiotherapy patients have also evidenced variations in microbial populations, and in some cases, the most considerable changes could be associated with episodes of post-irradiation diarrhea \citep{Kumagai201810}.

The spaceflight environment comprises several factors that can influence the GI microbiome, including microgravity and radiation \citep{Jones2020193}. An investigation at the International Space Station (ISS) evaluated if the GI tract, skin, nose, and tongue microbial communities changed in nine astronauts during a six-month mission. As a result, the main differences occurred in the GI tract. The researchers identified relative abundance and the acquisition or loss of bacteria species, besides the increase of some pro-inflammatory cytokines. On the other hand, nose, skin, and tongue microbiome composition presented fewer changes, which can be linked to astronauts’ hygiene habits. All the variations occurred very early during the space mission and were maintained for at least 60 days after the return to Earth \citep{Voorhies20199911}. These results are consistent with the NASA twin study. For this study, they compared a pair of identical twin astronauts from NASA, one of them spent 340 days on the ISS, and the other stayed on Earth. After this period, they evaluated changes in telomere length, gene regulation, gut microbiome composition, body weight, carotid artery dimensions, subfoveal choroidal and retinal thickness, serum metabolites, immune response, and cognitive performance \citep{GarrettBakelman2019364}. \citet{Siddiqui2020} suggest that space radiation is responsible for the gut microbiome alterations in astronauts since radiotherapy patients present similar symptoms after acute radiation exposure.

\subsection{The Impact of Ionizing Radiation on the Genetic Material}
\subsubsection{Reactive Oxygen Species}
After exposure to ionizing radiation, several oxidative events are prone to happen in human cells. The main biological consequences occur as a direct effect of electromagnetic waves on the cellular genetic information or due to the ionizing radiation-induced radiolysis of water \citep{Azzam201248}. This substance represents the majority of the intracellular environment \citep{Cooper2000}. There is also scientific evidence of specific mtDNA mutations linked to the alteration of the cellular oxidative metabolism causing a rise in ROS production \citep{Kawamura2018ii91,Hahn2019392}. Water can produce different reactive species when lysed by ionizing radiation and, considering an aerobic cellular environment in conditions of a physiological pH, the most relevant chemical compounds are O2-, OH, and H2O2 \citep[H2 tends to evaporate and does not have the same dangerous effects of the other cited chemical species, e.g.,][]{Azzam201248}. The combination of ROS, other organic radicals, and molecular hydrogen may lead to the synthesis of peroxides, responsible for lipid peroxidation. Ionizing radiation may also activate NO synthases \citep[NOS, e.g.,][]{Folkes201347} which produce nitric oxide, a chemical compound capable of interacting with superoxide radicals to produce peroxynitrite groups capable of interacting with other intracellular targets \citep{Jourdheuil200128799}.  ROS and RNS are often produced by activated M1 macrophages that make use of these chemical species to destroy pathogens, especially during chronic inflammations. There is strong evidence that chronic inflammations may lead to the onset of several human tumors \citep{Multhoff2012ChronicII,Singh20171} and, for this reason, the oxidative and nitrosative burst caused by ionizing radiation may represent a risk factor for cancer (discussed further in Section~\ref{sec:CancerRisk}).

In 1972, \citet{Harman1972145} proposed that mitochondria may represent a biological clock because they tend to accumulate genetic mutations, a process also accelerated by the relatively close distance from the Electron Transport Chain (ETC). mtDNA mutations can disrupt the structure of ETC complexes in the external mitochondrial membrane causing an undesired electronic leakage responsible for the conversion of O$_2$ into O$_{2-}$, especially considering complexes I and III. A well-known mtDNA mutation is called Common Mutation (CD), consisting of a 4.977 bp deletion that affects the genetic information for five different tRNA and several subunits of complexes I, III, and IV \citep{Nie20138}. There is evidence that CD can be induced by ionizing radiation \citep{Wang2007433}, and other genetic alterations have been observed using minor doses of EM waves (e.g., 4934del). An important substance involved in cellular redox homeostasis is glutathione (GSH), a tripeptide composed of glutamic acid, cysteine, and glycine. The enzyme glutathione peroxidase uses GSH as a coenzyme to convert a peroxide group in water and alcohol (2 GSH+ROOH $\rightarrow$ GSSG+ROH+H2O). For this reason, the ratio 2GSH/GSSG is usually used in medical biochemistry to evaluate the redox potential of the considered cell.  Also, GSH participates in several biological processes like sulfur-containing amino acids' metabolism, disulfide linkages' reduction, and the stabilization of ROS and RNS.  An critical research study led by \citet{Shimizu1998299} demonstrated the protective role of glutathione on the irradiated right hemicerebrum of white rabbits. Results showed that ionizing radiation causes an increase in the levels of GSH and $\gamma$-glutamylcysteine synthetase (y-GCS). X-rays with a dose rate of 3 Gy/min were used to irradiate the whole right hemicerebrum. With an intrathecal injection, S-methyl GSH (15.6 $\mu$mol/Kg) and buthionine sulphoximine (2.3 $\mu$mol/Kg), an inhibitor of y-GCS were administered in 400 $\mu$L of isotonic saline. After six hours from the administration and the radiation exposure, the brains were removed and examined. The quantitative method to assess DNA damage was analyzing the levels of 8-OHdG with HPLC and, with a 20 Gy dose, a peak was observed three hours following irradiation. 

Also, 8-OHdG seemed to be proportional to the radiation doses:

\begin{center}
    
	Normal control = 5.00 $\pm$ 3.12 fmol/mg DNA
	
	5 Gy = 10.5 $\pm$ 0.13 fmol/mg DNA
	
	10 Gy = 21.83 $\pm$ 0.13 fmol/mg DNA
	
	20 Gy = 31.85 $\pm$ 3.92 fmol/mg DNA

\end{center}
	
A Northern Blot was conducted to analyze the quantity of gamma-GCS mRNA, and the results showed an impressive increase 6 hrs following exposure to 20 Gy radiation (6600 $\pm$ 950 vs. 2200 $\pm$ 140 PSL, that is photostimulated luminescence). A radiation dose of 20 Gy resulted in a 173\% increase in GSH levels in the irradiated hemicerebrum, and a similar but lesser increase was observed in the non-irradiated left hemicerebrum. Finally, treatments only with S-methyl GSH 6 hrs following 20 Gy of radiation resulted in a 31\% decrease in the formation of 8-OHdG, while treatments only with BSO resulted in an increase of 8-OHdG by 193\%. 

\subsubsection{DNA Damage}
As previously discussed, astronauts are constantly exposed to ionizing radiation (IR), and this radiation can lead to several impairments of the human physiological systems. Regarding the cellular level, the type (high-LET or low-LET radiation) and IR energy critically affect how DNA is damaged. This, in turn, may influence the cell’s survival and how the damaged DNA is repaired. IR leads to breaks in the DNA, which can occur directly or indirectly \citep{Cannan20163}. DNA breaks are produced indirectly by low-LET radiation (gamma rays and X-rays), which distributes its energy in random motion around the DNA \citep{McMahon2019205}, splitting nearby water molecules and creating hydrogen and reactive oxygen species (ROSs). These highly reactive species may react with nearby DNA, producing closely opposed single-strand DNA breaks (SSBs). These lesions are referred to as “clustered damage”. \citet{Cannan20163} defined these lesions in their review as more than a single DNA lesion, created by a single track of radiation, that resides within one or two helical turns of DNA. Low or high LET radiation doses as low as 1 Gy (100 rad) can produce clustered lesions. Monte Carlo-based modeling of radiation tracks suggests that low and high LET radiation can generate up to 10 and 25 lesions per damage cluster, respectively [n]. It is believed that SSBs are independently created, and then they locate themselves in the same site on complementary strands, generating a DSB \citep{McMahon2019205}. The ROSs generate additional DNA damage, including oxidized bases and sites of base losses \citep{Cannan20163}.

High LET radiation is a particular risk for astronauts traveling beyond the protection of the Van Allen radiation belts and to cancer patients treated with heavy ion radiation therapy. This radiation deposits energy much more densely and leads to direct DNA damage with greater complexity, potentially in more significant amounts per unit dose, which the cell finds more challenging to repair \citep{McMahon2019205}.  Direct DNA damage entails a collision between a high-energy particle and a strand of DNA, breaking the phosphodiester backbone, which induces DSB. This leads to molecule fragmentation and gives rise to structural chromosomal aberrations \citep{Puerta202027}. DSBs are the most genotoxic events that can happen to the genetic material \citep{Chapman2012497}. In fact, the DNA repair system usually uses the complementary strand to insert the correct bases, but after a DSB the information is missing on both strands. Cells react to these genetic lesions with Holliday’s junctions or the NHEJ system, whose aberrant action has already been reported several times in cancerous cells. These mechanisms are highly mutagenic and may be responsible for the onset of genetic mutations causing diseases like cancers. 

Considering the previous image, showing the frequency and the number of DSB derived by a simulation with 10,000 100-MeV proton beams, the energy of the colliding particles is so high that in literature, there is evidence that they may cause non-random DSBs, less than 25 per considered cluster. The Van Allen Belt partially protects astronauts that stay in the terrestrial atmosphere, while astronauts who travel beyond may be exposed to higher doses of ionizing radiation, increasing the cancer risk too \citep{Schaefer1959631}. According to \cite{DahmDaphi200967}, the number of DSBs caused by ionizing radiation increases linearly with radiation in a range of several hundred Gy. When a DSB is created in human cells, a molecular complex called MRN binds to it and facilitates the activation of ATM. It is autophosphorylated, allowing the activation and the consequent phosphorylation of many targets in the surrounding chromatin \citep{Martin20102171}. A vital pattern in many early DSBs is the phosphorylation of the histone variant H2AX (also called $\gamma$-H2AX). When DNA Damage Response (DDR) proteins accumulate in the damaged sites, the formation of foci (also called IRIFs, Ionizing Radiation-Induced Foci) starts \citep{Goodarzi201239}. Usually, the biomolecular pathways started by DDR proteins lead to cell cycle arrest in G1/S or G2/M. Mammalian cells make use of two major pathways to repair DSBs, which are NHEJ and HR.

In the 1970s, sucrose gradient sedimentation was used to measure DSBs in irradiated mammals \citep{Vignard2013362} and was substituted by some electrophoretic protocols in the 1980s (e.g., PGFE and Comet Assay). Thanks to the advancements in genomics, it is possible to determine the non-random distribution of IR-induced DSBs across the genome and evaluate the heterogeneous cell response. Since proteins must accumulate to be detected, NHEJ is challenging to highlight because its proteins are usually expressed, and only 1-2 additional proteins are necessary to repair a DSB. Instead, HR can be detected thanks to the accumulation of proteins like RAD51, RAD52, or BRCA2, but it is not useful for non-cycling cells and makes up only 15\% of total DSBs. FRAP (fluorescence recovery after photobleaching) can be used to get information about the mobility and the binding rates of recruited proteins on DSBs. Also, it is possible to create miniaturized classic antibodies called scFvs (single-chain variable fragments) that consist of the variable regions of heavy and light chains linked to a peptide. Recently a phosphospecific scFv was created to detect IRIFs in living cells \citep{Vielemeyer200920791}. Using the CHO-K1 cell line and 1.0 Gy of 200 kVp X-rays, \citet{Mori2018253} managed to deduce the induction probability density of DSBs based on the measured DNA amount and the calculated dose per nucleus. This research considers the different phases of the cell cycle, and the variations of DNA quantities were analyzed via PI staining. Also, researchers analyzed the plateau and the logarithmic phase of the growth curve in the CHO-K1 cell culture. The number of IRIFs was estimated via fluorescence microscopy and flow cytometry, analyzing the presence of $\gamma$-H2AX with fluorescent primary antibodies. A Monte-Carlo simulation was then conducted to calculate the distribution of energy deposition per nucleus, deducing the DSB number per nucleus. 

A control group of CHO-K1 cells was cultured to deduce the background of DNA damage and the cell cycle distribution. In the Monte-Carlo simulation, to obtain the probability density of the number of DSBs per nucleus, the number j was assigned to each cell, and the following formula can approximate DSBs:

\begin{center}
        DSB$_{j}$=k $\times$ Dose$_{j}$ $\times$ DNA$_{j}$ $\times$ BG$_{j}$,
\end{center}

where k is the DSB induction yield ($\sim$30 [1/Gy/cell]), Dose$_{j}$ is the absorbed dose imparted to the nucleus [Gy], DNA$_{j}$ is the relative DNA amount, and BG$_{j}$ [1/cell] represents the number of DSB$_{j}$ before the irradiation. These two last parameters are derived from the quantitative analysis of the CHO-K1 cell control group. Also, to analyze the dependence of DSB induction on the cell cycle, two conditions were created:

\begin{center}

	Plateau phase = 78.9\% G0/G1, 12.6\% S and 8.48\% G2/M
	
	Logarithmic phase = 32.4\% G0/G1, 56.9\% S and 10.7\% G2/M
	
\end{center}
	
\citet{Mori2018253} showed that the energy deposition for the cell nucleus gave a Gaussian distribution with a peak at 1 Gy and a standard deviation of 0.22 Gy (plateau) and 0.23 (logarithmic). For the non-irradiated group, the number of foci by microscopy was 8.07$\pm$10.45, while for 1.0 Gy the irradiated group was 39.82$\pm$18.51. The samples are 1.0 $\times$ 10$^{5}$ and  2.0 $\times$ 10$^{5}$, respectively. When it comes to BGj, the distribution of the background DSB number per cell nucleus has two peaks through the analysis of DNA profile in the logarithmic growth phase: the lower is mainly composed of G1 and G2/M cells while the higher one was composed of S cells. According to their experiment the median value of DBS per cell was 37.31, and the lower and upper quartile values were 30.53 and 46.64 for the plateau phase, and 52.58, 41.98, and 66.14 in the logarithmic phase. The model strongly agrees with the experimental data for both the plateau and logarithmic growth conditions. New forms of quantitative analysis of DSBs have been developed. \citet{Li20114130} created a $\gamma$-H2AX/MDC1-luc 2 reporter gene inserted in lentiviral particles and vehiculated to cancer cells in order to determine the number of DNA breaks. The H2AX and BRCT domain of MDC1 were fused at the N-terminal and C-terminal of the firefly luciferase, which was separated in half. When a DSB occurs, H2AX is converted into $\gamma$-H2AX and, by interacting with MDC1, the two halves of the luc 2 enzyme are brought together, reconstituting the luminescent protein. To test the reporter gene, they exposed cells to different doses of ionizing radiation, and a clear dose-dependent response in reporter activities was found. 

Ionizing radiation exposure caused rapid activation of the reporter within 30 minutes of exposure, but after 24 hours, the activation dropped significantly thanks to the successful response of DSBs in the irradiated cells. Surprisingly, after 24 hours, the reporter activity started to grow again, reaching a peak at around day five and dropping again after day 12. It has been hypothesized that this second wave of reporter gene activation may be related to irradiated cell apoptosis (programmed cell death), during which whole strand breaks are routinely generated and $\gamma$-H2AX-MDC1 interactions could occur. Also, this period overlaps with an increase in caspase 3 and PARP levels, supporting the apoptotic theory. It cannot be excluded that it may represent a symptom of a persistent genetic instability found in cells after exposure to ionizing radiation. The main objective of this reporter system was to evaluate the number of DSBs in vivo in intact tissues. Its function was studied on tumor xenografts in nude mice, and the irradiation was a single dose of 6 Gy.

\subsection{The Impact of Ionizing Radiation on Cancer Risk}\label{sec:CancerRisk}
The day-to-day life on Earth exposes us to different types of radiation than what astronauts in the International Space Station (ISS) are exposed to. \citet{Cucinotta20149} compare the induced cancer risk from exposure to galactic cosmic rays containing HZE ions to that from exposure to ground-based radiation, including X-rays and gamma rays. Much of the oncological effects of HZE ions are unknown since no data is available. Astronauts in LEO undergo exposure to galactic cosmic rays, as will future astronauts embarking on interplanetary space travel, making developing this knowledge critical. Although there is no data on human beings, animal studies have been carried out on tumors in mice and rats to analyze the tumors induced by HZE ions. Results have shown that tumors induced by HZE ion exposure have higher rates of becoming metastatic (spreading throughout the body) and also show an increase in the tumor grade \citep{Cucinotta20149}. Tumor grading is a process used in medical pathology to describe the tumor's observed cell abnormality and growth rate deviations to define how dangerous the tumor is overall. Exposure to HZE particles has been observed to result in the production of reactive oxygen species, inflammation, and impairment of DNA signaling pathways, all of which play a role in cancer development \citep{Sridharan20151}. These experiments also show that tumors induced by HZE ions are more lethal in terms of increased initiation and promotion of the tumor compared to ground-based gamma rays \citep{Cucinotta20149}.

\citet{Cucinotta20149} discuss the radiation exposure limits set by NASA to protect astronauts from increased cancer risk; “NASA’s radiation limits sets a 3\% cancer fatality rate as the upper bound of acceptable risk and considers uncertainties in risk predictions using the upper 95\% confidence level of the assessment.” \citet{Cucinotta20149} also carried out mathematical analysis of the cancer risks experienced by astronauts in low Earth orbit, such as in the ISS. They found that female astronauts could exceed NASA’s radiation-induced cancer mortality limits after just 18 months in low Earth orbit (i.e., the ISS). Male astronauts may exceed these same limits in 24 months. Female astronauts are expected to exceed the limit in a shorter period due to an on-average lower body mass and several female-specific cancers, including breast cancer, ovarian cancer, and uterine cancer \citep{Cucinotta20149}. As with cardiovascular disease, \citet{Elgart20188480} did not observe an excess risk of cancer mortality in early NASA astronauts from radiation exposure. \citet{BarcellosHoff201592} discuss how the accuracy of predicting astronaut cancer risk from radiation exposure is limited by a lack of understanding of how space radiation could induce cancer growth.

Other factors play a role in the cancer risk following radiation exposure. \cite{Straume2015259} discuss how age influences radiation-induced lifetime cancer risk in the Handbook of Cosmic Hazards and Planetary Defense. The radiation-induced risk of developing cancer decreases with age due to the latency period of cancer, which can be two decades or longer. A younger individual exposed to radiation has more remaining lifetime for resulting cancer to develop, while an older individual is likely to present a different competing fatal-health condition before radiation-induced cancer develops. \cite{Straume2015259} also discuss NASA's 3\% radiation-exposure induced deaths from cancer career exposure limit and report on estimates of how long astronauts could embark on interplanetary space travel before exceeding these limits: “Assuming 20 g/cm$^2$ Al shielding and average solar minimum conditions, the maximum duration would be about 150-200 days, not sufficient for a mission to Mars.”

As has been previously mentioned, radiation-induced inflammation can instigate increased cancer risk. The experimental UV-irradiation of laboratory mice has shown increased angiogenesis\footnote{Angiogenesis is the formation of new blood vessels and is essential for delivering adequate nutrients to the cancerous region to sustain its growth.} and metastasis\footnote{Metastasis is the spread of cancer from a different part of the body from where it started.} due to the UV-induced inflammatory response of neutrophils \citep{Bald2014109}. The effects of UV radiation are particularly interesting for future Mars missions considering the lack of a protective ozone layer on the red planet \citep{Cockell2000343}. This UV-induced angiogenesis and metastasis were observed in melanoma\footnote{Melanoma is a type of skin cancer derived in the melanocytes (pigment-producing cells of the skin) \citep{Bald2014109}.} cells. Advanced angiogenesis and metastasis are characteristics of more pathological and life-threatening tumors. In addition to the ability of HZE particles to induce more potent tumors, the radiation-induced inflammatory response can potentially worsen the outcome further. The radiation exposure that astronauts traveling to Mars are estimated to undergo accompanies undeniable risks for cancer development. This shows the importance of further research on mitigating such potential harm.

\section{Modeling the Radiation Exposure}\label{sec:sec3}

The interaction of charged particles with the human body (and its shielding by, e.g., a spacecraft) usually is simulated using the GEneration ANd Tracking of particles version 4 \citep[GEANT4, ][]{agostinelli2003geant4} package. This Monte Carlo package is widely used to model the propagation of charged particles through matter, especially in high-energy physics, space sciences, and planetary physics, for example, Martian studies by \citep{matthiae2017radiation, atri2020modeling, rostel2020subsurface}. Since the code is used widely in high-energy physics, planetary science, and space medicine, it is calibrated with various experimental results. To compute the dose deposition in various human body organs, we use the Committee on Medical Internal Radiation Dose (MIRD) human phantom available as a part of the GEANT4 package, which was initially developed by \citep{guatelli2006geant4}. Although some organs were not implemented in the original package, we modified the code to incorporate all missing organs. Thereby, the organ masses of the missing organs were based on specifications from ICRP-89 \citep{icrp_89}. In addition, we imported appropriate physics models into the human phantom model to simulate high-energy charged particle interactions.

We calculate the radiation exposure during transit in interplanetary space, when the spacecraft is in the cruise phase (see Section~\ref{sec:CruisePhase}), and on the surface of Mars, where some shielding is provided by the thin CO$_2$-dominated atmosphere (see Section~\ref{sec:Surface}. In each of the two cases, we calculate the direct radiation exposure to astronauts with no shielding and 5 g cm$^{-2}$ of aluminum shielding, which is commonly used in spacecraft. 

\subsection{Galactic Cosmic Ray and Solar Energetic Particle Spectra}\label{sec:Spectra}
\begin{figure}[!t]
\centering
\includegraphics[width=\textwidth]{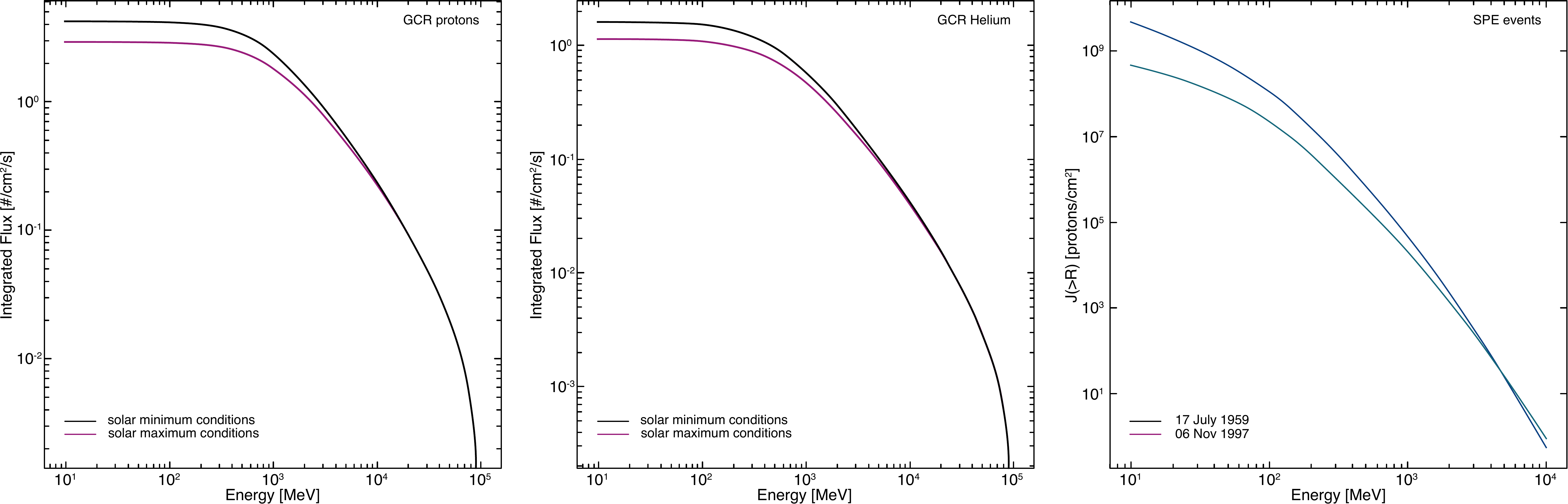}
\caption{Left panel: The spectra of GCR proton spectra during solar minimum (black) and maximum conditions (purple). Middle panel: The same for the Helium spectra. Right panel: The SPE spectra of the GLE events on 17 July 1959 (blue) and 6 November 1997 (orange). Data retrieved with the help of the BON10 model.}
\label{fig:spectra}
\end{figure}
To estimate the radiation exposure to astronauts, we need the energy spectra of GCRs to calculate background radiation exposure and the spectra of major SPEs to calculate the dose from episodic events. We use the Badhwar O'Neill (BON10) model \citep{o2010badhwar} to calculate the GCR spectra in outer space. Since the GCR spectrum varies considerably depending on solar modulation, we calculate the spectra for both solar minimum and maximum conditions to account for the two extremes. Heliospheric modulation changes with the distance from the Sun, but that effect is minor and can be ignored, so we assume that the GCR flux remains the same between 1 and 1.5 AU. GCR spectrum consists primarily of hydrogen and helium nuclei, which we have modeled in this paper and ignored the minor heavier species to simplify our calculations and save computing time. To simulate the impact of SPEs, we used two databases from which we obtained the energy spectra of various events \citep[e.g.,][]{atri2020modeling}. In addition, we study the impact of 60 solar particle events (SPEs) based on the catalog by \citet{Tylka-etal-2010}. The GCR spectra from the BON10 model and the two selected SPEs are shown in the panels of Fig. ~\ref{fig:spectra}. Note that for each SPE, 109 energy bins in the energy range of 10 MeV to 10 GeV have been used. Due to the lack of a magnetic field, particles were incident randomly from all over the hemisphere for each case. Thus, the human phantom was exposed to isotropic radiation to simulate deep-space conditions.
\subsection{The Simulation Setup}\label{sec:SimulationSetup}

The results shown in the following are based on the method described in \cite{atri2020modeling}; the results are based on the em-standard-opt3 and the G4HadronPhysics QGSP-BERT-HP physics lists to model secondary protons and neutrons, as recommended for incorporating neutron interactions better \citep{ivantchenko2012geant4}. The Mars Climate Database \citep[MCD, see][]{millour2018mars} was used to represent the atmosphere of Mars. The simulations were set up so that the incident radiation first interacts with the atmosphere, propagates down to the surface, and then interacts with the human phantom. 

In what follows, we investigate the impact of GCRs during both solar maxima and minima conditions and investigate the radiation exposure during two particular SPEs. Thereby, the simulations have been performed for:
\begin{enumerate}
    \item[(1)] Cruise phase with no shielding
    \item[(2)] Cruise phase with 5 g cm$^{-2}$ aluminum shielding
    \item[(3)] Martian surface with no shielding
    \item[(4)] Martian surface with 5 g cm$^{-2}$ aluminum shielding
\end{enumerate}

\noindent For these cases, the equivalent dose $H$ for the MIRD human phantom available as a part of the GEANT4 package \citep{guatelli2006geant4} was modeled. Thereby, $H$ is given by

\begin{equation}
H = \sum_j w_{R, j} \cdot D_j,
\end{equation}

\noindent the product of the radiation factor ($w_{R, j}$) defined by \citetalias{protection1991icrp}, accounting for the biological effectiveness of the radiation $j$ listed in Tab.~\ref{tab:w_j} and $D_j$, the modeled averaged dose rates. For neutrons, \citetalias{valentin2007icrp} defined a continuous function given by
\begin{equation}
    w_{R}=\begin{cases}
         2.5 + 18.2\cdot e^{\frac{-\left[ \ln\left(E\right)\right]^2}{6} },&\text{for } E < 1\,\text{MeV}  \\ 
         5.0 + 17.0\cdot e^{\frac{-\left[ \ln\left(2\cdot E\right)\right]^2}{6} },&\text{for } E\in\left[1\,\text{MeV}\--50\,\text{MeV}\right]  \\ 
         2.5 + 3.25\cdot e^{\frac{-\left[ \ln\left(0.04\cdot E\right)\right]^2}{6} },&\text{for } E> 50\,\text{MeV} 
    \end{cases}
    \label{eq:rwf}
,\end{equation}
with $E$ the kinetic energy of the neutron in MeV. For all other particles, $w_R$ is equal to $1$.

%+-+-+-+-+-+-+-+-+-+-+-+-+-+-+-+-+-+-+-+-+-+-+-+-+-+-+-+-+-+-+-+-+-+-+-+-
\begin{table}[!t]
\begin{center}
\label{tab:wj}
\begin{tabular}{|l|l|}
\hline
Radiation type $j$ & $w_{R,j}$ \\
\hline
Photons, electrons, muons & 1 \\
protons, charged pions & 2\\
$\alpha$, fission fragments, heavy ions & 20\\
neutrons & see Eq.~(\ref{eq:rwf})\\
\hline
\end{tabular}
\end{center}
\caption{Weighting factors $w_{R,j}$ defined by \citet{Petoussi-Henss-etal-2010}.}\label{tab:w_j}
\end{table}
%+-+-+-+-+-+-+-+-+-+-+-+-+-+-+-+-+-+-+-+-+-+-+-+-+-+-+-+-+-+-+-+-+-+-+-+-

Based on the simulation setup discussed above, the equivalent dose on a human body has been modeled. This study presents the equivalent dose rates in different human body organs for the first time. While the tables list the equivalent dose rates of over 40 vital organs of a female and male body, the figures only show selected features. 

\subsection{The Radiation Exposure during Transit in Interplanetary Space}\label{sec:CruisePhase}
\subsubsection{The GCR-induced equivalent dose rates}
\begin{figure}[!t]
\centering
\includegraphics[width=0.49\textwidth]{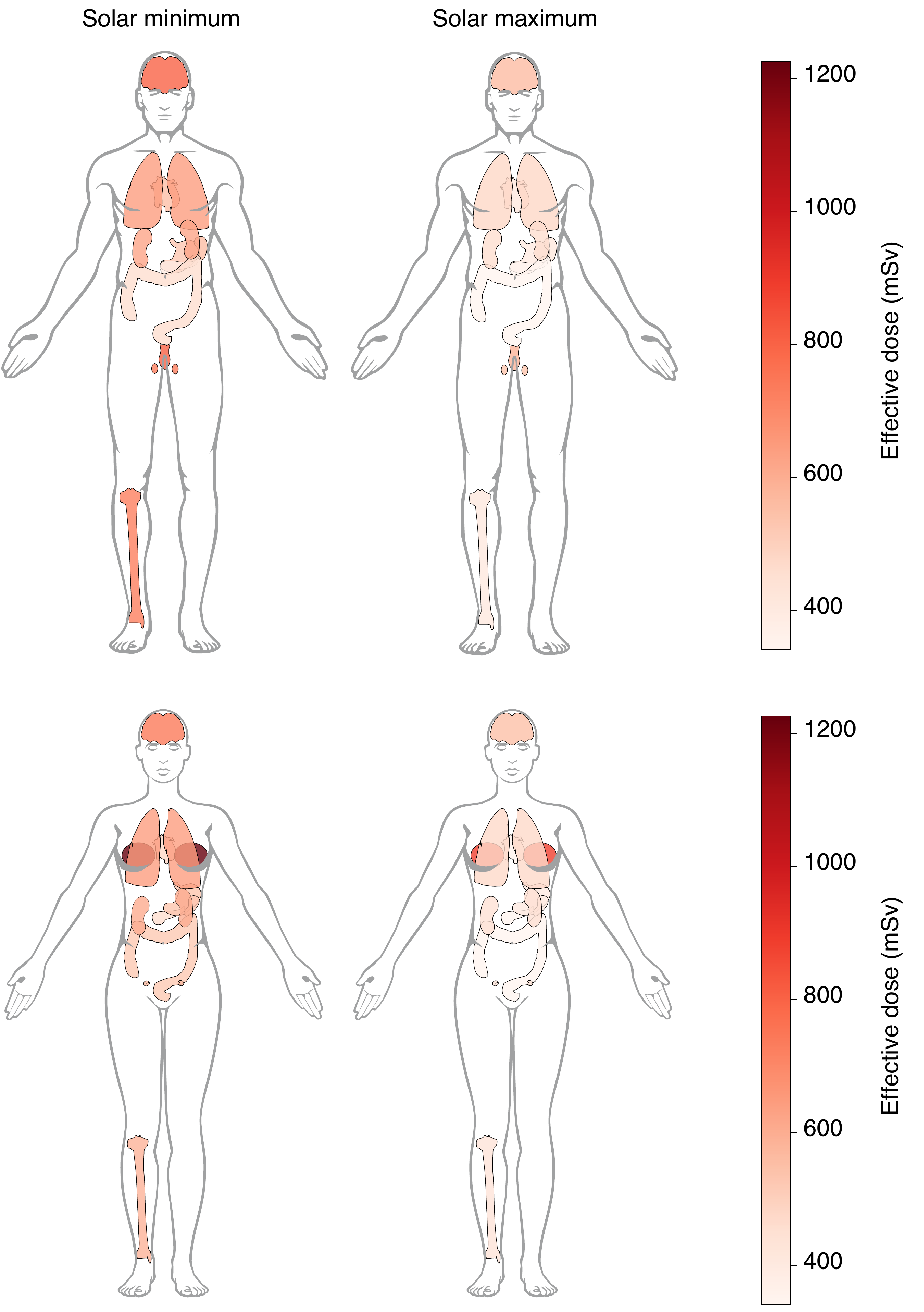}
\includegraphics[width=0.49\textwidth]{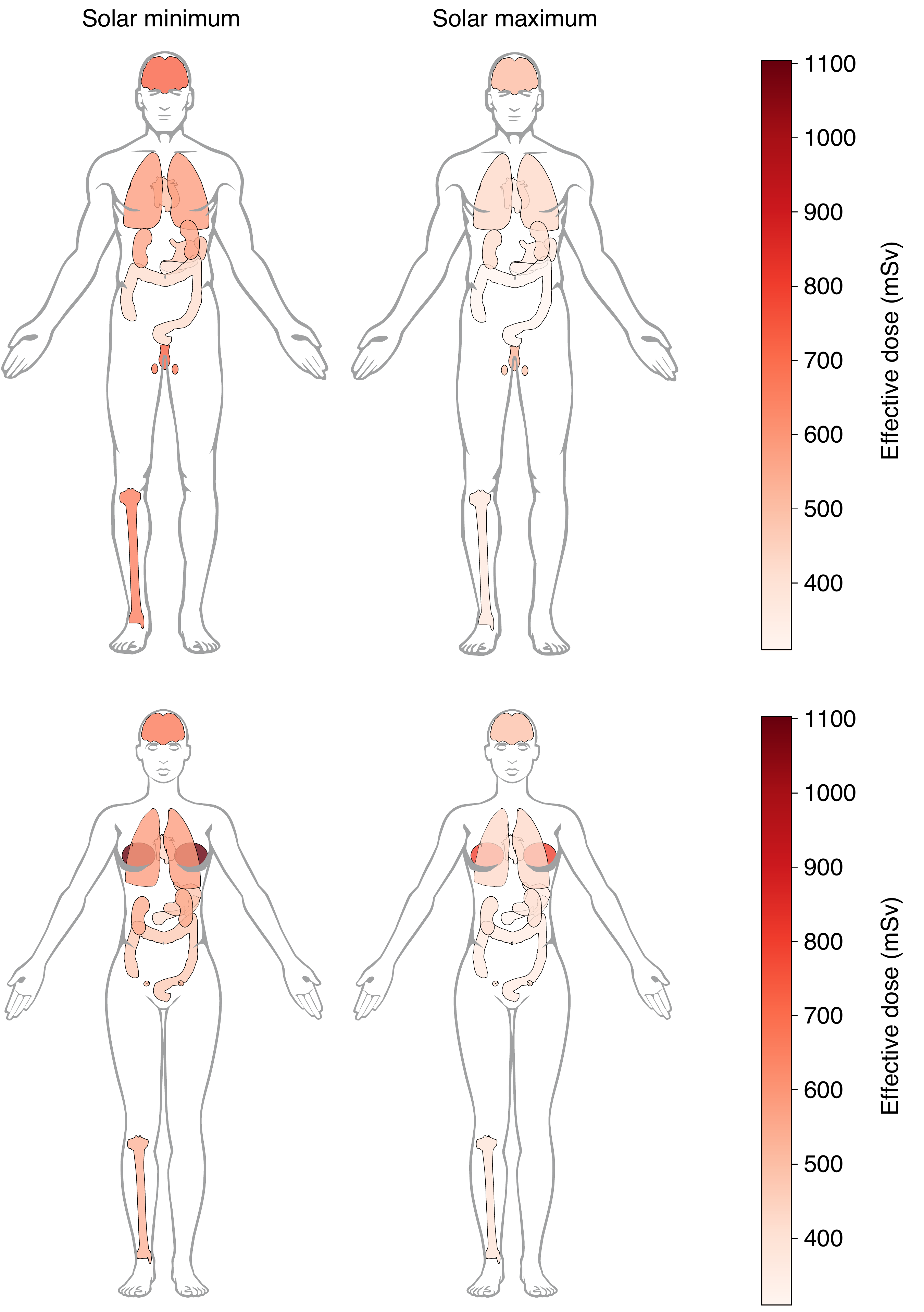}
\caption{Left panels: Equivalent dose (mSv) during cruise phase without shielding for key organs of a male human phantom (upper panels) and a female human phantom (lower panels) during solar minimum (left) and maximum (right) conditions. Right panels: Same as on the left with an additional 5 g cm$^{-2}$ aluminum shielding. The plots have been created with the python package \href{https://bitbucket.org/manuela_s/pyanatomogram/src/master/}{PyAnatomogram} \citep[see][]{PyAnatomogram}.}
\label{fig:a}
\end{figure}

The GCR-induced equivalent dose rates of GCR protons, $\alpha$ particles, and heavier particles during the cruise phase without shielding that the organs of a male and female astronaut encounter during solar minimum and maximum conditions are given in Tab.~3. Correspondingly, the left panels of Fig.~\ref{fig:a} show the total equivalent dose rates of the male (upper panels) and female (lower panels) brain, heart, lungs, kidneys, stomach, liver, left leg bone, and reproductive organs during solar minimum (left panels) and maximum (right panels) conditions.

\begin{filecontents*}{cruise_1_new.csv}
organ,m_h_min,m_h_max,m_he_min,m_he_max,m_hy_min,m_hy_max,f_h_min,f_h_max,f_he_min,f_he_max,f_hy_min,f_hy_max
Organ,Solar min., Solar max., Solar min., Solar max.,Solar min., Solar max.,Solar min., Solar max.,Solar min., Solar max.,Solar min., Solar max.
Brain,2.46E+02,1.76E+02,1.87E+02,1.45E+02,3.67E+02,2.63E+02,2.28E+02,1.65E+02,1.74E+02,1.36E+02,3.40E+02,2.45E+02
Head,1.36E+02,9.78E+01,1.33E+02,1.01E+02,2.03E+02,1.46E+02,1.30E+02,9.26E+01,1.27E+02,9.56E+01,1.94E+02,1.38E+02
Heart,1.65E+02,1.20E+02,1.40E+02,1.11E+02,2.45E+02,1.79E+02,1.43E+02,1.05E+02,1.21E+02,9.62E+01,2.13E+02,1.56E+02
LeftAdrenal,1.88E+02,1.36E+02,1.54E+02,1.20E+02,2.80E+02,2.03E+02,2.32E+02,1.69E+02,1.68E+02,1.32E+02,3.46E+02,2.52E+02
LeftArmBone,2.70E+02,1.92E+02,2.00E+02,1.53E+02,4.02E+02,2.87E+02,2.66E+02,1.90E+02,1.96E+02,1.50E+02,3.96E+02,2.83E+02
LeftBreast,0.00E+00,0.00E+00,0.00E+00,0.00E+00,0.00E+00,0.00E+00,3.48E+02,2.48E+02,3.18E+02,2.38E+02,5.19E+02,3.70E+02
LeftClavicle,2.40E+02,1.73E+02,1.76E+02,1.37E+02,3.58E+02,2.57E+02,2.28E+02,1.65E+02,1.79E+02,1.40E+02,3.40E+02,2.45E+02
LeftKidney,1.98E+02,1.43E+02,1.45E+02,1.15E+02,2.94E+02,2.13E+02,1.95E+02,1.41E+02,1.43E+02,1.12E+02,2.90E+02,2.10E+02
LeftLeg,2.40E+02,1.71E+02,2.06E+02,1.57E+02,3.58E+02,2.55E+02,2.38E+02,1.70E+02,2.06E+02,1.56E+02,3.55E+02,2.54E+02
LeftLegBone,1.75E+02,1.27E+02,1.46E+02,1.15E+02,2.60E+02,1.90E+02,1.71E+02,1.25E+02,1.42E+02,1.12E+02,2.55E+02,1.86E+02
LeftLung,1.99E+02,1.44E+02,1.56E+02,1.23E+02,2.97E+02,2.15E+02,1.95E+02,1.42E+02,1.56E+02,1.23E+02,2.91E+02,2.11E+02
LeftOvary,0.00E+00,0.00E+00,0.00E+00,0.00E+00,0.00E+00,0.00E+00,2.10E+02,1.55E+02,1.90E+02,1.51E+02,3.13E+02,2.31E+02
LeftScapula,2.52E+02,1.81E+02,1.87E+02,1.43E+02,3.75E+02,2.69E+02,2.86E+02,2.04E+02,2.10E+02,1.61E+02,4.26E+02,3.04E+02
LeftTeste,2.64E+02,1.90E+02,1.73E+02,1.33E+02,3.93E+02,2.83E+02,0.00E+00,0.00E+00,0.00E+00,0.00E+00,0.00E+00,0.00E+00
LowerLargeIntestine,1.37E+02,1.01E+02,1.15E+02,9.18E+01,2.05E+02,1.50E+02,1.67E+02,1.22E+02,1.38E+02,1.09E+02,2.49E+02,1.82E+02
MaleGenitalia,2.30E+02,1.65E+02,2.28E+02,1.71E+02,3.43E+02,2.46E+02,0.00E+00,0.00E+00,0.00E+00,0.00E+00,0.00E+00,0.00E+00
MiddleLowerSpine,1.55E+02,1.13E+02,1.23E+02,9.70E+01,2.31E+02,1.69E+02,1.53E+02,1.12E+02,1.22E+02,9.62E+01,2.28E+02,1.66E+02
Pancreas,1.17E+02,8.60E+01,9.84E+01,7.86E+01,1.74E+02,1.28E+02,1.36E+02,1.00E+02,1.20E+02,9.56E+01,2.03E+02,1.49E+02
Pelvis,1.60E+02,1.17E+02,1.25E+02,9.92E+01,2.39E+02,1.74E+02,1.57E+02,1.15E+02,1.22E+02,9.62E+01,2.34E+02,1.71E+02
RibCage,2.70E+02,1.93E+02,2.08E+02,1.59E+02,4.02E+02,2.88E+02,2.58E+02,1.85E+02,1.98E+02,1.51E+02,3.84E+02,2.76E+02
RightAdrenal,2.14E+02,1.54E+02,1.45E+02,1.14E+02,3.19E+02,2.30E+02,2.08E+02,1.51E+02,1.52E+02,1.20E+02,3.10E+02,2.26E+02
RightArmBone,2.78E+02,1.99E+02,2.06E+02,1.57E+02,4.14E+02,2.97E+02,2.66E+02,1.91E+02,1.98E+02,1.52E+02,3.96E+02,2.85E+02
RightBreast,0.00E+00,0.00E+00,0.00E+00,0.00E+00,0.00E+00,0.00E+00,3.74E+02,2.66E+02,3.36E+02,2.52E+02,5.57E+02,3.96E+02
RightClavicle,2.16E+02,1.56E+02,1.59E+02,1.25E+02,3.22E+02,2.32E+02,2.44E+02,1.75E+02,1.76E+02,1.37E+02,3.64E+02,2.60E+02
RightKidney,1.90E+02,1.38E+02,1.39E+02,1.09E+02,2.83E+02,2.06E+02,1.81E+02,1.32E+02,1.32E+02,1.03E+02,2.70E+02,1.96E+02
RightLeg,2.38E+02,1.70E+02,2.06E+02,1.56E+02,3.55E+02,2.54E+02,2.46E+02,1.76E+02,2.12E+02,1.61E+02,3.67E+02,2.62E+02
RightLegBone,1.67E+02,1.21E+02,1.37E+02,1.08E+02,2.49E+02,1.81E+02,1.76E+02,1.28E+02,1.45E+02,1.14E+02,2.63E+02,1.90E+02
RightLung,1.95E+02,1.41E+02,1.53E+02,1.20E+02,2.90E+02,2.10E+02,1.95E+02,1.41E+02,1.57E+02,1.23E+02,2.91E+02,2.10E+02
RightOvary,0.00E+00,0.00E+00,0.00E+00,0.00E+00,0.00E+00,0.00E+00,1.28E+02,9.34E+01,1.00E+02,7.94E+01,1.90E+02,1.39E+02
RightScapula,2.60E+02,1.87E+02,1.95E+02,1.48E+02,3.87E+02,2.78E+02,2.64E+02,1.88E+02,1.94E+02,1.48E+02,3.93E+02,2.80E+02
RightTeste,1.98E+02,1.43E+02,1.37E+02,1.07E+02,2.96E+02,2.13E+02,0.00E+00,0.00E+00,0.00E+00,0.00E+00,0.00E+00,0.00E+00
Skull,2.64E+02,1.90E+02,2.20E+02,1.68E+02,3.93E+02,2.83E+02,2.58E+02,1.84E+02,2.14E+02,1.63E+02,3.84E+02,2.75E+02
SmallIntestine,1.34E+02,9.84E+01,1.12E+02,8.88E+01,2.00E+02,1.47E+02,1.43E+02,1.05E+02,1.18E+02,9.40E+01,2.13E+02,1.56E+02
Spleen,1.68E+02,1.23E+02,1.34E+02,1.06E+02,2.51E+02,1.83E+02,1.59E+02,1.15E+02,1.31E+02,1.04E+02,2.36E+02,1.72E+02
Stomach,1.61E+02,1.18E+02,1.26E+02,1.00E+02,2.40E+02,1.75E+02,1.63E+02,1.19E+02,1.27E+02,1.01E+02,2.43E+02,1.77E+02
Thymus,1.56E+02,1.14E+02,1.24E+02,9.78E+01,2.33E+02,1.70E+02,1.54E+02,1.12E+02,1.23E+02,9.70E+01,2.30E+02,1.66E+02
Thyroid,1.74E+02,1.27E+02,1.42E+02,1.12E+02,2.60E+02,1.90E+02,1.61E+02,1.17E+02,1.21E+02,9.56E+01,2.40E+02,1.74E+02
Trunk,1.63E+02,1.18E+02,1.42E+02,1.09E+02,2.43E+02,1.75E+02,1.63E+02,1.18E+02,1.41E+02,1.08E+02,2.43E+02,1.75E+02
UpperLargeIntestine,1.38E+02,1.01E+02,1.15E+02,9.18E+01,2.06E+02,1.51E+02,1.48E+02,1.08E+02,1.23E+02,9.84E+01,2.20E+02,1.61E+02
UpperSpine,1.98E+02,1.43E+02,1.52E+02,1.20E+02,2.94E+02,2.13E+02,1.75E+02,1.27E+02,1.39E+02,1.09E+02,2.60E+02,1.90E+02
UrinaryBladder,1.62E+02,1.18E+02,1.26E+02,9.92E+01,2.41E+02,1.76E+02,1.75E+02,1.27E+02,1.34E+02,1.07E+02,2.60E+02,1.90E+02
Uterus,0.00E+00,0.00E+00,0.00E+00,0.00E+00,0.00E+00,0.00E+00,1.38E+02,1.01E+02,1.12E+02,8.96E+01,2.06E+02,1.51E+02
\end{filecontents*}
\DTLloaddb{cruise_1}{cruise_1_new.csv}
\noindent\adjustbox{max width=\textwidth,max totalheight=\textheight-6ex}{
    \begin{tabular}{|c|c|c|c|c|c|c|c|c|c|c|c|c|}
    \hline
    \multirow{3}{*}{} & \multicolumn{6}{|c|}{Male Phantom} & \multicolumn{6}{|c|}{Female Phantom} \\
     \cline{2-13}
     & \multicolumn{2}{|c|}{Hydrogen} & \multicolumn{2}{|c|}{Helium} & \multicolumn{2}{|c|}{Heavy} & \multicolumn{2}{|c|}{Hydrogen} & \multicolumn{2}{|c|}{Helium} & \multicolumn{2}{|c|}{Heavy}\\
     \cline{2-13}
     \hline
    \DTLforeach{cruise_1}
    {\organ=organ,\mhmin=m_h_min,\mhmax=m_h_max,\mhemin=m_he_min,\mhemax=m_he_max,\mhymin=m_hy_min,\mhymax=m_hy_max,\fhmin=f_h_min,\fhmax=f_h_max,\fhemin=f_he_min,\fhemax=f_he_max, \fhymin=f_hy_min,\fhymax=f_hy_max}
    {%\DTLiffirstrow{}{\\}%
    \organ & \mhmin & \mhmax & \mhemin & \mhemax & \mhymin & \mhymax & \fhmin & \fhmax & \fhemin & \fhemax & \fhymin & \fhymax%}\\
    \DTLiflastrow{}{\DTLiffirstrow{\\ \hline}{\\}}
    }\\
    \hline
    \label{tab:2}
    \end{tabular}
}
\captionof{table}{Equivalent dose (mSv) during cruise phase in different organs of male and female phantoms}

\begin{filecontents*}{5g_cruise_new.csv}
organ,m_h_min,m_h_max,m_he_min,m_he_max,m_hy_min,m_hy_max,f_h_min,f_h_max,f_he_min,f_he_max,f_hy_min,f_hy_max
Organ,Solar min., Solar max., Solar min., Solar max.,Solar min., Solar max.,Solar min., Solar max.,Solar min., Solar max.,Solar min., Solar max.
Brain,2.21E+02,1.59E+02,1.69E+02,1.31E+02,3.30E+02,2.37E+02,2.05E+02,1.48E+02,1.57E+02,1.22E+02,3.06E+02,2.21E+02
Head,1.22E+02,8.80E+01,1.20E+02,9.05E+01,1.82E+02,1.31E+02,1.17E+02,8.33E+01,1.14E+02,8.60E+01,1.74E+02,1.24E+02
Heart,1.48E+02,1.08E+02,1.26E+02,9.99E+01,2.21E+02,1.61E+02,1.29E+02,9.45E+01,1.09E+02,8.66E+01,1.92E+02,1.41E+02
Left Adrenal,1.69E+02,1.22E+02,1.38E+02,1.08E+02,2.52E+02,1.82E+02,2.09E+02,1.52E+02,1.51E+02,1.18E+02,3.11E+02,2.27E+02
Left Arm Bone,2.43E+02,1.73E+02,1.80E+02,1.38E+02,3.62E+02,2.58E+02,2.39E+02,1.71E+02,1.77E+02,1.35E+02,3.57E+02,2.54E+02
Left Breast,0.00E+00,0.00E+00,0.00E+00,0.00E+00,0.00E+00,0.00E+00,3.13E+02,2.23E+02,2.86E+02,2.14E+02,4.67E+02,3.33E+02
Left Clavicle,2.16E+02,1.55E+02,1.58E+02,1.23E+02,3.22E+02,2.31E+02,2.05E+02,1.48E+02,1.61E+02,1.26E+02,3.06E+02,2.21E+02
Left Kidney,1.78E+02,1.29E+02,1.31E+02,1.03E+02,2.65E+02,1.92E+02,1.75E+02,1.27E+02,1.28E+02,1.00E+02,2.61E+02,1.89E+02
Left Leg,2.16E+02,1.54E+02,1.85E+02,1.41E+02,3.22E+02,2.30E+02,2.14E+02,1.53E+02,1.85E+02,1.41E+02,3.19E+02,2.29E+02
Left Leg Bone,1.57E+02,1.14E+02,1.32E+02,1.03E+02,2.34E+02,1.71E+02,1.54E+02,1.12E+02,1.28E+02,1.00E+02,2.30E+02,1.67E+02
Left Lung,1.79E+02,1.30E+02,1.41E+02,1.10E+02,2.67E+02,1.93E+02,1.76E+02,1.28E+02,1.41E+02,1.10E+02,2.62E+02,1.90E+02
Left Ovary,0.00E+00,0.00E+00,0.00E+00,0.00E+00,0.00E+00,0.00E+00,1.89E+02,1.40E+02,1.71E+02,1.36E+02,2.82E+02,2.08E+02
Left Scapula,2.27E+02,1.63E+02,1.68E+02,1.28E+02,3.38E+02,2.42E+02,2.57E+02,1.84E+02,1.89E+02,1.45E+02,3.84E+02,2.74E+02
Left Teste,2.38E+02,1.71E+02,1.55E+02,1.20E+02,3.54E+02,2.54E+02,0.00E+00,0.00E+00,0.00E+00,0.00E+00,0.00E+00,0.00E+00
Lower Large Intestine,1.24E+02,9.05E+01,1.04E+02,8.26E+01,1.84E+02,1.35E+02,1.50E+02,1.10E+02,1.24E+02,9.85E+01,2.24E+02,1.64E+02
Male Genitalia,2.07E+02,1.49E+02,2.05E+02,1.54E+02,3.08E+02,2.22E+02,0.00E+00,0.00E+00,0.00E+00,0.00E+00,0.00E+00,0.00E+00
Middle Lower Spine,1.40E+02,1.02E+02,1.10E+02,8.73E+01,2.08E+02,1.52E+02,1.38E+02,1.00E+02,1.10E+02,8.66E+01,2.05E+02,1.50E+02
Pancreas,1.05E+02,7.74E+01,8.86E+01,7.07E+01,1.57E+02,1.15E+02,1.22E+02,9.00E+01,1.08E+02,8.60E+01,1.82E+02,1.34E+02
Pelvis,1.44E+02,1.05E+02,1.12E+02,8.93E+01,2.15E+02,1.57E+02,1.41E+02,1.03E+02,1.10E+02,8.66E+01,2.11E+02,1.54E+02
Rib Cage,2.43E+02,1.74E+02,1.87E+02,1.43E+02,3.62E+02,2.59E+02,2.32E+02,1.67E+02,1.79E+02,1.36E+02,3.46E+02,2.48E+02
Right Adrenal,1.93E+02,1.39E+02,1.30E+02,1.02E+02,2.87E+02,2.07E+02,1.87E+02,1.36E+02,1.37E+02,1.08E+02,2.79E+02,2.03E+02
Right Arm Bone,2.50E+02,1.79E+02,1.85E+02,1.41E+02,3.73E+02,2.67E+02,2.39E+02,1.72E+02,1.79E+02,1.37E+02,3.57E+02,2.56E+02
Right Breast,0.00E+00,0.00E+00,0.00E+00,0.00E+00,0.00E+00,0.00E+00,3.37E+02,2.39E+02,3.02E+02,2.27E+02,5.02E+02,3.57E+02
Right Clavicle,1.94E+02,1.40E+02,1.43E+02,1.12E+02,2.90E+02,2.09E+02,2.20E+02,1.57E+02,1.59E+02,1.24E+02,3.27E+02,2.34E+02
Right Kidney,1.71E+02,1.24E+02,1.25E+02,9.85E+01,2.55E+02,1.85E+02,1.63E+02,1.18E+02,1.18E+02,9.25E+01,2.43E+02,1.76E+02
Right Leg,2.14E+02,1.53E+02,1.85E+02,1.41E+02,3.19E+02,2.29E+02,2.21E+02,1.58E+02,1.91E+02,1.45E+02,3.30E+02,2.35E+02
Right Leg Bone,1.50E+02,1.09E+02,1.24E+02,9.72E+01,2.24E+02,1.63E+02,1.59E+02,1.15E+02,1.31E+02,1.02E+02,2.37E+02,1.71E+02
Right Lung,1.75E+02,1.27E+02,1.38E+02,1.08E+02,2.61E+02,1.89E+02,1.76E+02,1.27E+02,1.41E+02,1.11E+02,2.62E+02,1.89E+02
Right Ovary,0.00E+00,0.00E+00,0.00E+00,0.00E+00,0.00E+00,0.00E+00,1.15E+02,8.41E+01,9.00E+01,7.15E+01,1.71E+02,1.25E+02
Right Scapula,2.34E+02,1.68E+02,1.75E+02,1.34E+02,3.49E+02,2.50E+02,2.38E+02,1.69E+02,1.75E+02,1.33E+02,3.54E+02,2.52E+02
Right Teste,1.79E+02,1.29E+02,1.23E+02,9.59E+01,2.66E+02,1.92E+02,0.00E+00,0.00E+00,0.00E+00,0.00E+00,0.00E+00,0.00E+00
Skull,2.38E+02,1.71E+02,1.98E+02,1.51E+02,3.54E+02,2.54E+02,2.32E+02,1.66E+02,1.93E+02,1.47E+02,3.46E+02,2.47E+02
Small Intestine,1.21E+02,8.86E+01,1.00E+02,7.99E+01,1.80E+02,1.32E+02,1.29E+02,9.45E+01,1.06E+02,8.46E+01,1.92E+02,1.41E+02
Spleen,1.51E+02,1.10E+02,1.21E+02,9.52E+01,2.26E+02,1.64E+02,1.43E+02,1.04E+02,1.18E+02,9.32E+01,2.13E+02,1.55E+02
Stomach,1.45E+02,1.06E+02,1.14E+02,9.00E+01,2.16E+02,1.58E+02,1.47E+02,1.07E+02,1.14E+02,9.05E+01,2.19E+02,1.60E+02
Thymus,1.41E+02,1.02E+02,1.12E+02,8.80E+01,2.10E+02,1.53E+02,1.39E+02,1.00E+02,1.10E+02,8.73E+01,2.07E+02,1.50E+02
Thyroid,1.57E+02,1.14E+02,1.28E+02,1.01E+02,2.34E+02,1.71E+02,1.45E+02,1.05E+02,1.09E+02,8.60E+01,2.16E+02,1.57E+02
Trunk,1.47E+02,1.06E+02,1.28E+02,9.79E+01,2.19E+02,1.58E+02,1.47E+02,1.06E+02,1.27E+02,9.72E+01,2.19E+02,1.58E+02
Upper Large Intestine,1.24E+02,9.13E+01,1.04E+02,8.26E+01,1.85E+02,1.36E+02,1.33E+02,9.72E+01,1.11E+02,8.86E+01,1.98E+02,1.45E+02
Upper Spine,1.78E+02,1.29E+02,1.37E+02,1.08E+02,2.65E+02,1.92E+02,1.57E+02,1.14E+02,1.25E+02,9.79E+01,2.34E+02,1.71E+02
Urinary Bladder,1.45E+02,1.06E+02,1.13E+02,8.93E+01,2.17E+02,1.59E+02,1.57E+02,1.14E+02,1.21E+02,9.59E+01,2.34E+02,1.71E+02
Uterus,0.00E+00,0.00E+00,0.00E+00,0.00E+00,0.00E+00,0.00E+00,1.24E+02,9.13E+01,1.00E+02,8.06E+01,1.85E+02,1.36E+02
\end{filecontents*}
\DTLloaddb{5g_cruise}{5g_cruise_new.csv}
\noindent\adjustbox{max width=\textwidth,max totalheight=\textheight-6ex}{
    \begin{tabular}{|c|c|c|c|c|c|c|c|c|c|c|c|c|}
    \hline
    \multirow{3}{*}{} & \multicolumn{6}{|c|}{Male Phantom} & \multicolumn{6}{|c|}{Female Phantom} \\
     \cline{2-13}
     & \multicolumn{2}{|c|}{Hydrogen} & \multicolumn{2}{|c|}{Helium} & \multicolumn{2}{|c|}{Heavy} & \multicolumn{2}{|c|}{Hydrogen} & \multicolumn{2}{|c|}{Helium} & \multicolumn{2}{|c|}{Heavy}\\
     \cline{2-13}
     \hline
    \DTLforeach{cruise_1}
    {\organ=organ,\mhmin=m_h_min,\mhmax=m_h_max,\mhemin=m_he_min,\mhemax=m_he_max,\mhymin=m_hy_min,\mhymax=m_hy_max,\fhmin=f_h_min,\fhmax=f_h_max,\fhemin=f_he_min,\fhemax=f_he_max, \fhymin=f_hy_min,\fhymax=f_hy_max}
    {%\DTLiffirstrow{}{\\}%
    \organ & \mhmin & \mhmax & \mhemin & \mhemax & \mhymin & \mhymax & \fhmin & \fhmax & \fhemin & \fhemax & \fhymin & \fhymax%}\\
    \DTLiflastrow{}{\DTLiffirstrow{\\ \hline}{\\}}}\\
    \hline
    \label{tab:3}
    \end{tabular}
}
\captionof{table}{Equivalent dose (mSv) during cruise phase with 5 g cm$^{-2}$ Al shielding in different organs}

\vspace{1cm}
Because the modulation of low-energy GCRs is much less during solar minimum conditions, the effective dose rates of male and female astronauts, on average, are about 30\% higher than those during solar maximum conditions. As can be seen, organs like the brain, lungs, and reproductive organs are most affected by the influence of GCRs during the cruising phase. Thereby, female astronauts would suffer more from higher dose rates (equivalent dose rates of up to 1200 mSv during solar minimum conditions could be reached) and, thus, would have a higher risk of developing breast and ovarian cancer. 
\\
\\
Adding a shielding of 5 g cm$^{-2}$ aluminum reduces the equivalent dose rates in the order of 10\%. The organ-dependent equivalent dose rates are listed in Tab.~4. At the same time, the right panels of Fig.~\ref{fig:a} show a visualization of the dose rates in the key organs of the male and female phantom during solar minimum (left) and maximum (right) conditions. 
%\\
%\\
%\textcolor{red}{TBA: How do the equivalent dose rates compare to those at the terrestrial surface? Are there any other publications to compare the values with?} 
%
%\subsubsection{The SPE-induced equivalent dose rates}
%
%\textcolor{red}{TBA!!!}
\subsection{The Radiation Exposure on the Martian Surface}\label{sec:Surface}
\subsubsection{The GCR-induced equivalent dose rates}

\begin{figure}[!t]
\centering
\includegraphics[width=0.49\textwidth]{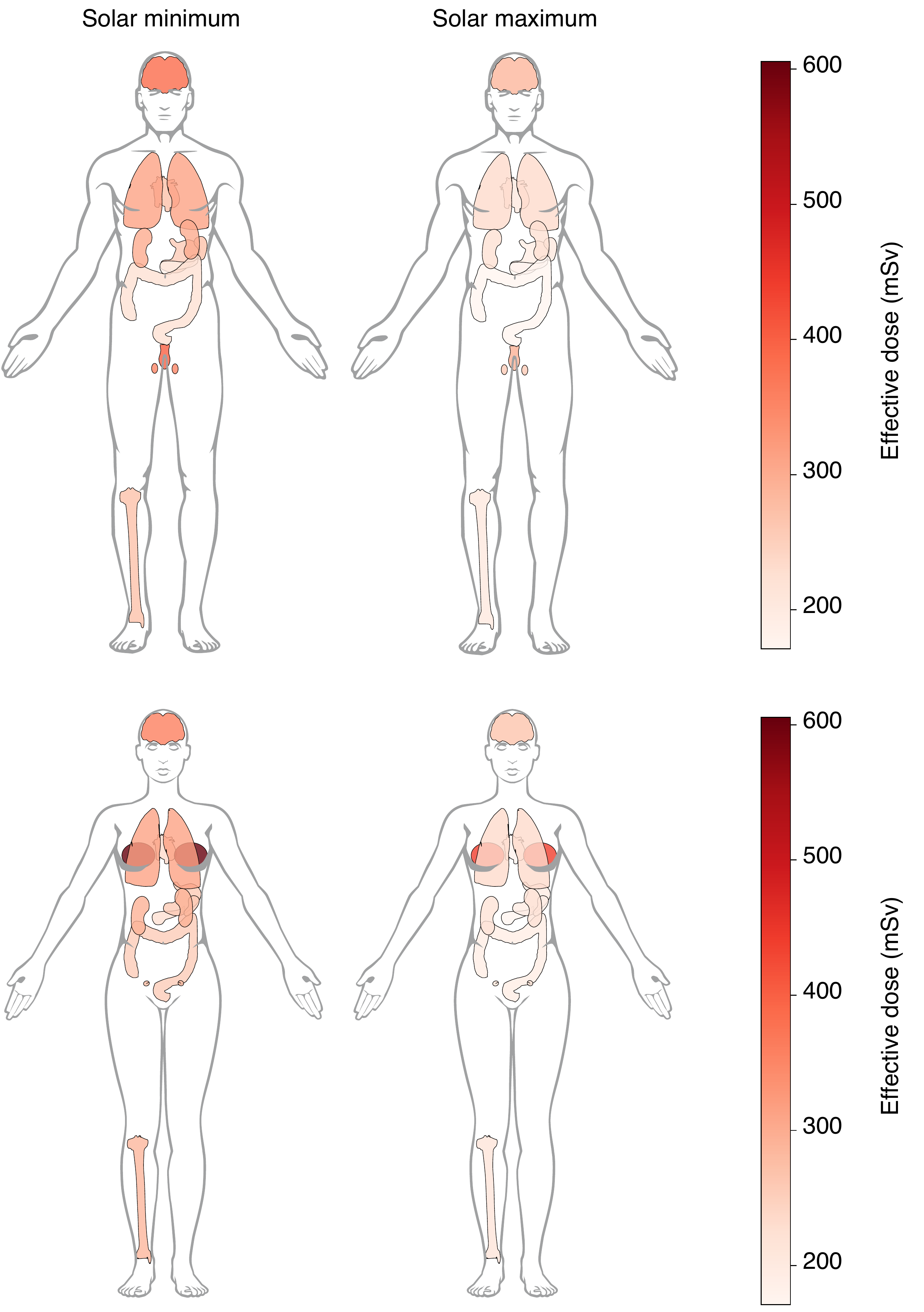}
\includegraphics[width=0.49\textwidth]{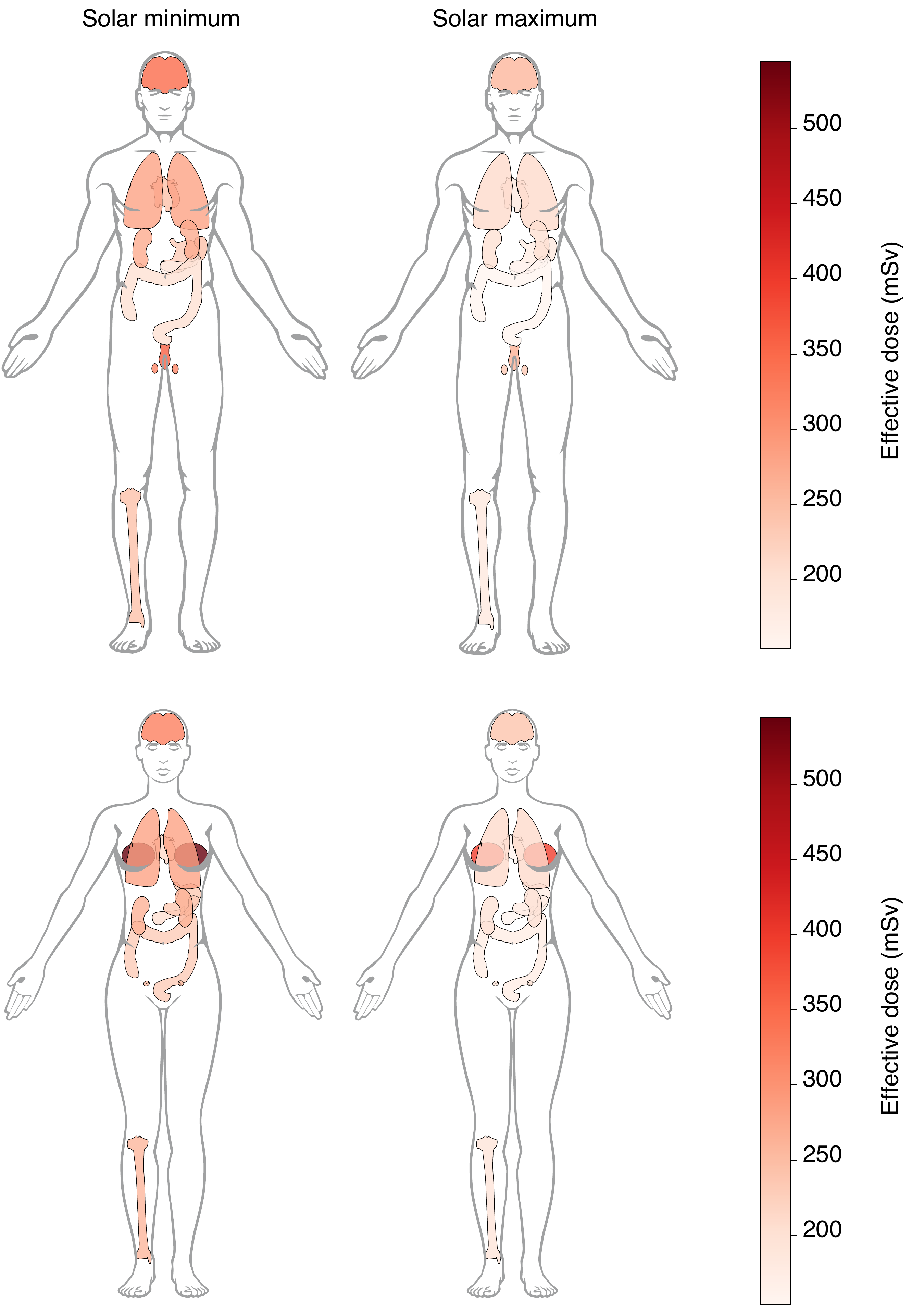}
\caption{Left panels: Equivalent dose (mSv) on the non-shielded Martian surface for key organs of a male human phantom (upper panel) and a female human phantom (lower panels) during solar minimum (left) and maximum (right) conditions. Right panels: Same as in the left panels, here for 5 g cm$^{-2}$ Al shielding. The plots have been created with the python package \href{https://bitbucket.org/manuela_s/pyanatomogram/src/master/}{PyAnatomogram} \citep[see][]{PyAnatomogram}.}
\label{fig:b}
\end{figure}

Tables~5 and~6 list the particle-dependent GCR-induced equivalent dose rates during solar minimum and maximum conditions of the male and female phantoms at the Martian surface without shielding and within a 5 g cm$^{-2}$ aluminum shielding, respectively. The impact of GCRs on the martian surface on the key organs of both phantoms and cases is further displayed in the panels of Fig.~\ref{fig:b}.
\\
\\
Comparing the results to Fig.~\ref{fig:a} shows that the (un-)shielded equivalent dose rates at the Martian surface are about a factor of 2.5 lower than those during the cruise phase within 5 g cm$^{-2}$ Al shielding, a consequence of the thin CO$_2$-dominated atmosphere that is shielding GCRs more effectively. Figure~\ref{fig:b} further shows that a shelter consisting of 5 g cm$^{-2}$ Al (right panels) will reduce the equivalent dose rate values only slightly (by about 13\%). As in the cruise phase, female astronauts' organs show a stronger response to the GCR-induced radiation exposure. Thereby, the most affected organs are the uterus and breasts. 
\\
\\
%In the case of SEPs, just like in the case of GCRs, there is a significant difference between dose rates during the cruise phase and on the ground. However, the effect of shielding is much greater as compared to GCRs because the energy of particles is much lower. Also, organ-to-organ variation in dose rates is much higher in the case of SEPs because of the lower energy particles, whereas in GCRs, it is more uniform due to higher energy particles. 
According to \citet{rostel2020subsurface}, for solar maximum conditions, the surface equivalent dose of the ICRU phantom reduces by about 49 - 54\% compared to that of solar minimum conditions. However, our study gives a more detailed insight into the radiation exposure of human organs. Compared to \citet{rostel2020subsurface}, we find a slightly lesser reduction of the equivalent dose rates during solar maximum conditions; depending on the organ, the values are reduced by 39 - 49\%. However, note that the soil composition also impacts the radiation environment on the Martian surface due to a potentially enhanced secondary particle flux caused by the interaction of cosmic rays with the soil. \citet{rostel2020subsurface} found that the equivalent dose rates reduce by about 45\% at regions with a water content of 50\% H$_2$O in comparison to the often used dry andesite rock\footnote{The andesite rock surface is composed of mass fractions of 44\% O, 27\% Si, 12\% Fe, and 17 \% other elements with a mass density of 2.8 g/cm$^3$.}. Thus, regions with underground water, as recently discovered by Mars Express, have the potential to provide indirect shielding against cosmic rays and might be the perfect ground for human landing sites.

\begin{filecontents*}{surface_new.csv}
organ,m_h_min,m_h_max,m_he_min,m_he_max,m_hy_min,m_hy_max,f_h_min,f_h_max,f_he_min,f_he_max,f_hy_min,f_hy_max
Organ,Solar min., Solar max., Solar min., Solar max.,Solar min., Solar max.,Solar min., Solar max.,Solar min., Solar max.,Solar min., Solar max.
Brain,1.72E+02,1.23E+02,1.31E+02,1.02E+02,8.43E+01,6.05E+01,1.60E+02,1.15E+02,1.22E+02,9.52E+01,7.81E+01,5.64E+01
Head,9.52E+01,6.85E+01,9.31E+01,7.04E+01,4.66E+01,3.35E+01,9.10E+01,6.48E+01,8.90E+01,6.69E+01,4.46E+01,3.17E+01
Heart,1.15E+02,8.43E+01,9.83E+01,7.77E+01,5.64E+01,4.13E+01,1.00E+02,7.35E+01,8.48E+01,6.73E+01,4.91E+01,3.60E+01
LeftAdrenal,1.32E+02,9.52E+01,1.08E+02,8.43E+01,6.44E+01,4.66E+01,1.62E+02,1.18E+02,1.17E+02,9.21E+01,7.95E+01,5.79E+01
LeftArmBone,1.89E+02,1.35E+02,1.40E+02,1.07E+02,9.25E+01,6.59E+01,1.86E+02,1.33E+02,1.37E+02,1.05E+02,9.12E+01,6.50E+01
LeftBreast,0.00E+00,0.00E+00,0.00E+00,0.00E+00,0.00E+00,0.00E+00,2.44E+02,1.74E+02,2.23E+02,1.67E+02,1.19E+02,8.50E+01
LeftClavicle,1.68E+02,1.21E+02,1.23E+02,9.56E+01,8.22E+01,5.92E+01,1.60E+02,1.15E+02,1.25E+02,9.77E+01,7.81E+01,5.64E+01
LeftKidney,1.38E+02,1.00E+02,1.02E+02,8.02E+01,6.77E+01,4.91E+01,1.36E+02,9.87E+01,9.98E+01,7.81E+01,6.67E+01,4.83E+01
LeftLeg,1.68E+02,1.20E+02,1.44E+02,1.10E+02,8.22E+01,5.87E+01,1.67E+02,1.19E+02,1.44E+02,1.09E+02,8.16E+01,5.84E+01
LeftLegBone,1.22E+02,8.90E+01,1.02E+02,8.02E+01,5.99E+01,4.36E+01,1.20E+02,8.74E+01,9.93E+01,7.81E+01,5.87E+01,4.28E+01
LeftLung,1.39E+02,1.01E+02,1.09E+02,8.58E+01,6.83E+01,4.93E+01,1.37E+02,9.93E+01,1.09E+02,8.58E+01,6.70E+01,4.86E+01
LeftOvary,0.00E+00,0.00E+00,0.00E+00,0.00E+00,0.00E+00,0.00E+00,1.47E+02,1.09E+02,1.33E+02,1.06E+02,7.20E+01,5.31E+01
LeftScapula,1.76E+02,1.27E+02,1.31E+02,9.98E+01,8.64E+01,6.20E+01,2.00E+02,1.43E+02,1.47E+02,1.13E+02,9.80E+01,6.99E+01
LeftTeste,1.85E+02,1.33E+02,1.21E+02,9.31E+01,9.05E+01,6.50E+01,0.00E+00,0.00E+00,0.00E+00,0.00E+00,0.00E+00,0.00E+00
LowerLargeIntestine,9.62E+01,7.04E+01,8.08E+01,6.43E+01,4.71E+01,3.45E+01,1.17E+02,8.54E+01,9.67E+01,7.66E+01,5.72E+01,4.18E+01
MaleGenitalia,1.61E+02,1.16E+02,1.60E+02,1.20E+02,7.88E+01,5.67E+01,0.00E+00,0.00E+00,0.00E+00,0.00E+00,0.00E+00,0.00E+00
MiddleLowerSpine,1.09E+02,7.92E+01,8.58E+01,6.79E+01,5.31E+01,3.88E+01,1.07E+02,7.81E+01,8.54E+01,6.73E+01,5.24E+01,3.82E+01
Pancreas,8.18E+01,6.02E+01,6.89E+01,5.50E+01,4.00E+01,2.95E+01,9.52E+01,7.00E+01,8.39E+01,6.69E+01,4.66E+01,3.43E+01
Pelvis,1.12E+02,8.18E+01,8.74E+01,6.94E+01,5.49E+01,4.00E+01,1.10E+02,8.02E+01,8.54E+01,6.73E+01,5.39E+01,3.93E+01
RibCage,1.89E+02,1.35E+02,1.46E+02,1.11E+02,9.25E+01,6.62E+01,1.81E+02,1.30E+02,1.39E+02,1.06E+02,8.84E+01,6.35E+01
RightAdrenal,1.50E+02,1.08E+02,1.01E+02,7.97E+01,7.33E+01,5.28E+01,1.46E+02,1.06E+02,1.06E+02,8.39E+01,7.13E+01,5.19E+01
RightArmBone,1.95E+02,1.39E+02,1.44E+02,1.10E+02,9.53E+01,6.83E+01,1.86E+02,1.34E+02,1.39E+02,1.06E+02,9.12E+01,6.55E+01
RightBreast,0.00E+00,0.00E+00,0.00E+00,0.00E+00,0.00E+00,0.00E+00,2.62E+02,1.86E+02,2.35E+02,1.76E+02,1.28E+02,9.12E+01
RightClavicle,1.51E+02,1.09E+02,1.12E+02,8.74E+01,7.40E+01,5.34E+01,1.71E+02,1.22E+02,1.23E+02,9.62E+01,8.36E+01,5.99E+01
RightKidney,1.33E+02,9.67E+01,9.72E+01,7.66E+01,6.52E+01,4.74E+01,1.27E+02,9.21E+01,9.21E+01,7.20E+01,6.22E+01,4.51E+01
RightLeg,1.67E+02,1.19E+02,1.44E+02,1.09E+02,8.16E+01,5.84E+01,1.72E+02,1.23E+02,1.48E+02,1.13E+02,8.43E+01,6.02E+01
RightLegBone,1.17E+02,8.48E+01,9.62E+01,7.56E+01,5.72E+01,4.15E+01,1.23E+02,8.95E+01,1.02E+02,7.97E+01,6.05E+01,4.38E+01
RightLung,1.36E+02,9.87E+01,1.07E+02,8.39E+01,6.67E+01,4.83E+01,1.37E+02,9.87E+01,1.10E+02,8.64E+01,6.70E+01,4.83E+01
RightOvary,0.00E+00,0.00E+00,0.00E+00,0.00E+00,0.00E+00,0.00E+00,8.95E+01,6.54E+01,7.00E+01,5.56E+01,4.38E+01,3.20E+01
RightScapula,1.82E+02,1.31E+02,1.36E+02,1.04E+02,8.91E+01,6.39E+01,1.85E+02,1.32E+02,1.36E+02,1.03E+02,9.05E+01,6.44E+01
RightTeste,1.39E+02,1.00E+02,9.56E+01,7.46E+01,6.80E+01,4.91E+01,0.00E+00,0.00E+00,0.00E+00,0.00E+00,0.00E+00,0.00E+00
Skull,1.85E+02,1.33E+02,1.54E+02,1.18E+02,9.05E+01,6.50E+01,1.81E+02,1.29E+02,1.50E+02,1.14E+02,8.84E+01,6.32E+01
SmallIntestine,9.41E+01,6.89E+01,7.81E+01,6.22E+01,4.61E+01,3.37E+01,1.00E+02,7.35E+01,8.27E+01,6.58E+01,4.91E+01,3.60E+01
Spleen,1.18E+02,8.58E+01,9.41E+01,7.41E+01,5.76E+01,4.20E+01,1.11E+02,8.08E+01,9.16E+01,7.25E+01,5.44E+01,3.95E+01
Stomach,1.13E+02,8.23E+01,8.85E+01,7.00E+01,5.51E+01,4.03E+01,1.14E+02,8.33E+01,8.90E+01,7.04E+01,5.59E+01,4.08E+01
Thymus,1.09E+02,7.97E+01,8.69E+01,6.85E+01,5.36E+01,3.90E+01,1.08E+02,7.81E+01,8.58E+01,6.79E+01,5.28E+01,3.82E+01
Thyroid,1.22E+02,8.90E+01,9.93E+01,7.87E+01,5.97E+01,4.36E+01,1.13E+02,8.18E+01,8.48E+01,6.69E+01,5.51E+01,4.00E+01
Trunk,1.14E+02,8.23E+01,9.93E+01,7.62E+01,5.59E+01,4.03E+01,1.14E+02,8.23E+01,9.87E+01,7.56E+01,5.59E+01,4.03E+01
UpperLargeIntestine,9.67E+01,7.10E+01,8.08E+01,6.43E+01,4.74E+01,3.47E+01,1.03E+02,7.56E+01,8.64E+01,6.89E+01,5.06E+01,3.70E+01
UpperSpine,1.38E+02,1.00E+02,1.06E+02,8.39E+01,6.77E+01,4.91E+01,1.22E+02,8.90E+01,9.72E+01,7.62E+01,5.99E+01,4.36E+01
UrinaryBladder,1.13E+02,8.27E+01,8.79E+01,6.94E+01,5.54E+01,4.05E+01,1.22E+02,8.90E+01,9.41E+01,7.46E+01,5.99E+01,4.36E+01
Uterus,0.00E+00,0.00E+00,0.00E+00,0.00E+00,0.00E+00,0.00E+00,9.67E+01,7.10E+01,7.81E+01,6.27E+01,4.74E+01,3.47E+01
\end{filecontents*}
\DTLloaddb{surface}{surface_new.csv}
\noindent\adjustbox{max width=\textwidth,max totalheight=\textheight-6ex}{
    \begin{tabular}{|c|c|c|c|c|c|c|c|c|c|c|c|c|}
    \hline
    \multirow{3}{*}{} & \multicolumn{6}{|c|}{Male Phantom} & \multicolumn{6}{|c|}{Female Phantom} \\
     \cline{2-13}
     & \multicolumn{2}{|c|}{Hydrogen} & \multicolumn{2}{|c|}{Helium} & \multicolumn{2}{|c|}{Heavy} & \multicolumn{2}{|c|}{Hydrogen} & \multicolumn{2}{|c|}{Helium} & \multicolumn{2}{|c|}{Heavy}\\
     \cline{2-13}
     \hline
    \DTLforeach{cruise_1}
    {\organ=organ,\mhmin=m_h_min,\mhmax=m_h_max,\mhemin=m_he_min,\mhemax=m_he_max,\mhymin=m_hy_min,\mhymax=m_hy_max,\fhmin=f_h_min,\fhmax=f_h_max,\fhemin=f_he_min,\fhemax=f_he_max, \fhymin=f_hy_min,\fhymax=f_hy_max}
    {%\DTLiffirstrow{}{\\}%
    \organ & \mhmin & \mhmax & \mhemin & \mhemax & \mhymin & \mhymax & \fhmin & \fhmax & \fhemin & \fhemax & \fhymin & \fhymax%}\\
    \DTLiflastrow{}{\DTLiffirstrow{\\ \hline}{\\}}
    }\\
    \hline
    \label{tab:4}
    \end{tabular}\\
}
\captionof{table}{Equivalent dose (mSv) in different organs of a human phantom on the Martian surface without shielding.}

\begin{filecontents*}{surface_5g_new.csv}
organ,m_h_min,m_h_max,m_he_min,m_he_max,m_hy_min,m_hy_max,f_h_min,f_h_max,f_he_min,f_he_max,f_hy_min,f_hy_max
Organ,Solar min., Solar max., Solar min., Solar max.,Solar min., Solar max.,Solar min., Solar max.,Solar min., Solar max.,Solar min., Solar max.

Brain,1.55E+02,1.11E+02,1.18E+02,9.16E+01,7.59E+01,5.44E+01,1.44E+02,1.04E+02,1.10E+02,8.57E+01,7.03E+01,5.08E+01
Head,8.57E+01,6.16E+01,8.38E+01,6.34E+01,4.19E+01,3.02E+01,8.19E+01,5.83E+01,8.01E+01,6.02E+01,4.01E+01,2.86E+01
Heart,1.04E+02,7.59E+01,8.85E+01,6.99E+01,5.08E+01,3.71E+01,9.02E+01,6.62E+01,7.64E+01,6.06E+01,4.42E+01,3.24E+01
LeftAdrenal,1.18E+02,8.57E+01,9.68E+01,7.59E+01,5.80E+01,4.19E+01,1.46E+02,1.06E+02,1.06E+02,8.29E+01,7.16E+01,5.21E+01
LeftArmBone,1.70E+02,1.21E+02,1.26E+02,9.63E+01,8.33E+01,5.93E+01,1.68E+02,1.19E+02,1.24E+02,9.44E+01,8.20E+01,5.85E+01
LeftBreast,0.00E+00,0.00E+00,0.00E+00,0.00E+00,0.00E+00,0.00E+00,2.19E+02,1.56E+02,2.00E+02,1.50E+02,1.07E+02,7.65E+01
LeftClavicle,1.51E+02,1.09E+02,1.11E+02,8.61E+01,7.40E+01,5.32E+01,1.44E+02,1.04E+02,1.13E+02,8.79E+01,7.03E+01,5.08E+01
LeftKidney,1.24E+02,9.02E+01,9.16E+01,7.22E+01,6.09E+01,4.42E+01,1.23E+02,8.88E+01,8.98E+01,7.03E+01,6.00E+01,4.35E+01
LeftLeg,1.51E+02,1.08E+02,1.30E+02,9.90E+01,7.40E+01,5.28E+01,1.50E+02,1.07E+02,1.30E+02,9.85E+01,7.34E+01,5.26E+01
LeftLegBone,1.10E+02,8.01E+01,9.21E+01,7.22E+01,5.39E+01,3.92E+01,1.08E+02,7.86E+01,8.93E+01,7.03E+01,5.28E+01,3.85E+01
LeftLung,1.25E+02,9.07E+01,9.85E+01,7.72E+01,6.14E+01,4.44E+01,1.23E+02,8.93E+01,9.85E+01,7.72E+01,6.03E+01,4.37E+01
LeftOvary,0.00E+00,0.00E+00,0.00E+00,0.00E+00,0.00E+00,0.00E+00,1.32E+02,9.77E+01,1.19E+02,9.54E+01,6.48E+01,4.78E+01
LeftScapula,1.59E+02,1.14E+02,1.18E+02,8.98E+01,7.77E+01,5.58E+01,1.80E+02,1.29E+02,1.32E+02,1.01E+02,8.82E+01,6.29E+01
LeftTeste,1.66E+02,1.19E+02,1.09E+02,8.38E+01,8.14E+01,5.85E+01,0.00E+00,0.00E+00,0.00E+00,0.00E+00,0.00E+00,0.00E+00
LowerLargeIntestine,8.66E+01,6.34E+01,7.27E+01,5.78E+01,4.24E+01,3.10E+01,1.05E+02,7.69E+01,8.71E+01,6.89E+01,5.14E+01,3.76E+01
MaleGenitalia,1.45E+02,1.04E+02,1.44E+02,1.08E+02,7.09E+01,5.10E+01,0.00E+00,0.00E+00,0.00E+00,0.00E+00,0.00E+00,0.00E+00
MiddleLowerSpine,9.77E+01,7.13E+01,7.72E+01,6.11E+01,4.78E+01,3.49E+01,9.63E+01,7.03E+01,7.69E+01,6.06E+01,4.71E+01,3.44E+01
Pancreas,7.36E+01,5.42E+01,6.20E+01,4.95E+01,3.60E+01,2.65E+01,8.57E+01,6.30E+01,7.55E+01,6.02E+01,4.19E+01,3.08E+01
Pelvis,1.01E+02,7.36E+01,7.86E+01,6.25E+01,4.94E+01,3.60E+01,9.90E+01,7.22E+01,7.69E+01,6.06E+01,4.85E+01,3.53E+01
RibCage,1.70E+02,1.22E+02,1.31E+02,9.99E+01,8.33E+01,5.96E+01,1.63E+02,1.17E+02,1.25E+02,9.54E+01,7.96E+01,5.71E+01
RightAdrenal,1.35E+02,9.71E+01,9.12E+01,7.17E+01,6.60E+01,4.76E+01,1.31E+02,9.54E+01,9.58E+01,7.55E+01,6.42E+01,4.67E+01
RightArmBone,1.75E+02,1.25E+02,1.30E+02,9.90E+01,8.57E+01,6.14E+01,1.68E+02,1.20E+02,1.25E+02,9.58E+01,8.20E+01,5.89E+01
RightBreast,0.00E+00,0.00E+00,0.00E+00,0.00E+00,0.00E+00,0.00E+00,2.36E+02,1.68E+02,2.12E+02,1.59E+02,1.15E+02,8.20E+01
RightClavicle,1.36E+02,9.82E+01,1.00E+02,7.86E+01,6.66E+01,4.81E+01,1.54E+02,1.10E+02,1.11E+02,8.66E+01,7.53E+01,5.39E+01
RightKidney,1.20E+02,8.71E+01,8.74E+01,6.89E+01,5.87E+01,4.26E+01,1.14E+02,8.29E+01,8.29E+01,6.48E+01,5.59E+01,4.06E+01
RightLeg,1.50E+02,1.07E+02,1.30E+02,9.85E+01,7.34E+01,5.26E+01,1.55E+02,1.11E+02,1.34E+02,1.01E+02,7.59E+01,5.42E+01
RightLegBone,1.05E+02,7.64E+01,8.66E+01,6.80E+01,5.14E+01,3.74E+01,1.11E+02,8.05E+01,9.16E+01,7.17E+01,5.44E+01,3.94E+01
RightLung,1.23E+02,8.88E+01,9.63E+01,7.55E+01,6.00E+01,4.35E+01,1.23E+02,8.88E+01,9.90E+01,7.77E+01,6.03E+01,4.35E+01
RightOvary,0.00E+00,0.00E+00,0.00E+00,0.00E+00,0.00E+00,0.00E+00,8.05E+01,5.88E+01,6.30E+01,5.00E+01,3.94E+01,2.88E+01
RightScapula,1.64E+02,1.18E+02,1.23E+02,9.35E+01,8.02E+01,5.76E+01,1.66E+02,1.18E+02,1.22E+02,9.30E+01,8.14E+01,5.80E+01
RightTeste,1.25E+02,9.02E+01,8.61E+01,6.72E+01,6.12E+01,4.42E+01,0.00E+00,0.00E+00,0.00E+00,0.00E+00,0.00E+00,0.00E+00
Skull,1.66E+02,1.19E+02,1.39E+02,1.06E+02,8.14E+01,5.85E+01,1.63E+02,1.16E+02,1.35E+02,1.03E+02,7.96E+01,5.69E+01
SmallIntestine,8.47E+01,6.20E+01,7.03E+01,5.59E+01,4.15E+01,3.03E+01,9.02E+01,6.62E+01,7.45E+01,5.92E+01,4.42E+01,3.24E+01
Spleen,1.06E+02,7.72E+01,8.47E+01,6.67E+01,5.19E+01,3.78E+01,9.99E+01,7.27E+01,8.24E+01,6.53E+01,4.89E+01,3.56E+01
Stomach,1.01E+02,7.41E+01,7.96E+01,6.30E+01,4.96E+01,3.63E+01,1.03E+02,7.50E+01,8.01E+01,6.34E+01,5.03E+01,3.67E+01
Thymus,9.85E+01,7.17E+01,7.82E+01,6.16E+01,4.82E+01,3.51E+01,9.71E+01,7.03E+01,7.72E+01,6.11E+01,4.76E+01,3.44E+01
Thyroid,1.10E+02,8.01E+01,8.93E+01,7.08E+01,5.37E+01,3.92E+01,1.01E+02,7.36E+01,7.64E+01,6.02E+01,4.96E+01,3.60E+01
Trunk,1.03E+02,7.41E+01,8.93E+01,6.85E+01,5.03E+01,3.63E+01,1.03E+02,7.41E+01,8.88E+01,6.80E+01,5.03E+01,3.63E+01
UpperLargeIntestine,8.71E+01,6.39E+01,7.27E+01,5.78E+01,4.26E+01,3.13E+01,9.30E+01,6.80E+01,7.77E+01,6.20E+01,4.55E+01,3.33E+01
UpperSpine,1.24E+02,9.02E+01,9.58E+01,7.55E+01,6.09E+01,4.42E+01,1.10E+02,8.01E+01,8.74E+01,6.85E+01,5.39E+01,3.92E+01
UrinaryBladder,1.02E+02,7.45E+01,7.91E+01,6.25E+01,4.98E+01,3.65E+01,1.10E+02,8.01E+01,8.47E+01,6.72E+01,5.39E+01,3.92E+01
Uterus,0.00E+00,0.00E+00,0.00E+00,0.00E+00,0.00E+00,0.00E+00,8.71E+01,6.39E+01,7.03E+01,5.64E+01,4.26E+01,3.13E+01
\end{filecontents*}
\DTLloaddb{surface_5g}{surface_5g_new.csv}
\noindent\adjustbox{max width=\textwidth,max totalheight=\textheight-6ex}{
    \begin{tabular}{|c|c|c|c|c|c|c|c|c|c|c|c|c|}
    \hline
    \multirow{3}{*}{} & \multicolumn{6}{|c|}{Male Phantom} & \multicolumn{6}{|c|}{Female Phantom} \\
     \cline{2-13}
     & \multicolumn{2}{|c|}{Hydrogen} & \multicolumn{2}{|c|}{Helium} & \multicolumn{2}{|c|}{Heavy} & \multicolumn{2}{|c|}{Hydrogen} & \multicolumn{2}{|c|}{Helium} & \multicolumn{2}{|c|}{Heavy}\\
     \cline{2-13}
     \hline
    \DTLforeach{cruise_1}
    {\organ=organ,\mhmin=m_h_min,\mhmax=m_h_max,\mhemin=m_he_min,\mhemax=m_he_max,\mhymin=m_hy_min,\mhymax=m_hy_max,\fhmin=f_h_min,\fhmax=f_h_max,\fhemin=f_he_min,\fhemax=f_he_max, \fhymin=f_hy_min,\fhymax=f_hy_max}
    {%\DTLiffirstrow{}{\\}%
    \organ & \mhmin & \mhmax & \mhemin & \mhemax & \mhymin & \mhymax & \fhmin & \fhmax & \fhemin & \fhemax & \fhymin & \fhymax%}\\
    \DTLiflastrow{}{\DTLiffirstrow{\\ \hline}{\\}}
    }\\
    \hline
    \label{tab:5}
    \end{tabular}\\
}
\captionof{table}{Equivalent dose (mSv) on Mars surface with 5 g cm$^{-2}$ Al shielding in different organs}

\clearpage

\subsubsection{The SPE-induced equivalent dose rates}\label{sec:SPE-inducedEquiDose}
\begin{figure}[!t]
\centering
\includegraphics[width=0.8\textwidth]{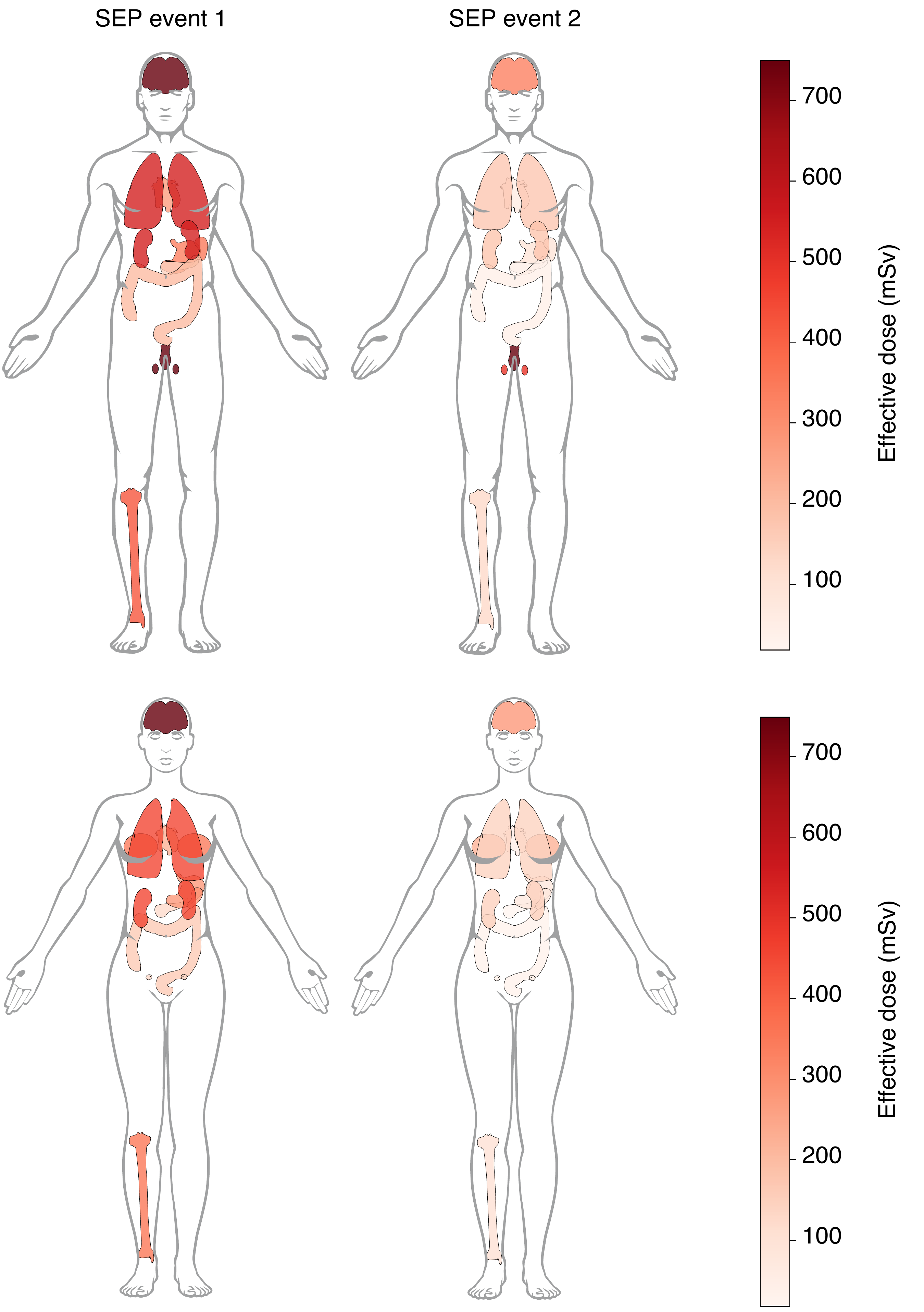}
\caption{Equivalen dose in male (upper panels) and female (lower panels) organs induced by SEP events 1 (left panels) and 2 (right panels). The plots have been created with the python package \href{https://bitbucket.org/manuela_s/pyanatomogram/src/master/}{PyAnatomogram} \citep[see][]{PyAnatomogram}.}
\label{fig:d}
\end{figure}

Since the start of the neutron monitor era, 72 solar energetic particle events that were energetic enough to be detected on the terrestrial surface have occurred. As mentioned before, since the end of 2012, at least five such events also have been detected at the Martian surface. 
\\
\\
To estimate the risk that strong SPEs could pose to the health of future astronauts, we study the impact of 60 SPE events on the equivalent dose rates on human organs. While the corresponding total equivalent dose rates are listed in Tabs. 11 to 16 in the Appendix, Fig.~\ref{fig:c} shows a heat map of the impact of all 60 events on the organs of a male phantom. Thereby, the darker the color, the higher the equivalent dose. 
\begin{figure}
\centering
\includegraphics[width=\textwidth]{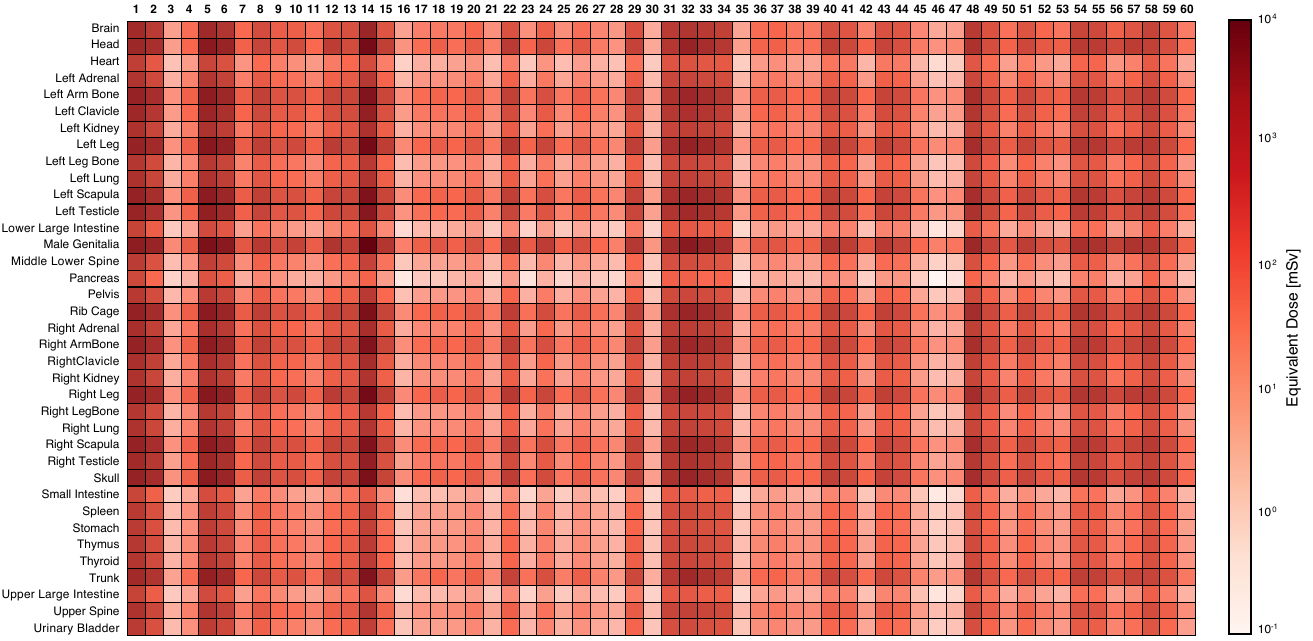}
\caption{Heat map of the equivalent dose rates of male phantom organs during the 60 selected energetic particle events [1-60]. Figure based on the tables given in the Appendix.}
\label{fig:c}
\end{figure}
\\
\\
As can be seen, the SPE-induced equivalent dose rates often are much higher than those induced by GCRs during solar minimum and maximum conditions. While the highest GCR-induced dose rates were around 600 mSv, some SPE-induced values show rates of up to 10$^{4}$ mSv. However, the organ-to-organ variation of the dose rates is much higher, most likely caused by the higher low-energy particle flux. Figure~\ref{fig:c} shows that an event like the event that occurred on the 4th of September 1972 (event 14) would have led to the strongest radiation exposure, and fifteen organs would have been exposed to dose rates around 10$^4$ mSv (10 Sv). Without proper shielding, this event would have catastrophic consequences since a single dose of 1 Sv would cause radiation sickness (nausea, vomiting, and hemorrhaging), and a single dose of 5 Sv would already kill about half of the astronauts exposed to it within the first month after the event. Thus, it is fair to assume that unshielded, a single dose of 10 Sv, like caused in event 14, would be life-threatening to all future astronauts stationed at Mars. Events 1, 5, and 32 would lead to dose rates up to 3 Sv, while events 2, 31, 33, 34, and 48 would lead to values around 1 Sv.
\\
\\
Figure~\ref{fig:d} shows the equivalent dose rates of the male (upper panels) and female (lower panels) phantom during the SEP event of 17 July 1959 (left panels) and 06 November 1997 (right panels). Surprisingly, our results suggest that during SPEs, the male phantom organs suffer higher equivalent doses than the female phantom organs. Thus, while female astronauts are most likely more affected by GCRs, they are less affected by the lower-energetic solar particles.
\\
\\
The RAD instrument onboard the Curiosity rover has measured the radiation dose on the Martian surface since late 2012. According to \citet{hassler2014mars} the average radiation dose at Gale Crater is 0.64 $\pm$ 0.12 mSv/day. For the standard ICRU sphere, we modeled an average dose rate of 0.59 mSv/day, which (within the uncertainties) agrees with the measurements.

%In the case of SEPs, just like in the case of GCRs, there is a significant difference between dose rates during the cruise phase and on the ground. However, the effect of shielding is much greater as compared to GCRs because the energy of particles is much lower. 

\subsection{The GCR-induced Radiation Exposure During a Mission to Mars}

The impact of radiation on astronaut health in crewed missions has always been a concern on numerous missions to the ISS and the Lunar surface. Most of these missions were short-duration, posing little risk to astronauts from GCRs or major events emitting solar-origin particles. Long-duration missions to the ISS do provide insight into the possible impact of radiation on astronauts, although their duration has been much shorter compared to planned missions to Mars, and their close proximity to the Earth makes them better shielded compared to interplanetary missions. 

\begin{figure}
    \centering
    \includegraphics[width=\textwidth]{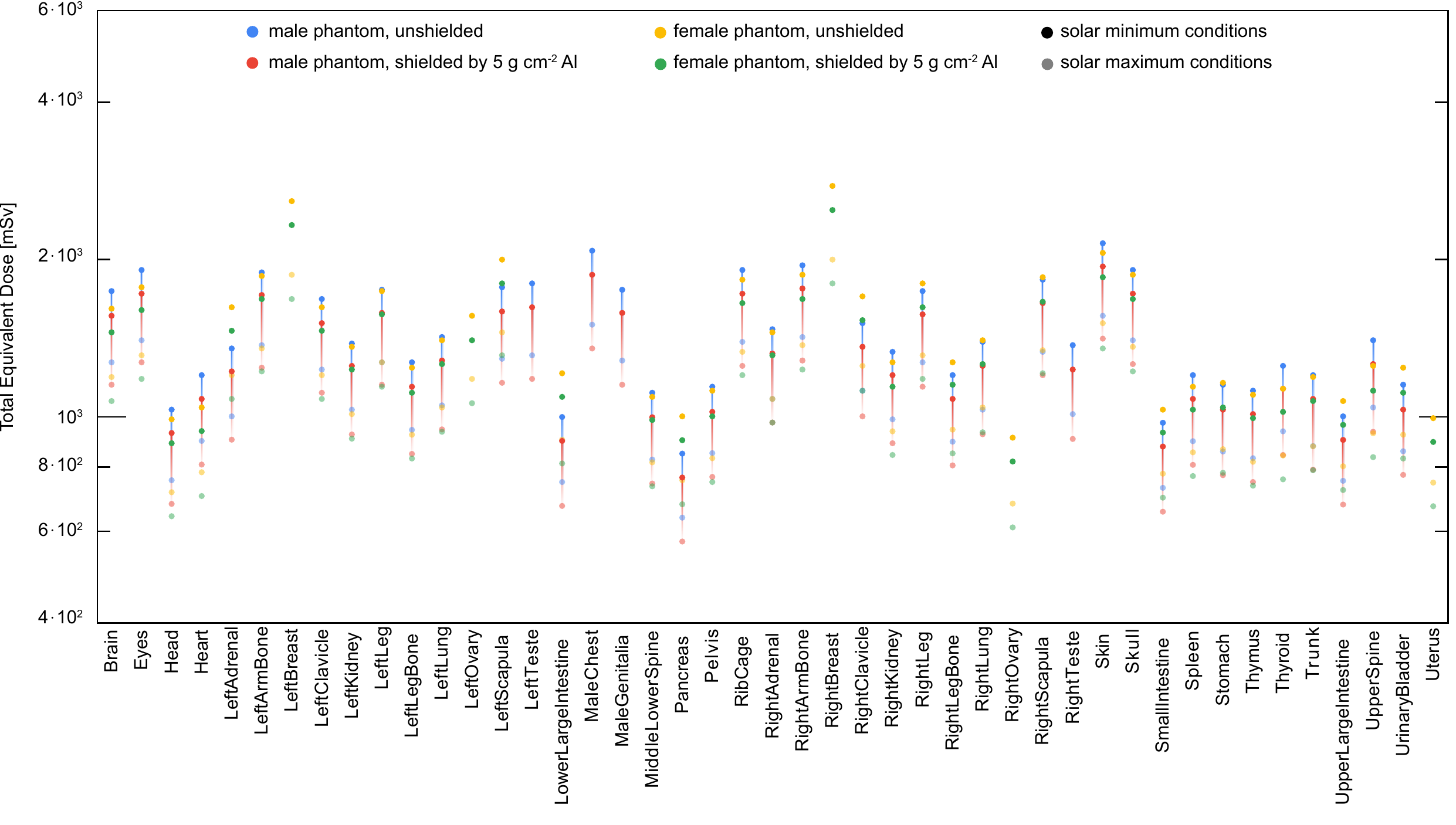}
    \caption{Total GCR-induced equivalent dose rates during a crewed mission to Mars (600 days on cruise and 400 days on the surface) during solar minimum (colored symbols) and maximum (shaded symbols) conditions.}
    \label{fig:mission}
\end{figure}

Because of the distance and orbital considerations, the planned crewed missions to Mars are long-duration missions, exposing astronauts to enhanced radiation from GCRs and SEPs up to a level we have never witnessed before in human spaceflight. While SEPs are hard to predict and their intensity can vary drastically due to the lower energy range of incident particles, typical radiation shields such as the 5 g cm$^{-2}$ aluminum shield are very effective in protecting astronauts from their impact. GCRs, on the other hand, are continuous background radiation, and their flux can be estimated very accurately. Nevertheless, the energy of the particles involved is much higher, and typical shielding used in spacecraft does not provide much protection. Secondary productions from GCRs, especially thermal neutrons, threaten astronauts on missions to Mars. 
\newpage
\begin{filecontents*}{total2_new.csv}
organ,m_h_min,m_h_max,m_he_min,m_he_max,f_h_min,f_h_max,f_he_min,f_he_max
Organ,Solar min., Solar max., Solar min., Solar max.,Solar min., Solar max.,Solar min., Solar max.
Brain,1.74E+03,1.27E+03,1.56E+03,1.15E+03,1.61E+03,1.19E+03,1.45E+03,1.07E+03
Eyes,1.91E+03,1.40E+03,1.72E+03,1.27E+03,1.77E+03,1.31E+03,1.60E+03,1.18E+03
Head,1.03E+03,7.54E+02,9.29E+02,6.79E+02,9.87E+02,7.15E+02,8.88E+02,6.43E+02
Heart,1.20E+03,8.97E+02,1.08E+03,8.08E+02,1.04E+03,7.81E+02,9.37E+02,7.03E+02
LeftAdrenal,1.35E+03,1.00E+03,1.22E+03,9.02E+02,1.62E+03,1.20E+03,1.46E+03,1.08E+03
LeftArmBone,1.89E+03,1.37E+03,1.71E+03,1.24E+03,1.86E+03,1.35E+03,1.68E+03,1.22E+03
LeftBreast,0.00E+00,0.00E+00,0.00E+00,0.00E+00,2.59E+03,1.87E+03,2.33E+03,1.68E+03
LeftClavicle,1.68E+03,1.23E+03,1.51E+03,1.11E+03,1.62E+03,1.20E+03,1.46E+03,1.08E+03
LeftKidney,1.38E+03,1.03E+03,1.25E+03,9.23E+02,1.36E+03,1.01E+03,1.23E+03,9.06E+02
LeftLeg,1.75E+03,1.27E+03,1.58E+03,1.15E+03,1.74E+03,1.27E+03,1.57E+03,1.14E+03
LeftLegBone,1.27E+03,9.42E+02,1.14E+03,8.47E+02,1.24E+03,9.22E+02,1.11E+03,8.30E+02
LeftLung,1.42E+03,1.05E+03,1.28E+03,9.44E+02,1.40E+03,1.04E+03,1.26E+03,9.34E+02
LeftOvary,0.00E+00,0.00E+00,0.00E+00,0.00E+00,1.56E+03,1.18E+03,1.40E+03,1.06E+03
LeftScapula,1.77E+03,1.29E+03,1.59E+03,1.16E+03,2.00E+03,1.45E+03,1.80E+03,1.31E+03
LeftTeste,1.80E+03,1.31E+03,1.62E+03,1.18E+03,0.00E+00,0.00E+00,0.00E+00,0.00E+00
LowerLargeIntestine,9.97E+02,7.48E+02,8.97E+02,6.73E+02,1.21E+03,9.02E+02,1.09E+03,8.12E+02
MaleChest,2.08E+03,1.50E+03,1.87E+03,1.35E+03,0.00E+00,0.00E+00,0.00E+00,0.00E+00
MaleGenitalia,1.75E+03,1.28E+03,1.58E+03,1.15E+03,0.00E+00,0.00E+00,0.00E+00,0.00E+00
MiddleLowerSpine,1.11E+03,8.26E+02,9.96E+02,7.43E+02,1.09E+03,8.16E+02,9.84E+02,7.34E+02
Pancreas,8.48E+02,6.39E+02,7.63E+02,5.75E+02,1.00E+03,7.54E+02,9.00E+02,6.78E+02
Pelvis,1.14E+03,8.50E+02,1.02E+03,7.65E+02,1.12E+03,8.31E+02,1.00E+03,7.48E+02
RibCage,1.91E+03,1.39E+03,1.72E+03,1.25E+03,1.83E+03,1.33E+03,1.65E+03,1.20E+03
RightAdrenal,1.47E+03,1.08E+03,1.32E+03,9.73E+02,1.45E+03,1.08E+03,1.31E+03,9.73E+02
RightArmBone,1.95E+03,1.42E+03,1.76E+03,1.28E+03,1.87E+03,1.37E+03,1.68E+03,1.23E+03
RightBreast,0.00E+00,0.00E+00,0.00E+00,0.00E+00,2.77E+03,2.00E+03,2.49E+03,1.80E+03
RightClavicle,1.51E+03,1.12E+03,1.36E+03,1.00E+03,1.70E+03,1.25E+03,1.53E+03,1.12E+03
RightKidney,1.33E+03,9.87E+02,1.20E+03,8.88E+02,1.27E+03,9.36E+02,1.14E+03,8.43E+02
RightLeg,1.74E+03,1.27E+03,1.57E+03,1.14E+03,1.80E+03,1.31E+03,1.62E+03,1.18E+03
RightLegBone,1.20E+03,8.94E+02,1.08E+03,8.05E+02,1.27E+03,9.43E+02,1.15E+03,8.49E+02
RightLung,1.39E+03,1.03E+03,1.25E+03,9.24E+02,1.40E+03,1.04E+03,1.26E+03,9.32E+02
RightOvary,0.00E+00,0.00E+00,0.00E+00,0.00E+00,9.10E+02,6.80E+02,8.19E+02,6.12E+02
RightScapula,1.83E+03,1.33E+03,1.65E+03,1.20E+03,1.85E+03,1.34E+03,1.66E+03,1.21E+03
RightTeste,1.37E+03,1.01E+03,1.23E+03,9.05E+02,0.00E+00,0.00E+00,0.00E+00,0.00E+00
Skin,2.15E+03,1.56E+03,1.94E+03,1.41E+03,2.06E+03,1.51E+03,1.85E+03,1.35E+03
Skull,1.91E+03,1.40E+03,1.72E+03,1.26E+03,1.87E+03,1.36E+03,1.68E+03,1.22E+03
SmallIntestine,9.72E+02,7.29E+02,8.75E+02,6.56E+02,1.03E+03,7.76E+02,9.31E+02,6.98E+02
Spleen,1.20E+03,8.96E+02,1.08E+03,8.07E+02,1.14E+03,8.53E+02,1.03E+03,7.68E+02
Stomach,1.15E+03,8.56E+02,1.03E+03,7.71E+02,1.16E+03,8.65E+02,1.04E+03,7.79E+02
Thymus,1.12E+03,8.31E+02,1.01E+03,7.48E+02,1.10E+03,8.18E+02,9.92E+02,7.36E+02
Thyroid,1.25E+03,9.36E+02,1.13E+03,8.43E+02,1.13E+03,8.41E+02,1.02E+03,7.57E+02
Trunk,1.20E+03,8.77E+02,1.08E+03,7.90E+02,1.19E+03,8.76E+02,1.07E+03,7.88E+02
UpperLargeIntestine,1.00E+03,7.52E+02,9.01E+02,6.77E+02,1.07E+03,8.02E+02,9.63E+02,7.22E+02
UpperSpine,1.40E+03,1.04E+03,1.26E+03,9.34E+02,1.25E+03,9.28E+02,1.12E+03,8.35E+02
UrinaryBladder,1.15E+03,8.57E+02,1.03E+03,7.72E+02,1.24E+03,9.22E+02,1.11E+03,8.30E+02
Uterus,0.00E+00,0.00E+00,0.00E+00,0.00E+00,9.92E+02,7.46E+02,8.93E+02,6.72E+02
\end{filecontents*}
\DTLloaddb{total}{total2_new.csv}
\noindent\adjustbox{max width=\textwidth,max totalheight=\textheight-6ex}{
{\tiny\renewcommand{\arraystretch}{.8}
\resizebox{!}{.35\paperheight}{%
    \begin{tabular}{|c|c|c|c|c|c|c|c|c|}
    \hline
    \multirow{3}{*}{} & \multicolumn{4}{|c|}{Male Phantom} & \multicolumn{4}{|c|}{Female Phantom} \\
     \cline{2-9}
     & \multicolumn{2}{|c|}{No shield} & \multicolumn{2}{|c|}{Shield} & \multicolumn{2}{|c|}{No shield} & \multicolumn{2}{|c|}{Shield} \\
     \cline{2-9}
%     & Solar Min. & Solar Max. & Solar Min. & Solar Max. & Solar Min. & Solar Max. & Solar Min. & Solar Max. \\
     \hline
    \DTLforeach{total}
    {\organ=organ,\mhmin=m_h_min,\mhmax=m_h_max,\mhemin=m_he_min,\mhemax=m_he_max,\fhmin=f_h_min,\fhmax=f_h_max,\fhemin=f_he_min,\fhemax=f_he_max}
    {\DTLiffirstrow{}{\\}%
    \organ & \mhmin & \mhmax & \mhemin & \mhemax & \fhmin & \fhmax & \fhemin & \fhemax}\\
    \hline
    \end{tabular}\label{tab:total_GCRs}\\
    }}}
\captionof{table}{Total Equivalent Dose (mSv) from GCRs in various organs, with and without shielding, in a crewed Mars mission of 600 days in cruise and 400 days on the surface}
\vspace{0.5cm}
We further studied the integrated equivalent dose deposited in various organs of the human body during the entirety of a mission to Mars. Therefore, we consider a mission that will take 600 days in the cruise phase and a 400 days stay on the surface of Mars, assuming both un-shielded and shielded conditions, respectively. The corresponding results are listed in Tab.~7 and shown in Fig.~\ref{fig:mission}. Here the total GCR-induced equivalent dose rates of the male and female  phantom during unshielded (blue/yellow dots) and shielded conditions (red/green dots) are displayed. The results have been performed for solar minimum (fully colored symbols) and maximum (shaded colored symbols) conditions. As can be seen, the GCR-induced total dose rates received during the entire mission vary only within about 35\% for different solar activity phases and within about 17\% when a 5 g cm$^{-2}$ aluminum shielding is assumed.
The International Commission on Radiological Protection (ICRP) guides space agencies on the appropriate dose limits for astronauts and cosmonauts. The European Space Agency (ESA), Russian Space Agency (RSA), and Japanese Space Agency (JAXA) set their dose limits based on this guidance. The ESA and RSA have set career dose limits of 1 Sv for their astronauts and cosmonauts \citep{Cucinotta2010ARS}. However, as can be seen, many of the radiation dose limits %set by these ESA, RSA, JAXA, and NASA 
would be exceeded in a Mars mission. Many individual organs, however, are expected to receive a radiation dose above 1 Sv on the Mars mission. 
\\
\\
Based on the latest report published by NASA regarding the limits of radiation exposure, the total career effective radiation dose due to space missions should not exceed 600 mSv \citep{NASATec2022}. Considering the results of our study, both in the absence and presence of shielding structures, many internal organs would be potentially irradiated with higher doses, leading to severe health hazards that should be addressed in preparing for future Mars missions. NASA suggests implementing \textit{as low as reasonably achievable} (ALARA) principles  to guarantee specific standards for radiation safety based on the duration of radiation exposure, the distance from the radiation source, and the shielding material and structures. 
\\
\\
In addition to career dose limits, ESA has also set annual dose limits for specific body parts, such as 0.5 Sv, 1.0 Sv, and 3.0 Sv for the blood-forming organs, the eye, and the skin, respectively \citep{Cucinotta2010ARS}. The bone marrow of the spine is one of the key parts of the body for the formation of blood and blood cells. Our results show that the lower and upper spine can be expected to receive radiation doses ranging from 0.734 Sv to 1.40 Sv. Note, however, that the dose rate limits set by the ESA for, e.g., the blood-forming organs are annual dose limits. Thus, with an annual limit of 0.5 Sv, the allowed dosage on a 1000-day is 1.37 Sv, slightly below our upper limit of 1.40 Sv.

Some radiation doses derived from our calculations are consistent with nuclear accidents or exposure to radiotherapy, which are the only reliable sources of comparison in the existing literature to define an objective correlation between ionizing radiation and the onset of related pathologies in the human body. It is also necessary to consider that organs are differently radiosensible, which means that similar dose rates might trigger cellular events with different gravity in the tissues and systems. 
\\
\\
There are notable physiological systems that are affected by low doses of radiation. For example, mice exposed to doses as small as 0.1 or 0.25 Gy have displayed deficiencies in recognition memory nine months after the exposure took place \citep{Frederico2009278}. Mice receiving doses as small as 0.1 Gy have also shown damages associated with neurodegenerative diseases, such as amyloid accumulation and changes to behavior and cognition \citep{Cherry2012e53275}. The above table shows that the brain can receive radiation doses from 1.07 Sv to 1.74 Sv. With such small doses associated with accumulating brain damage in mice, it is worth considering if these radiation doses associated with space travel could have long-term effects on astronauts after returning to Earth. Other organs affected by low doses of radiation include the thyroid gland and the cardiovascular system. Exposure of the thyroid gland to radiation doses as low as 50-100mGy has been associated with increased thyroid malignancies in children \citep{Sinnott2010756}. Likewise, studies of Chernobyl survivors have shown that exposure to as little as 0.15 Gy of radiation can increase an individual's chances of developing radiation-induced cardiovascular disease \citep{Hughson2018167}.
\\
\\
Our results further show that the left and right lungs are predicted to receive radiation doses ranging from 0.00287 Sv to 0.561 Sv. These are low doses. However, the Martian environment poses other threats to pulmonary health, such as harmful perchlorates in the Martian soil \citep{Wadsworth20174662}. The skin is another organ that has been observed to be sensitive to low doses of radiation. Irradiation of the face and neck with 0.1 to 0.5 Gy in patients with tinea capitis has been strongly correlated with the development of basal cell carcinoma \citep{MyungHee20061798}. Additionally, the digestive system is reported to exhibit reduced gastric motility after irradiation at doses as low as 1 Gy \citep{Jones202039}.
\subsection{The SPE-induced Radiation Exposure During Cruise Phase and on the surface of Mars}
\begin{figure}
    \centering
    \includegraphics[width=\textwidth]{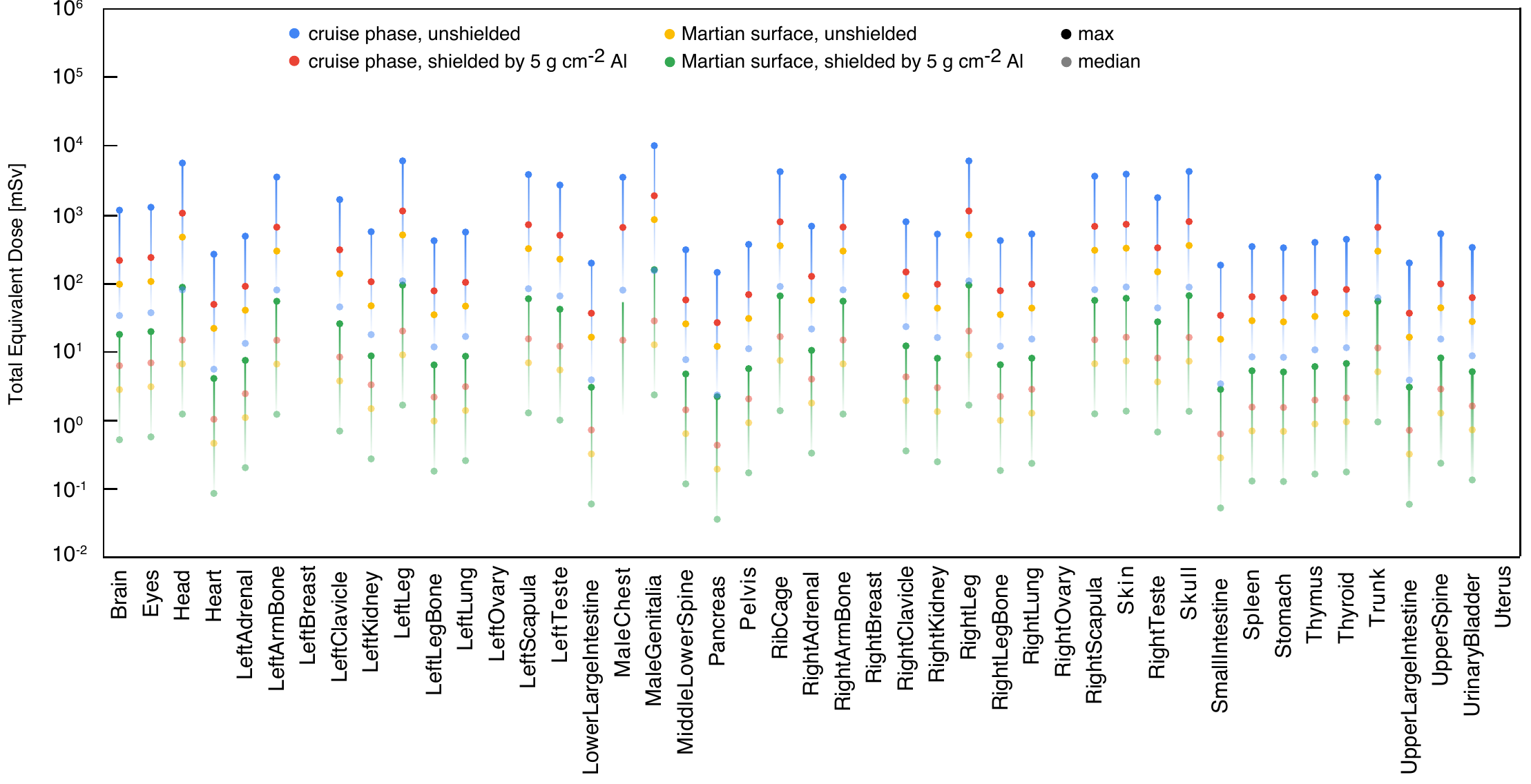}
    \caption{Maximum and median equivalent dose (mSv) of all SEP events in various organs, with and without shielding.}
    \label{fig:mission2}
\end{figure}

\begin{filecontents*}{seps_summary_m2_new.csv}
organ,m_h_min,m_h_max,m_he_min,m_he_max,f_h_min,f_h_max,f_he_min,f_he_max
Brain,3.41E+01,1.17E+03,6.34E+00,2.17E+02,2.84E+00,9.74E+01,5.28E-01,1.81E+01
Eyes,3.75E+01,1.29E+03,6.97E+00,2.39E+02,3.13E+00,1.07E+02,5.81E-01,1.99E+01
Head,8.08E+01,5.69E+03,1.50E+01,1.06E+03,6.73E+00,4.74E+02,1.25E+00,8.81E+01
Heart,5.63E+00,2.67E+02,1.05E+00,4.96E+01,4.69E-01,2.22E+01,8.72E-02,4.13E+00
LeftAdrenal,1.34E+01,4.89E+02,2.48E+00,9.10E+01,1.11E+00,4.08E+01,2.07E-01,7.58E+00
LeftArmBone,8.04E+01,3.56E+03,1.49E+01,6.62E+02,6.70E+00,2.97E+02,1.24E+00,5.51E+01
LeftBreast,0.00E+00,0.00E+00,0.00E+00,0.00E+00,0.00E+00,0.00E+00,0.00E+00,0.00E+00
LeftClavicle,4.56E+01,1.67E+03,8.48E+00,3.10E+02,3.80E+00,1.39E+02,7.07E-01,2.59E+01
LeftKidney,1.80E+01,5.68E+02,3.34E+00,1.06E+02,1.50E+00,4.73E+01,2.78E-01,8.80E+00
LeftLeg,1.09E+02,6.12E+03,2.03E+01,1.14E+03,9.11E+00,5.10E+02,1.69E+00,9.48E+01
LeftLegBone,1.19E+01,4.20E+02,2.21E+00,7.81E+01,9.90E-01,3.50E+01,1.84E-01,6.50E+00
LeftLung,1.69E+01,5.61E+02,3.14E+00,1.04E+02,1.41E+00,4.68E+01,2.62E-01,8.70E+00
LeftOvary,0.00E+00,0.00E+00,0.00E+00,0.00E+00,0.00E+00,0.00E+00,0.00E+00,0.00E+00
LeftScapula,8.39E+01,3.87E+03,1.56E+01,7.19E+02,6.99E+00,3.22E+02,1.30E+00,5.99E+01
LeftTeste,6.58E+01,2.72E+03,1.22E+01,5.05E+02,5.49E+00,2.26E+02,1.02E+00,4.21E+01
LowerLargeIntestine,3.94E+00,1.98E+02,7.33E-01,3.69E+01,3.28E-01,1.65E+01,6.11E-02,3.07E+00
MaleChest,8.00E+01,3.53E+03,1.49E+01,6.56E+02,0.00E+00,0.00E+00,0.00E+00,0.00E+00
MaleGenitalia,1.54E+02,1.02E+04,2.85E+01,1.90E+03,1.28E+01,8.53E+02,2.38E+00,1.59E+02
MiddleLowerSpine,7.77E+00,3.10E+02,1.44E+00,5.76E+01,6.48E-01,2.58E+01,1.20E-01,4.80E+00
Pancreas,2.37E+00,1.45E+02,4.40E-01,2.69E+01,1.97E-01,1.21E+01,3.66E-02,2.24E+00
Pelvis,1.12E+01,3.71E+02,2.08E+00,6.89E+01,9.34E-01,3.09E+01,1.74E-01,5.74E+00
RibCage,9.05E+01,4.26E+03,1.68E+01,7.91E+02,7.54E+00,3.55E+02,1.40E+00,6.59E+01
RightAdrenal,2.17E+01,6.84E+02,4.04E+00,1.27E+02,1.81E+00,5.70E+01,3.37E-01,1.06E+01
RightArmBone,8.08E+01,3.57E+03,1.50E+01,6.63E+02,6.73E+00,2.97E+02,1.25E+00,5.52E+01
RightBreast,0.00E+00,0.00E+00,0.00E+00,0.00E+00,0.00E+00,0.00E+00,0.00E+00,0.00E+00
RightClavicle,2.35E+01,7.93E+02,4.36E+00,1.47E+02,1.96E+00,6.61E+01,3.63E-01,1.23E+01
RightKidney,1.63E+01,5.24E+02,3.03E+00,9.75E+01,1.36E+00,4.37E+01,2.52E-01,8.12E+00
RightLeg,1.09E+02,6.11E+03,2.03E+01,1.14E+03,9.09E+00,5.09E+02,1.69E+00,9.47E+01
RightLegBone,1.22E+01,4.22E+02,2.26E+00,7.85E+01,1.01E+00,3.52E+01,1.88E-01,6.54E+00
RightLung,1.55E+01,5.26E+02,2.87E+00,9.78E+01,1.29E+00,4.38E+01,2.39E-01,8.15E+00
RightOvary,0.00E+00,0.00E+00,0.00E+00,0.00E+00,0.00E+00,0.00E+00,0.00E+00,0.00E+00
RightScapula,8.12E+01,3.66E+03,1.51E+01,6.80E+02,6.77E+00,3.05E+02,1.26E+00,5.67E+01
Skin,8.89E+01,3.92E+03,1.65E+01,7.29E+02,7.41E+00,3.27E+02,1.38E+00,6.08E+01
RightTeste,4.41E+01,1.78E+03,8.20E+00,3.31E+02,3.68E+00,1.48E+02,6.84E-01,2.76E+01
Skull,8.84E+01,4.29E+03,1.64E+01,7.98E+02,7.37E+00,3.58E+02,1.37E+00,6.65E+01
SmallIntestine,3.45E+00,1.85E+02,6.42E-01,3.43E+01,2.88E-01,1.54E+01,5.35E-02,2.86E+00
Spleen,8.52E+00,3.45E+02,1.58E+00,6.41E+01,7.10E-01,2.87E+01,1.32E-01,5.34E+00
Stomach,8.40E+00,3.31E+02,1.56E+00,6.15E+01,7.00E-01,2.76E+01,1.30E-01,5.13E+00
Thymus,1.08E+01,3.97E+02,2.00E+00,7.38E+01,8.96E-01,3.31E+01,1.67E-01,6.15E+00
Thyroid,1.16E+01,4.40E+02,2.15E+00,8.18E+01,9.65E-01,3.67E+01,1.79E-01,6.82E+00
Trunk,6.20E+01,3.55E+03,1.15E+01,6.59E+02,5.17E+00,2.96E+02,9.60E-01,5.50E+01
UpperLargeIntestine,3.91E+00,1.99E+02,7.27E-01,3.69E+01,3.26E-01,1.65E+01,6.06E-02,3.08E+00
UpperSpine,1.55E+01,5.30E+02,2.89E+00,9.86E+01,1.29E+00,4.42E+01,2.40E-01,8.22E+00
UrinaryBladder,8.84E+00,3.35E+02,1.64E+00,6.22E+01,7.37E-01,2.79E+01,1.37E-01,5.18E+00
Uterus,0.00E+00,0.00E+00,0.00E+00,0.00E+00,0.00E+00,0.00E+00,0.00E+00,0.00E+00
\end{filecontents*}
\DTLloaddb{seps_summary_m2}{seps_summary_m2_new.csv}
\noindent\adjustbox{max width=\textwidth,max totalheight=\textheight-6ex}{
{\tiny\renewcommand{\arraystretch}{.8}
\resizebox{!}{.35\paperheight}{%
    \begin{tabular}{|c|c|c|c|c|c|c|c|c|}
    \hline
    \multirow{3}{*}{Organ} & \multicolumn{4}{|c|}{Cruise} & \multicolumn{4}{|c|}{Surface} \\
     \cline{2-9}
     & \multicolumn{2}{|c|}{No shield} & \multicolumn{2}{|c|}{Shield} & \multicolumn{2}{|c|}{No shield} & \multicolumn{2}{|c|}{Shield} \\
     \cline{2-9}
    & median & max & median & max & median & max & median & max \\
     \hline
    \DTLforeach{seps_summary_m2}
    {\organ=organ,\mhmin=m_h_min,\mhmax=m_h_max,\mhemin=m_he_min,\mhemax=m_he_max,\fhmin=f_h_min,\fhmax=f_h_max,\fhemin=f_he_min,\fhemax=f_he_max}
    {\DTLiffirstrow{}{\\}%
    \organ & \mhmin & \mhmax & \mhemin & \mhemax & \fhmin & \fhmax & \fhemin & \fhemax}\\
    \hline

    \end{tabular} \label{tab:total_SPEs}\\
  }  }}
\captionof{table}{Maximum and median equivalent dose (mSv) from all SEP events in various organs, with and without shielding}
\vspace{5mm}
It could further be the case that SPEs occur during long-term missions. Therefore, we also studied the impact of SPEs. The total SPE-induced equivalent dose rates of the male phantom organs are given in Tab.~8 and displayed in Fig.~\ref{fig:mission2}. Here the values of the event with the strongest impact on the equivalent dose rates (i.e., the event of 4th of September 1972, denoted as max) and the median of all 60 SPE events are shown for the 600-day cruise phase and the 400-day Martian surface phase (both without and with shielding) are given.
\\
\\
As can be seen, an event like the one that occurred on the 4th of September 1972 drastically increases the radiation exposure of a 1000-day Mars mission. Thereby, the event poses a bigger threat during the cruise phase than at the surface of Mars. However, an event reflecting the median of all 60 investigated events would only pose a minor risk to crewed missions.
\section{Radiation Mitigation Strategies}
\label{sec:sec6}
As science studies more about the severity and impact of the radiation astronauts will face in future manned missions, there is still much to learn about possible methods of minimizing both the overall exposure and biological damage. Protection from radiation for space missions gets broken up into three categories: monitoring, medical supplies, and shielding \citep{dubey_radiation_1997}. Currently, space agencies like NASA’s Johnson Space Center track the severity of solar radiation in addition to the constant stream of galactic cosmic radiation \citep{tran_how_2019}. Scientists are reviewing how dietary supplements, such as antioxidants, have the potential to arm the body against the damage caused by space radiation \citep{brown_antioxidant_2010}. Teams are working on passive and active shielding methods designed to absorb and deflect a percentage of particles that would otherwise harm astronauts in short and long-term doses. And while all these methods may protect astronauts on their journey to places like Mars, research is still being performed on what in-situ resources, such as Martian soil, can be used for habitat shielding while on the surface. Even though the astronauts’ health is the first priority of manned space missions, what they take with them is still limited to the mass the spacecraft can carry and, ultimately, the cost. While there is still so much to learn regarding the effectiveness of these and other methods studied for the protection of astronaut health and safety, it is important to take a step back and review where we are now before we can understand where we have to go tomorrow, and - more importantly - how we get there. 
\subsection{Radiation Environment Tracking}
In order to limit the threat of radiation, the simplest prevention method often involves tracking potentially harmful exposure levels before reaching astronauts. This involves monitoring solar particle events (SPEs) as well as galactic cosmic radiation (GCRs) \citep[e.g.,][]{tran_how_2019}. Tracking current radiation environments and accurately predicting future threats can help identify the safest times to schedule missions and send astronauts who are already in interstellar space sufficient warnings to use extra shielding \citep{dubey_radiation_1997,tran_how_2019}. Solar activity fluctuates in 11-year cycles where radiation spikes occur as solar particle events more often at the “solar maximum” of the cycle \citep{tran_how_2019,r_battiston_active_2012}. SPEs are relatively lower energy and typically happen during solar flares \citep{r_battiston_active_2012,meyer_cosmic_1974}. NASA’s Johnson Space Center’s models for solar activity are still in the early stages of development, but the goal is to build a predictive map of when SPEs are likely to occur \citep{tran_how_2019}. Galactic cosmic radiation occurs as a stream of high energy particles that is generally only affected by solar activity; for example, GCRs are generally lower during solar maximum \cite{tran_how_2019,r_battiston_active_2012}. In addition to varying energy levels, the plethora of space radiation also involves different base chemicals and particles \citep{r_battiston_active_2012,meyer_cosmic_1974}. These are just some elements that make effective space radiation monitoring both difficult and critical to space missions. Mapping the radiation environment not only helps planning and warning but can also model astronauts' expected dose on space missions. Predicting solar flares, more accurate mapping of the type and energy of space particles, and understanding astronauts’ long-term dosing limits are all necessary for understanding the effectiveness of all types of radiation mitigation strategies.

\subsection{Medicine and Dietary Strategies}
% May be worth including a small intro paragraph?

\subsubsection{Probiotics}
This review has presented the different pathological impacts of ionizing radiation on the physiological systems and genetic material of the human body. A potential mitigation strategy to reduce radiation-induced harm is related to the dietary use of probiotics, antioxidants, and vitamins. Probiotics are live microorganisms that confer a health benefit on the host when ingested in adequate amounts \citep{WHOFAO2006}. Previously, studies on Earth have shown that probiotics provide several benefits to human health, including competition against pathogens, treatment for dysbiosis in the GI tract, reduction of gastrointestinal distress, production of beneficial metabolites, interactions with host cells that promote immune and psychological health, and protection from infection. Some studies have shown that probiotics can help alleviate some human illnesses associated with space flight conditions; for example, on US crew space flights, there have been some reports of antibiotic-associated diarrhea (DAA) due to prescribed antibiotics for the crew \citep{Douglas20171}. A study by \citet{Ouwehand2014458} demonstrates that using capsules with three different strains of probiotics reduces the incidence of DAA in patients receiving antibiotic treatment. Cases of respiratory infections have also been reported in space crews \citep{Douglas20171}. The administration of probiotics in tablets or capsules to patients with the same symptoms on Earth has been shown to reduce the symptoms, duration, and fever during infection \citep{Vrese2005481}. Probiotics are also a promising strategy to protect our microbiome and normal tissues from radiation. On Earth, radiotherapy is one of the most common sources of IR, and its human health impairments have been evaluated during and after the treatment \citep{Abdollahi2014495}. A paper from  \citet{Kumar201229} found that a diet supplemented with probiotics may have therapeutic potential to decrease the risk of cardiovascular disease in rats with induced hypercholesterolemia. Therefore, a programmed regimen of probiotics could reduce the incidence of radiation-induced cardiovascular disease (RICVD) presented in this review. 

\citet{Timko2010201} carried out a clinical trial with 42 radio-oncology patients who had undergone adjuvant postoperative radiation therapy (RT) after abdominal and pelvic cancer. They were given a probiotic supplement starting on the first day and lasting until the end of the RT. At the end of the study, it was concluded that prophylactic probiotic therapy could prevent radiation-induced diarrhea described previously. An extensive review by G.L. Douglas and A.A. Voorhies presents that probiotics could help alleviate some other conditions associated with space flight, such as dermatitis, rashes, and psychological distress \citep{Douglas20171}. Despite the positive health benefits, probiotics must be carefully selected and evaluated for spaceflight based on their strain-specific benefits and relevance to likely crew conditions.

\subsubsection{Antioxidants and Vitamins}
\citet{Brown2010462} carried out an experiment where they exposed laboratory mice of the strain C57BL/6 to a whole-body radiation dose of 8 Gy. The mice were 7-8 weeks old and underwent radiation exposure either on a normal diet or supplemented with antioxidants. The antioxidant supplementation diet included antioxidants L-selenomethionine, sodium ascorbate, N-acetyl cysteine, alpha-lipoic acid, alpha-tocopherol succinate, and co-enzyme Q10. There were distinct groups of 14-20 mice each that were started on the antioxidant diet either immediately, 12 hours after or 48 hours after radiation exposure. The study found an overarching conclusion that in a group of 18 mice, 78\% survived what would have been a lethal whole-body irradiation when on an antioxidant-supplemented diet. Antioxidants are believed to counteract the cellular damage induced by producing free radicals and reactive oxygen species imparted by ionizing radiation.

The effects of two antioxidants (beta-carotene and alpha-tocopherol) on the acute side effects of radiation exposure were investigated in patients with stage I or II squamous cell carcinoma of the head and neck. Beta-carotene is a provitamin A (a substance that the body can convert into vitamin A) found in many orange-colored fruits and vegetables, including cantaloupe, pumpkin, and carrots. Alpha-tocopherols are compounds with vitamin E activity that are sourced mainly from olives and sunflower oils. In the U.S.A., consumption of the gamma-tocopherol counterpart is more common from dietary sources, including soybeans. The squamous cell carcinoma patients were divided into two groups. A group of 273 patients formed the supplementation arm, and 267 patients formed the placebo arm. All patients were over 18 years old. In the placebo group (those not receiving supplementation of B-carotene or a-tocopherol), 24.8\% reported severe side effects from the radiation therapy treatment. Comparatively, only 19.2\% of the group receiving antioxidant supplementation reported severe side effects. The positive protective effect of antioxidants was only observed when both B-carotene and a-tocopherol were administered, but not when a-tocopherol was administered alone. When the two antioxidants were administered in combination, the larynx was one of the most protected tissues from the side effects of radiotherapy. Common side effects of radiotherapy for the head and neck are those that affect the larynx, such as sore throat and difficulty swallowing. Those patients who received the combination antioxidant treatment also reported fewer sleep disturbances than those patients not receiving any antioxidant supplementation \citep{Bairati20165813}.

Although antioxidant supplementation is shown through this study to reduce some of the adverse side effects of radiotherapy, other studies suggest that antioxidants can interfere with the efficacy of radiotherapy treatment. Patients treated for oral cavity cancer via radiotherapy participated in a study investigating the antioxidant effects of a vitamin E rinse before and after each 2 Gy fraction of radiotherapy. The rinsing solution contained 400 mg of vitamin E. At the 2-year follow-up period, the overall survival of those patients was observed to be reduced in the group who had done the vitamin E rinse compared to the placebo group. This was attributed to an interference of the vitamin E with the treatment, however; an overall 36\% reduction in the side effect of symptomatic mucositis was also observed \citep{Ferreira2004313}. This shows a requirement for clinicians to weigh up the risks and benefits of using antioxidants to prevent radiotherapy-induced harm.

\citet{Burns2007195} carried out experimental models of rat skin exposure to space radiation with acute doses ranging from 0.5 to 10 Gy. Inflammation was observed to play a key role in cancer induction by radiation exposure. One of the main radiation types used in the experiment was iron radiation (which is an abundant component of space radiation). The researchers found that vitamin A had protective abilities against 47\% of radiation-induced neoplasms (abnormal/excessive growths). Vitamin A was found to suppress the expression of numerous (80\%) genes involved in the inflammatory response and thereby suppress the acute inflammatory response. This research supports the importance of a high beta-carotene (provitamin A) abundance in the diet of future astronauts. 

The interference of antioxidants with radiotherapy is not of much concern for the safety of astronauts, who are presumably not being treated for cancer whilst in spaceflight. The protective effects of antioxidants against radiation damage are of higher interest for space travel science. It is promising that a diet high in fruits and vegetables and the utilization of antioxidant supplements could mitigate the harmful effects of space radiation on astronaut health. Using male C57BL/6J mice, it has been demonstrated that dried plums may show a radioprotective role against IR. In particular, researchers have tested the antioxidant properties of dried plums against low-LET gamma-ray radiation and a combination of protons and HZE ions, mimicking space radiation. In the simulation, after 11 days of irradiation, mice showed a reduction in bone loss, usually induced by radiation. It has been postulated the natural polyphenols (gallic acid, caffeoyl-quinic acids, coumaric acid, and rutin) present in dried plums may contribute to the reduction of IR effects on the human bone \citep{Schreurs201621343}, and it would be interesting to analyze if there would be similar results for bone cells in microgravity conditions.

\subsection{Mitigation Hardware}
To this day, the easiest form of engineering countermeasures against radiation is considered to be the passive method of putting enough absorbing mass between the astronaut and space, and hence, passive shielding is acknowledged to be the most feasible method for protecting against space radiation \citep{langley_space_1970}. However, the biggest challenges every space mission has to overcome are mass requirements and the available technology. Only some mass can be sent to space. The more massive a spacecraft is, the more difficult and expensive it is to launch, or the fewer critical items it can take with it. Engineers have used this opportunity to explore active radiation shielding concepts driven by ideas that reduce the cost and mass required for equipment that keeps radiation exposure within sustainable limits. In this Section, we analyze both of these radiation mitigation methods, along with the proposed strategies to reduce energy levels during the human stay on the surface of Mars.

\subsubsection{Active Shielding}
Active radiation protection includes hardware that actively blocks or deflects GCRs or solar particles. Since the 1970s, three such types of shielding have been explored to actively protect against space radiation: electrostatic, plasma, and magnetic \citep{langley_space_1970}. Few studies have been conducted on active shielding concepts that use electrostatic methods. Electrostatic shields use electrically charged spheres to direct charged particles around the craft and create a “safe zone” that is minimally affected by radiation \citep{barthel_review_2019,metzger_asymmetric_2004,townsend_critical_2005}. Many studies of electrostatic shields measure effectiveness by how much of the particles’ flux is changed with the field’s voltage \citep{metzger_asymmetric_2004}.

In the 1980s, it was reported that designing shields that used electrostatic fields to deflect protons was difficult because they had very high voltage requirements and were “inherently insatiable” because concepts were hard to design and required particles to be oppositely charged to the electrostatic spheres  \citep{townsend_galactic_1984}. Later in 2005, electrostatic shields were re-examined, and calculations showed that the order of voltage and sphere radii magnitudes were too large or unrealistic to make any difference in deflecting galactic cosmic radiation ion energies \citep{townsend_critical_2005}. In 2004, a study on the use of novel asymmetric electrostatic shields was published, which was hypothesized to address many of the shortcomings in previous research on electrostatic shielding\citep{metzger_asymmetric_2004,townsend_critical_2005}. The effectiveness of asymmetric electrostatic fields was examined against both low and high-energy particles, including iron; essentially, the fields were more effective at changing the flux of low-energy particles than they were at changing high-energy particles, and the field’s effectiveness increased as the field’s voltage increased \citep{metzger_asymmetric_2004}. 

A later example of the most developed electrostatic shield concepts uses electrostatically inflated membrane structures (EIMS) made of aluminized Mylar to create a charge flux strong enough to defect solar radiation \citep{townsend_critical_2005,tripathi_meeting_nodate}. The EIMS had 10 kV of charging, 5 keV of energy for electron flux, and 5 mA of current. Again, it was found that low-energy particles were completely blocked, and the fluxes of high-energy particles were only reduced; this meant that the shield blocked solar particle events and was 70\% more effective than the best hydrogen-rich passive shielding against galactic cosmic radiation \citep{tripathi_meeting_nodate}. It was also concluded that electrostatic fields are very power-hungry for the voltages needed for adequate absorption of particles, but the issue can be overcome by designing the electrostatic fields to deflect particles instead of stopping them. Additionally, due to the differences in particle energies, the fields were significantly more effective at deflecting solar particle energy than galactic cosmic radiation. \citep{tripathi_meeting_nodate}. This design was a modular implementation of electrostatic fields and comprised lightweight Gossamer structures that had the potential for easy deployment on spacecraft. Some of the biggest drawbacks to the design were unpredicted vibrations in the structures that showed more study of EIMS is required before they can effectively be used for protecting a spacecraft on a journey to Mars\citep{tripathi_meeting_nodate}. Many studies regarding electrostatic field designs do not have the same heritage of research, materials, and development as other active shielding architectures do. This lack of heritage affects the concept’s ability to become a robust design for spaceworthiness. Much of the technology required for electrostatic shields to be feasible for interplanetary missions is still in development \citep{barthel_review_2019}.

Plasma shields use a combination of electrostatic and magnetic fields where an electrostatic charge creates a pocket of electrons (the plasma) held in place by a magnetic field \citep{townsend_critical_2005,levy_plasma_1968}. Much like the electrostatic shields, in plasma shields, voltage plays a big role in the effectiveness of the electron cloud around the spacecraft. A magnetic field generated by superconducting coils would have to be strong enough to contain the plasma around the craft. The biggest issues with implementing such a concept are the voltage outputs, the plasma stability, and how the plasma would affect the rest of the spacecraft systems \citep{levy_plasma_1968}.

It is possible to use space plasma for both propulsion and radiation protection by creating a “mini-magnetosphere” that can shield up to energies 200 times greater than the magnet itself with a bending power in the order of 1 Tm \citep{winglee_advances_2004,jia_study_2013}. Plasma shields show a lot of potential for architectures that can be multipurpose and modular but lack the development and study needed for spaceflight. Much like the electrostatic shields, implementations of plasma shields for radiation protection require more robust study before they can be used for a mission to Mars. Magnetic shields are currently the most popular and developed concept for active radiation protection. Much like how the magnetosphere protects the Earth’s surface from the most harmful radiation, the idea is to use power to generate a magnetic field that surrounds the spacecraft. Given enough energy in the right configuration for the spacecraft, instead of using charged particles in the form of plasma, the magnetic field will either absorb or deflect oncoming particles before they reach the spacecraft. While it is difficult to protect against all radiation particles, the goal is to deflect enough particles to minimize the effects of harmful radiation dosing on astronauts and equipment for long space missions.

The International Space Station (ISS) uses a device called the Alpha Magnetic Spectrometer (AMS-02), a 5-by-4-by-3-meter superconducting magnet that detects and deflects cosmic particles in its path \citep{noauthor_ams_nodate,r_battiston_active_2012,rainey_alpha_2015}. While the AMS-02 generates a field strength of 0.14 T \citep{noauthor_ams_nodate}, it is isolated to the AMS to prevent interactions with Earth’s magnetosphere and causing navigation issues with the ISS. The AMS-02 measures the deflection of galactic cosmic radiation and solar particle events the ISS encounters \citep{rainey_alpha_2015}. Not only did the AMS-02 help develop our understanding of the cosmic particles we will encounter on manned space missions, but it has also provided excellent groundwork for conceptualizing improvements to active shielding for such missions.

Starting from the AMS-02, different magnet geometries and field strengths were simulated to model the effects those fields have on the reduction of radiation dosing for long missions \citep{r_battiston_active_2012}. One of the best-developed designs is the “Double Helix”, presenting several superconducting magnetic cylindrical solenoids with two layers of helical cable windings tilted opposite to each other that would generate a toroidal magnetic field surrounding the spacecraft.

In particular, two different Double Helix magnet strengths, measured in Tesla (T), were evaluated: the 2T, 4Tm Double Helix coils, with a field strength of 2 T and a bending power of 4 Tm;
and the 8 T, 16 Tm Double Helix, with a field strength of 8 T and a bending power of 16 Tm. This was then compared to both freespace and the presence of a passive spacecraft shield of aluminum \citep{r_battiston_active_2012}. Table 9 shows the annual dosing (at solar minimum) for skin, blood-forming organs (BFO), and the body.

% ADD: Table of results
\begin{table}[t]
    \centering
    \begin{tabular}{|c|c|c||c|c|c||c|c|c|} % number of columns
    \hline
% Headers\\
    \multicolumn{9}{|c|}{Effective Reduction in Annual Radiation Dosing}\\
    \hline
    \multicolumn{3}{|c||}{Passive Shield$^{*}$} & \multicolumn{3}{|c|}{2T, 4Tm Double Helix} & \multicolumn{3}{||c|}{8T, 16Tm Double Helix}\\
     \hline\hline
     % DATA & DATA & DATA & ETC \\
37\% & 28\% & 29\% & 50\% & 41\% & 40\% & 68\% & 61\% & 58\% \\
     \hline
\multicolumn{3}{|c||}{With passive shielding} & 21\% & 18\% & 16\% & 50\% & 46\% & 41\% \\
\hline
    \end{tabular}
    \caption{The percent reduction of the annual radiation dosing from freespace using passive shielding, 2T, 4Tm Double Helix, and 8T, 16Tm Double Helix; includes the additional reduction in dosing both Double Helix designs offer in addition to the spacecraft's passive shield(Passive shield was defined as the "spacecraft" with a 1.5 cm aluminum shielding. BFO - Blood Forming Organs).}
    \label{table:DoubleHelixReults1}
\end{table}

As it results, the Double Helix is more effective than passive shielding. While the 2T, 4Tm Double Helix could reduce radiation doses by 40\%, the 8T, 16Tm Double Helix configuration could reduce them by 60\% at the cost of more power and mass. In addition, the shield would effectively reduce the constant level of galactic cosmic radiation during an interplanetary mission to doses of 26-28 rem/y at solar maximum and 37 rem/y at solar minimum; both annual doses well below the maximum of 50 rem/y allowed for astronauts. Based on the energy of the particles the shield was simulated to deflect, the Double Helix would also be very effective in protecting against random solar particle events, should they arise during a several-year voyage to Mars.

Since the study was published in 2012, any future Mars missions using this or similar concepts would have at least eight years of new heritage technology at their disposal. Reviewing past architectures is often the first step in designing new and innovative technologies. One of the biggest challenges in choosing the best type of active shielding for long-duration space missions has much to do with the lack of concepts in the architecture space. For example, the lack of studies using alternative types of active shielding to magnetic fields hints that magnetic field-based designs comprise the majority of concepts. The maturity and continued study of this superconducting magnet design shows great potential for realistically making long-lasting, semi-modular designs for long space missions \citep{r_battiston_active_2012}. Other designs can show engineers the potential to use multiple shielding types to meet the redundancies necessary for a manned exploration mission or even more easily repaired with in-situ resources \citep{tripathi_meeting_nodate}. Much further research and developments are needed to make space-worthy active shielding alternatives robust enough to help us safely send astronauts to Mars. In the meantime, superconducting magnet designs are a great place to start modeling the effectiveness of dosing reduction by using updated radiation environment simulations.

\subsubsection{Passive Shielding}

Passive radiation shielding is one of the countermeasures to exposure to space radiation. This protection includes using a sufficient amount of material to absorb the energy from cosmic radiation. Earlier, research studies on passive shielding focused on using additional mass for the primary purpose of space radiation protection. However, using additional mass just for the purpose of radiation shielding is impractical due to the limited launch capability of rockets and the mass required to sustain astronauts for extended periods of time for missions beyond Low Earth Orbits. Furthermore, adding extra thickness does not reduce much the incoming energy. From up to a point, energy levels are nearly constant, or the extra thickness even increases the dose in some cases because of the creation of secondary particles. Therefore, there had been suggestions about alternate passive shielding options like the use of advanced structural materials, food, or astronaut waste. 

Some general statements influencing the selection criteria of candidate materials in shielding are the following. Firstly, due to human health and safety as a primary concern, shielding designs need to be conservative and adhere to the ALARA. It is a safety principle designed to minimize radiation doses.  Secondly, materials must be lightweight and should have desirable characteristics such as structural, mechanical, and thermal, along with their space radiation shielding properties. The common materials which are considered with regard to the space radiation shielding experiments are aluminum, copper, polyethylene, water, and graphite. These materials are frequently present in current spacecraft designs. They are chemically well-defined and have the required physical properties. 

The effectiveness of a material as a radiation shield generally increases with decreasing atomic number \citep{guetersloh2006polyethylene}. Materials with the most electrons per unit mass, least mean excitation energy and least tight binding energies are the best energy absorbers \citep{wilson1995hzetrn}. These conditions make hydrogen the most favored material. A series of measurements were conducted using a particle beam, 1 GeV/nuc \textsuperscript{56}Fe, which is a representative of the heavy ion component of GCRs, and various shielding materials including composites, light elements, and heavy elements in \citep{zeitlin2006measurements}. Results from these measurements confirmed that hydrogen is the most effective shielding material. Since hydrogen is highly effective, polyethylene ($CH_2$), which has two hydrogen atoms and one carbon atom molecule per molecule, is also an effective shielding material. Apart from these, its availability, non-toxicity, and chemically stable nature under typical conditions make polyethylene the most convenient reference material for shielding tests. 

Graphite also occupied a special significance in these tests as carbon is present in the molecular formula of polyethylene, and the atomic mass of this carbon component is divisible by 4. Another reason behind the selection of graphite is that it is a component of carbon composites. The mechanical and radiation shielding properties of carbon fibers and carbon composites enable them to function as the multifunctional superstructure of future spacecraft or planetary surface habitats \citep{national2008managing}. Dosimetric measurements were carried out for each one of these materials at shielding depths ranging from 5-20 g/cm\textsuperscript{2} in 5 g/cm\textsuperscript{2} intervals, finding a similar response with aluminum composites.

The shielding effectiveness of any given material at a specific depth can be compared to the reference value of 1.000 corresponding to polyethylene \citep[e.g.,][]{dewitt2020shielding}). A shielding effectiveness value is the ability of a given material to reduce the absorbed dose and dose equivalent at a given depth x. In other words, it quantifies how well a material can absorb or deflect the radiation it encounters. For a set of values at a given depth, the value nearest or exceeding 1.000 means that the corresponding material performs well relative to polyethylene. In the above study, two important depths were tested, 5 g/cm\textsuperscript{2} (bulkhead) and 20 g/cm\textsuperscript{2} (storm shelter). Results have shown that water is the most effective space radiation shielding material relative to polyethylene, whether based on absorbed dose or dose equivalent. Shielding properties observed for water could be of practical significance as water is a necessary consumable and the crew will require large quantities of it in long interplanetary missions. As for the other materials tested in \citep{dewitt2020shielding}, copper had the worst behavior, followed by aluminum and graphite.

Despite so many merits, the main drawback of polyethylene is its poor mechanical properties. Therefore, one of the main priorities of space agencies today is to develop a highly efficient radiation shielding material that is also capable of withstanding the typical loads occurring during a mission. Nanoparticles of carbon and boron can play a major role in designing such materials as they have a high modulus and great electrical and thermal properties. According to  \citet{kanagaraj2007mechanical}, carbon nanotubes enhance the mechanical properties of polyethylene and provide a good load transfer to the matrix. Other studies like \citet{nambiar2012polymer}  have shown that carbon nanotubes and nanoclays have the potential to reduce the harmful effects of ionizing radiation and to create multifunctional composite materials. Other than that, boron-based composites are found to be reliable materials for radiation shielding. This is due to the neutron absorbing ability of \textsuperscript{10}B isotope, which has a wide cross-section.

In \citet{Laurenzi:2020}, numerical analysis was performed with the HZETRN2015 code on typical aerospace materials such as Kapton, Aluminium, PPS, PEEK, and RTM6 to highlight the performance of polyethylene in radiation protection. This was followed by the investigation of the shielding properties of polyethylene-based nanocomposites at different percentages of fillers, considering the three types of radiation: GCR, SPE, and particles in the LEO environment. Results have shown that Medium Density Polyethylene (MDP) is the most effective polymer in the reduction of equivalent doses from the three sources of radiation. It was confirmed in the simulations that high hydrogen content is important in limiting the damage by both protons and heavy nuclei. Aerospace-grade epoxy RTM6 has also shown good radiation protection properties. This numerical analysis also suggested that graphene oxide (GO) could be a useful reinforcement of polyethylene matrices for the fabrication of composites that are multifunctional and have radiation-shielding properties in the space environment.

In general, the radiation shielding performances of a given material are studied in three steps. Firstly, Monte Carlo simulations are used to validate the shielding capabilities of a material. If results are promising, the material properties are further evaluated by ion irradiation particle accelerator facilities. It is expected that ions in these tests are the most abundant ones in space. Lastly, materials that perform the best in both these steps undergo a characterization in space. Currently, the ISS is the best available laboratory to test material response in space radiation. Even though it is within the protection of Earth’s magnetic field, the radiation spectrum in ISS at high altitudes replicates the spectrum of outer space radiation \citep{narici2015radiation}. Measurements were performed in the ISS during the ALTEA shield project on Polyethylene and Kevlar to investigate shielding effectiveness. Three detectors of ALTEA were used, which could merge radiation measurements with ISS position \citep{zaconte2008altea}. Therefore, it was possible to select measurements in the orbital track that best represent the radiation expected in a deep space environment. Moreover, as these three detectors were identical, they could be used concurrently. Results have shown that Kevlar has a good radiation shielding performance and hence is comparable to polyethylene. It is also resistant to impacts, which is important for debris shielding, making it a good candidate. Moreover, since it is available as a fabric, it may be adapted for use in Extra Vehicular Activity (EVA) suits or for extra shielding in some specific locations of habitat, like the crew sleeping quarters. These features make Kevlar an optimal candidate for a future integrated approach in space missions.

In space applications, every kilogram of mass has a significant impact on the mission cost and feasibility. While dense materials or thick layers of material are great at attenuating the energy levels of incoming radiation, they also contribute to increasing the amount of mass. Additionally, in space, the goal is to keep radiation from getting into the area where the astronauts are located. For long-duration space missions, the habitat of astronauts can be a bit larger than most Earth-based radiation sources, so the volume enclosed by the shield would be much larger. This will require even more shielding material. So, passive shielding can be prohibitively heavy when used as the sole method of radiation protection for long-duration space flights. These studies and results are proof that future research on passive shielding should not be just towards “better materials”, but the focus should be an integrated and harmonious approach to the shielding issue. This approach will consider different passive elements, the use of multipurpose materials and possibly active shielding as well along with pharmaceutical countermeasures.  

\subsubsection{Martian Habitats}

The next step of human exploration into space is establishing a human presence on planet Mars with a permanent habitat on the surface, but this adventure comes with its challenges. One of these is the danger derived from the absence of a magnetic field that, just like on Earth, would protect the surface from GCRs and solar energetic particles (SEPs). The Martian atmosphere is a natural low‐energy cutoff for incoming particles and only SEP events with a strong high‐energy component can be observed on the surface. GCRs are the main contribution to the surface radiation and are modulated by the solar activity variation in time \citep{guo2017dependence}. GCRs are present on Mars as a continuous radiation background and the long exposure makes them one of the main risks for future exploration missions \citep{cucinotta2006evaluating}.

Mars is characterized by an extremely rarefied atmosphere (the average surface pressure is 610 Pa) that is <1\% thinner than that of Earth and, as analyzed by \citep{guo2015modeling}, the surface radiation environment depends inversely on the atmospheric pressure: the measured dose rate decreases when the pressure increases; and in general, the column depth in the vertical direction is much smaller than toward the horizon \citep{guo2017dependence}. It’s also worth mentioning how the flux of GCR particles changes in anti-correlation with solar activity due to the solar wind. The surface dose rate is in fact anti-correlated with the atmospheric depth for solar modulation potentials smaller than $\sim$900-1000 MV; while for stronger solar activities, the anti-correlation vanishes, and the shielding effect disappears \citep{guo2017dependence}. This means that at weaker solar activities, more lower-energy primary GCR particles are more affected by atmospheric shielding. Therefore it is reasonable to consider the atmospheric shielding effect on the surface of Mars during the periods of solar minimum. By considering a Mars elevation map (e.g., \cite{Smith:1999}), we can consider the northern plains, as well as the Hellas Planitia, as the best settlement areas, in terms of radiation shielding, where the atmosphere can work as an additional protection to astronauts during extravehicular activities.

Since carrying shielding material from Earth to Mars could highly affect the mission costs, one of the most favored shielding solutions that can be employed on the surface of Mars is the usage of in-situ resources as a building material for a habitat: in this case surface regolith. In conformity to this, it is also worth mentioning the possibility of setting up a settlement underground, inside caves or lava tubes, which will be explored later. 
It has been demonstrated that the absorbed dose and equivalent dose rate induced by GCR particles vary with height above and below the Martian surface and with the subsurface composition. In particular, to the end of \citep{rostel2020subsurface} that considers the possibility of employing surface regolith as a shielding material, different depths below the surface were taken into account (which could be intended as different thickness values and how they mitigate radiation). The two most reasonable options to employ on Mars are:
\begin{enumerate}
\item A homogeneous mixture of 50\% water and 50\% basaltic andesite rock by weight, presumably realistic at the Martian North Pole (bulk density of 1.4 g/cm3);
\item An iron-rich sandstone (dry regolith) as analyzed by “ChemCam” on board Curiosity (bulk density of 2.2 g/cm3);
\end{enumerate}

Results show the 50\% water mixture is a lot more effective than dry regolith, making a water-rich material more recommended in the construction of a Mars habitat. Considering that the equivalent dose limit of radiation that astronauts can receive on the surface of Mars has been calculated to be 500 mSv/year, the walls of the habitat in question are recommended to present at least a 10-cm thick portion of this material. It must be taken into consideration that the Mars environment is also home to induction - in other words, secondary radiation - produced in the atmosphere and local surface, so using only in-situ resources might not be enough or could even turn out negative. The issue can be controlled by adding polymer binders to the regolith, but this kind of research still requires further development \citep{tripathi2006risk}.

Lava tubes have been identified in the northern shield of Alba Mons by the Viking orbiter, in the Tartarus Colles by the HiRISE camera on MRO, and on Olympus Mons, based on MGS/MOC images. According to studies conducted on Earth, caves tend to maintain stable environmental conditions \citep{leveille2010lava}, resulting in less stress on equipment than the wide diurnal swings on the surface. They have roofs tens of meters thick - around 20 meters on Mars \citep{walden1998utility} - making the environment relatively safe from solar radiation, cosmic rays, wind, dust storms, and micrometeorites. For these reasons, they are a favorable solution to setting up a habitat. Constructions inside a lava tube could be simple inflatable structures, making them lighter and faster to set up, and maintain more easily than surface structures.

When designing a Mars habitat, it will also be necessary to consider the psychological health of the astronauts other than their safety against radiation. For example, surface habitats built in regolith will be unable to have windows, and a completely closed environment may play an important role in the psychology of the crew during long-term missions. Similarly, being able to relax and work efficiently right under tens of meters of rock shouldn’t be underestimated, regardless of the unlikelihood of cave-ins in lava caverns that have survived for thousands, millions, or billions of years.

It seems that the best solutions to creating a habitat to effectively shield radiation on Mars are two, both involving in-situ regolith. The best one, in regards to blocking out the most radiation, is setting up the habitat inside an underground cave, thus exploiting the shielding properties of Mars’ surface itself. In fact, data shows that a depth of ten meters would already be able to entirely dissolve the effect of radiation. On the other hand, if the mission profile requires an on-surface habitat, the best solution would then be to use a mixture of regolith and water as a building material, with a minimum thickness of 10 cm in order to keep the received radiation dose below the limit that astronauts can receive in a year.

Nevertheless, other materials may be useful too. Simulations suggested that hydrogen-based materials are indeed more effective in terms of radiation in the Martian environment too. Specifically, materials with similar properties to polyethylene, like cyclohexane, PMMA, Mylar, and Kevlar, appeared to have almost identical behavior against cosmic rays, whereas liquid hydrogen outweighed by far any other material. Other alternatives, like carbon fiber or aluminum, still reduce dose levels at a sufficient percent though \citep{Gakis:2022}. It is then advocated to combine materials such as Martian regolith and aluminum in order to employ the in-situ resources and the spacecraft carrying astronauts. 

Considering the great shielding properties of water, it could also be interesting to consider placing water storage containers strategically on the walls and ceiling of the habitat and have them continuously refilled by the recycling systems. Previous research suggests that a layer of water 7 cm thick reduces the ionizing radiation transmitted through it by half. However, a drawback to water is that ionizing cosmic radiation is mostly gamma rays and protons that, if captured by the water, become H+ ions which could acidify the shielding water in time, making it unsafe to drink. 

A similar concept has been talked about using human wastes \citep{Falck:2017} as, in fact, urine and water present the same shielding effectiveness, and feces are able to shield even when removing water for recycling purposes, making it doubly useful. The same was considered for plants too, and fungi, in particular, were tested on the International Space Station, revealing that 1.7 mm could already decrease radiation levels by 1.82-5.04\%  \citep{Shunk:2021}.

Other materials – like polyethylene, as discussed in the passive shielding section – present the main drawback of weight. Should they be chosen to build the structure of the habitat, in fact, they would need to be carried from Earth to Mars in great quantity, meaning enormous costs. However, another and more interesting application of these materials could be found in spacesuits for extravehicular activities on the surface or in the structure of pressurized rovers, where regolith would be hardly applicable. A concept of a personal radiation shielding suit using water and human waste liquids has already been studied \citep{Baiocco:2018}.

Considering the addition of weight to the spacecraft as a consequence of using passive shielding, an active shielding solution may be more fitting to protect the astronauts during transit from Earth to Mars and back. The usage of active shielding for Mars habitats can also be considered by implementing a toroidal configuration \citep{musenich2013magnesium}. If the habitat presents a semi-cylindrical shape, large barrel toroidal magnets can surround it and be able to block a significant portion of ionizing radiation. However, given the non-negligible weight of a magnetic-field generating system, further studies and simulations should be carried out in order to assess the efficiency of both shielding solutions compared to their weight. Additionally, active shielding is not convenient on the surface of Mars, in terms of energy consumption, compared to passive. The only energy consumption derived from passive shielding is that required by the set-up phase – if we consider the energy needed by robotic 3D printers to build a habitat in regolith above the surface or the one needed to inflate modules below the surface. On the other hand, for active shielding, the energy required continuously in time should also be considered. Superconductors require a large amount of power and must also support a cooling system \citep{sargent2020magnetic}. These concepts make active shielding more expensive and with roughly the same shielding result. Consequently, to assess its feasibility, further research should focus on power management and the transportation of the system from Earth to Mars and its set-up on the surface.

\section{Summary}
\label{sec:sec7}

The radioprotection of the human reproductive system represents a fundamental and critical step to preserve the ability to procreate and lead to a healthy and successful pregnancy, especially in terms of potential future colonization. There is evidence of an increased risk for male gonadic cells to develop testicular germ cell tumor (TGCT) in patients who undergo diagnostic scans with doses that might exceed 20 mSv of direct and indirect radiation \citep{Nead201836}. Considering that our simulations account for doses of 1800 mSv and 1370 mSv, for the left and right testis, respectively, during a 1000-day mission, it might be assumed that the exposure to an even higher radiation dose is correlated with the reported cancer development. In addition, the shielding strategies evaluated in the simulation are not strong enough to significantly impact the radiation dose delivered to the male reproductive system. 

Similar conclusions might be derived by analyzing the female reproductive system in our computational system, considering that ovaries, uterus, and female breasts would be exposed to radiation doses that largely exceed the limit recently imposed by NASA for space missions. There is limited evidence in the literature of an association between radiation exposure and the onset of ovarian cancer but based on the results deriving from a cohort of 62,534 female nuclear bomb survivors, a positive but not significant correlation between ovarian cancer and ionizing radiation can be pointed out \citep{Utada2020195}. A decreasing trend in relative cancer risk with increased age at exposure was evidenced, along with a stronger correlation with type 2 ovarian cancers (serous and undifferentiated carcinomas), which happen to develop de novo with chromosomal instabilities and a more aggressive clinical phenotype. Considering the same cohort of nuclear bomb survivors, there is no significant correlation between ionizing radiation and cervical cancer, but there is suggestive evidence of a radiation-related increase in corpus cancer rates, especially in females irradiated before the age of 20 years when physiological changes take place at the endometrial tissue \citep{Utada20182}. Based on the higher radiosensitivity exhibited by, the younger female population, it would be reasonable to furtherly investigate the correlation between age and vulnerability to ionizing radiation in the future to better identify the most suitable candidates for long-term Mars space missions. 

Finally, in the same experimental cohort, there is an increased prevalence of female breast cancer in subjects exposed to nuclear radiation before the age of 20 years, and the age of menarche appears to be statistically correlated with a higher vulnerability to the biological effects of ionizing radiation and the onset or related breast cancer \citep{Brenner2018190}. We also found a significant radiation dose for male breast cancer, which could be a significant dose for male astronauts based on the chosen minimum cut-offs. Despite this consideration, it is necessary to report an eventual risk of male breast cancer-derived by the exposure to ionizing radiation, as evidenced by studies conducted with patients who underwent X-ray chest treatments, with an increased risk from 20 to 35 years after the initial exposure and a decline after three or four decades after the last registered exposure \citep{Thomas19945}. 

The evident discrepancy between radiation doses used in medical studies and the radiation doses that astronauts are expected to encounter during a Mars mission represents a research gap to be filled in the field of space medicine. With the central nervous system, thyroid gland, cardiovascular system, skin, digestive system, and reproductive systems all shown to be affected by low doses of radiation, we advocate that it is important to understand the potential effects of the radiation doses predicted for a Mars mission. This includes the long-term effects, such as alterations to DNA, after astronauts have returned to Earth. 

Several mitigation strategies have also been discussed. Medicine and dietary strategies include probiotics, antioxidants, and vitamins. The development of active shielding hardware is an active area of research, although none of these approaches have been certified as safe ratings for human spaceflight by any agencies. Passive shielding techniques are more developed, and both data and simulations provide many options for consideration in these long-duration missions. In general, hydrogen-rich materials provide the best shielding. Finally, various types of habitats which will provide ample shielding from the harsh radiation environment have been discussed. The best options are using a mixture of regolith and water for constructing walls and construction inside lava tubes, which will provide a safe environment from surface radiation, dust storms, and micrometeorites.

\section{Appendix}
\begin{table}
    \centering
    \begin{tabular}{|c|c|c|c|}
    \hline
Event & Year & Month & Day\\
    \hline
1 & 1956 & Feb & 23\\
2 & 1959 & Jul & 17\\
3 & 1960 & May & 4\\
4 & 1960 & Sep & 3\\
5 & 1960 & Nov & 12\\
6 & 1960 & Nov & 15\\
7 & 1960 & Nov & 20\\
8 & 1961 & Jul & 18\\
9 & 1967 & Jan & 28\\
10 & 1968 & Nov & 18\\
11 & 1969 & Mar & 30\\
12 & 1971 & Jan & 24\\
13 & 1971 & Sep & 1\\
14 & 1972 & Aug & 4\\
15 & 1972 & Aug & 7\\
16 & 1973 & Apr & 29\\
17 & 1976 & Apr & 30\\
18 & 1977 & Sep & 19\\
19 & 1977 & Sep & 24\\
20 & 1977 & Nov & 22\\
21 & 1978 & May & 7\\
22 & 1978 & Sep & 23\\
23 & 1981 & May & 10\\
24 & 1981 & Oct & 12\\
25 & 1982 & Nov & 26\\
26 & 1982 & Dec & 8\\
27 & 1984 & Feb & 16\\
28 & 1989 & Jul & 25\\
29 & 1989 & Aug & 16\\
30 & 1989 & Sep & 29\\
31 & 1989 & Oct & 19\\
32 & 1989 & Oct & 19\\
33 & 1989 & Oct & 22\\
34 & 1989 & Oct & 24\\
35 & 1989 & Nov & 15\\
36 & 1990 & May & 21\\
37 & 1990 & May & 24\\
38 & 1990 & May & 26\\
39 & 1990 & May & 28\\
40 & 1991 & Jun & 11\\
41 & 1991 & Jun & 15\\
42 & 1992 & Jun & 25\\
43 & 1992 & Nov & 2\\
44 & 1997 & Nov & 6\\
45 & 1998 & May & 2\\
46 & 1998 & May & 6\\
47 & 1998 & Aug & 24\\
48 & 2000 & Jul & 14\\
49 & 2001 & Apr & 15\\
50 & 2001 & Apr & 18\\
51 & 2001 & Nov & 4\\
52 & 2001 & Dec & 26\\
53 & 2002 & Aug & 24\\
54 & 2003 & Oct & 28\\
55 & 2003 & Oct & 29\\
56 & 2003 & Nov & 2\\
57 & 2005 & Jan & 17\\
58 & 2005 & Jan & 20\\
59 & 2006 & Dec & 13\\
60 & 2012 & May & 17\\
\hline
    \end{tabular}
    \caption{List of the 60 SPEs used to model the SPE-induced radiation doses.}
    \label{tab:1}
\end{table}
\newpage
\begin{filecontents*}{spe_events_10_1.csv}
organ,e1,e2,e3,e4,e5,e6,e7,e8,e9,e10,
Organ,Event 1, Event 2, Event 3, Event 4, Event 5, Event 6, Event 7, Event 8, Event 9, Event 10, 
Brain,8.99E+02,3.32E+02,3.53E+00,1.88E+01,9.72E+02,4.81E+02,2.19E+01,1.16E+02,4.18E+01,3.74E+01,
Head,1.08E+03,7.00E+02,3.77E+00,2.42E+01,2.40E+03,1.36E+03,3.09E+01,1.97E+02,6.15E+01,1.20E+02,
Heart,2.67E+02,5.70E+01,1.13E+00,4.10E+00,1.86E+02,8.97E+01,4.96E+00,2.14E+01,1.04E+01,6.05E+00,
LeftAdrenal,4.89E+02,1.42E+02,2.00E+00,8.99E+00,4.31E+02,2.10E+02,1.06E+01,5.15E+01,2.09E+01,1.54E+01,
LeftArmBone,1.48E+03,7.09E+02,5.58E+00,3.38E+01,2.08E+03,1.07E+03,4.00E+01,2.31E+02,7.52E+01,8.85E+01,
LeftBreast,0.00E+00,0.00E+00,0.00E+00,0.00E+00,0.00E+00,0.00E+00,0.00E+00,0.00E+00,0.00E+00,0.00E+00,
LeftClavicle,1.03E+03,4.22E+02,4.00E+00,2.26E+01,1.22E+03,6.11E+02,2.63E+01,1.44E+02,4.96E+01,4.86E+01,
LeftKidney,5.64E+02,1.86E+02,2.26E+00,1.12E+01,5.50E+02,2.70E+02,1.30E+01,6.62E+01,2.52E+01,2.04E+01,
LeftLeg,1.55E+03,8.84E+02,5.62E+00,3.51E+01,2.85E+03,1.56E+03,4.33E+01,2.64E+02,8.38E+01,1.35E+02,
LeftLegBone,4.20E+02,1.24E+02,1.71E+00,7.76E+00,3.76E+02,1.84E+02,9.15E+00,4.46E+01,1.81E+01,1.35E+01,
LeftLung,5.61E+02,1.78E+02,2.26E+00,1.09E+01,5.30E+02,2.59E+02,1.28E+01,6.37E+01,2.48E+01,1.93E+01,
LeftOvary,0.00E+00,0.00E+00,0.00E+00,0.00E+00,0.00E+00,0.00E+00,0.00E+00,0.00E+00,0.00E+00,0.00E+00,
LeftScapula,1.51E+03,7.51E+02,5.64E+00,3.51E+01,2.19E+03,1.12E+03,4.14E+01,2.44E+02,7.76E+01,9.38E+01,
LeftTeste,1.30E+03,5.90E+02,4.93E+00,2.93E+01,1.71E+03,8.70E+02,3.44E+01,1.96E+02,6.46E+01,7.11E+01,
LowerLargeIntestine,1.98E+02,3.91E+01,8.52E-01,2.90E+00,1.30E+02,6.28E+01,3.53E+00,1.48E+01,7.53E+00,4.13E+00,
MaleGenitalia,2.01E+03,1.31E+03,7.07E+00,4.64E+01,4.34E+03,2.43E+03,5.87E+01,3.74E+02,1.15E+02,2.14E+02,
MiddleLowerSpine,3.10E+02,8.33E+01,1.28E+00,5.41E+00,2.57E+02,1.25E+02,6.42E+00,3.03E+01,1.29E+01,9.00E+00,
Pancreas,1.45E+02,2.18E+01,6.40E-01,1.80E+00,7.95E+01,3.81E+01,2.27E+00,8.48E+00,5.12E+00,2.31E+00,
Pelvis,3.71E+02,1.14E+02,1.51E+00,6.92E+00,3.43E+02,1.68E+02,8.16E+00,4.05E+01,1.61E+01,1.26E+01,
RibCage,1.59E+03,8.00E+02,5.92E+00,3.68E+01,2.35E+03,1.21E+03,4.36E+01,2.58E+02,8.20E+01,1.02E+02,
RightAdrenal,6.78E+02,2.27E+02,2.70E+00,1.36E+01,6.69E+02,3.28E+02,1.59E+01,8.10E+01,3.05E+01,2.48E+01,
RightArmBone,1.50E+03,7.13E+02,5.64E+00,3.41E+01,2.09E+03,1.07E+03,4.03E+01,2.33E+02,7.58E+01,8.89E+01,
RightBreast,0.00E+00,0.00E+00,0.00E+00,0.00E+00,0.00E+00,0.00E+00,0.00E+00,0.00E+00,0.00E+00,0.00E+00,
RightClavicle,6.54E+02,2.31E+02,2.60E+00,1.32E+01,6.79E+02,3.36E+02,1.55E+01,8.08E+01,2.98E+01,2.59E+01,
RightKidney,5.24E+02,1.70E+02,2.11E+00,1.03E+01,5.04E+02,2.47E+02,1.20E+01,6.06E+01,2.33E+01,1.86E+01,
RightLeg,1.55E+03,8.83E+02,5.61E+00,3.50E+01,2.84E+03,1.56E+03,4.32E+01,2.64E+02,8.37E+01,1.34E+02,
RightLegBone,4.22E+02,1.27E+02,1.72E+00,7.90E+00,3.83E+02,1.88E+02,9.30E+00,4.56E+01,1.83E+01,1.39E+01,
RightLung,5.26E+02,1.63E+02,2.13E+00,1.01E+01,4.89E+02,2.39E+02,1.18E+01,5.87E+01,2.31E+01,1.77E+01,
RightOvary,0.00E+00,0.00E+00,0.00E+00,0.00E+00,0.00E+00,0.00E+00,0.00E+00,0.00E+00,0.00E+00,0.00E+00,
RightScapula,1.49E+03,7.29E+02,5.59E+00,3.45E+01,2.12E+03,1.09E+03,4.07E+01,2.38E+02,7.61E+01,9.01E+01,
RightTeste,8.92E+02,3.93E+02,3.41E+00,1.97E+01,1.14E+03,5.81E+02,2.32E+01,1.31E+02,4.38E+01,4.72E+01,
Skull,1.49E+03,7.58E+02,5.53E+00,3.39E+01,2.27E+03,1.19E+03,4.06E+01,2.40E+02,7.69E+01,9.99E+01,
SmallIntestine,1.85E+02,3.35E+01,8.02E-01,2.56E+00,1.15E+02,5.51E+01,3.15E+00,1.28E+01,6.85E+00,3.55E+00,
Spleen,3.45E+02,9.16E+01,1.43E+00,6.04E+00,2.83E+02,1.37E+02,7.15E+00,3.36E+01,1.43E+01,9.81E+00,
Stomach,3.31E+02,9.02E+01,1.37E+00,5.84E+00,2.77E+02,1.35E+02,6.92E+00,3.28E+01,1.39E+01,9.74E+00,
Thymus,3.97E+02,1.16E+02,1.62E+00,7.41E+00,3.52E+02,1.71E+02,8.70E+00,4.22E+01,1.71E+01,1.25E+01,
Thyroid,4.40E+02,1.27E+02,1.80E+00,8.18E+00,3.84E+02,1.87E+02,9.60E+00,4.63E+01,1.89E+01,1.36E+01,
Trunk,8.81E+02,4.97E+02,3.21E+00,1.94E+01,1.63E+03,8.99E+02,2.42E+01,1.47E+02,4.73E+01,7.77E+01,
UpperLargeIntestine,1.99E+02,3.86E+01,8.55E-01,2.87E+00,1.29E+02,6.23E+01,3.51E+00,1.46E+01,7.50E+00,4.09E+00,
UpperSpine,5.30E+02,1.65E+02,2.14E+00,1.02E+01,4.92E+02,2.41E+02,1.19E+01,5.93E+01,2.33E+01,1.79E+01,
UrinaryBladder,3.35E+02,9.35E+01,1.38E+00,5.95E+00,2.86E+02,1.40E+02,7.05E+00,3.38E+01,1.41E+01,1.02E+01,
Uterus,0.00E+00,0.00E+00,0.00E+00,0.00E+00,0.00E+00,0.00E+00,0.00E+00,0.00E+00,0.00E+00,0.00E+00,
,,,,,,,,,,,
,,,,,,,,,,,
\end{filecontents*}
\DTLloaddb{spe_events_1}{spe_events_10_1.csv}\label{tab:6}
\noindent\adjustbox{max width=\textwidth,max totalheight=\textheight-6ex}{
    \begin{tabular}{|c|c|c|c|c|c|c|c|c|c|c|}
    \hline
    \DTLforeach{spe_events_1}
    {\organ=organ,\eone=e1,\etwo=e2,\ethree=e3,\efour=e4,\efive=e5, \esix=e6, \eseven=e7, \eeight=e8, \enine=e9, \eten=e10}
    {%\DTLiffirstrow{}{\\}%
    \organ & \eone & \etwo & \ethree & \efour & \efive & \esix & \eseven & \eeight & \enine & \eten
    \DTLiflastrow{}{\DTLiffirstrow{\\ \hline}{\\}}
    }
    \\ \hline
    \end{tabular}\\
}
\captionof{table}{Radiation dose deposition in different organs for SEP events}

\begin{filecontents*}{spe_events_10_2.csv}
organ,e1,e2,e3,e4,e5,e6,e7,e8,e9,e10,
Organ,Event 11, Event 12, Event 13, Event 14, Event 15, Event 16, Event 17, Event 18, Event 19, Event 20, 
Brain,1.86E+01,7.52E+01,9.52E+01,1.17E+03,6.77E+01,3.26E+00,1.03E+01,1.24E+01,1.30E+01,2.41E+01,
Head,2.25E+01,2.85E+02,1.18E+02,5.69E+03,2.51E+02,5.22E+00,1.91E+01,3.74E+01,1.83E+01,4.77E+01,
Heart,4.38E+00,1.18E+01,2.25E+01,1.14E+02,1.07E+01,6.74E-01,2.04E+00,2.16E+00,3.27E+00,5.25E+00,
LeftAdrenal,9.19E+00,3.05E+01,4.69E+01,3.90E+02,2.76E+01,1.52E+00,4.67E+00,5.27E+00,6.56E+00,1.13E+01,
LeftArmBone,3.25E+01,1.86E+02,1.67E+02,3.56E+03,1.66E+02,6.27E+00,2.06E+01,2.85E+01,2.31E+01,4.77E+01,
LeftBreast,0.00E+00,0.00E+00,0.00E+00,0.00E+00,0.00E+00,0.00E+00,0.00E+00,0.00E+00,0.00E+00,0.00E+00,
LeftClavicle,2.20E+01,9.88E+01,1.13E+02,1.67E+03,8.87E+01,3.99E+00,1.27E+01,1.59E+01,1.54E+01,2.95E+01,
LeftKidney,1.12E+01,4.05E+01,5.72E+01,5.68E+02,3.66E+01,1.90E+00,5.92E+00,6.87E+00,7.88E+00,1.41E+01,
LeftLeg,3.31E+01,3.07E+02,1.72E+02,6.12E+03,2.72E+02,7.07E+00,2.47E+01,4.24E+01,2.54E+01,5.98E+01,
LeftLegBone,7.91E+00,2.69E+01,4.05E+01,3.52E+02,2.43E+01,1.31E+00,4.05E+00,4.62E+00,5.66E+00,9.84E+00,
LeftLung,1.10E+01,3.83E+01,5.63E+01,5.15E+02,3.47E+01,1.84E+00,5.72E+00,6.56E+00,7.76E+00,1.37E+01,
LeftOvary,0.00E+00,0.00E+00,0.00E+00,0.00E+00,0.00E+00,0.00E+00,0.00E+00,0.00E+00,0.00E+00,0.00E+00,
LeftScapula,3.34E+01,1.97E+02,1.71E+02,3.87E+03,1.76E+02,6.56E+00,2.16E+01,3.00E+01,2.38E+01,4.98E+01,
LeftTeste,2.83E+01,1.47E+02,1.45E+02,2.72E+03,1.32E+02,5.34E+00,1.73E+01,2.30E+01,2.00E+01,4.01E+01,
LowerLargeIntestine,3.14E+00,8.03E+00,1.61E+01,7.19E+01,7.34E+00,4.76E-01,1.43E+00,1.49E+00,2.37E+00,3.75E+00,
MaleGenitalia,4.30E+01,4.98E+02,2.26E+02,1.02E+04,4.40E+02,9.89E+00,3.57E+01,6.66E+01,3.44E+01,8.75E+01,
MiddleLowerSpine,5.60E+00,1.77E+01,2.86E+01,2.17E+02,1.61E+01,9.08E-01,2.78E+00,3.10E+00,4.05E+00,6.87E+00,
Pancreas,2.05E+00,4.45E+00,1.05E+01,2.89E+01,4.10E+00,2.96E-01,8.65E-01,8.69E-01,1.62E+00,2.42E+00,
Pelvis,7.03E+00,2.51E+01,3.59E+01,3.52E+02,2.27E+01,1.18E+00,3.67E+00,4.26E+00,5.03E+00,8.86E+00,
RibCage,3.51E+01,2.15E+02,1.80E+02,4.26E+03,1.92E+02,6.93E+00,2.29E+01,3.25E+01,2.52E+01,5.31E+01,
RightAdrenal,1.36E+01,4.92E+01,6.97E+01,6.84E+02,4.45E+01,2.32E+00,7.21E+00,8.35E+00,9.55E+00,1.71E+01,
RightArmBone,3.27E+01,1.87E+02,1.68E+02,3.57E+03,1.67E+02,6.31E+00,2.07E+01,2.86E+01,2.33E+01,4.81E+01,
RightBreast,0.00E+00,0.00E+00,0.00E+00,0.00E+00,0.00E+00,0.00E+00,0.00E+00,0.00E+00,0.00E+00,0.00E+00,
RightClavicle,1.32E+01,5.19E+01,6.74E+01,7.93E+02,4.68E+01,2.29E+00,7.20E+00,8.63E+00,9.30E+00,1.70E+01,
RightKidney,1.03E+01,3.69E+01,5.28E+01,5.08E+02,3.33E+01,1.75E+00,5.43E+00,6.27E+00,7.29E+00,1.29E+01,
RightLeg,3.30E+01,3.07E+02,1.72E+02,6.11E+03,2.72E+02,7.06E+00,2.47E+01,4.23E+01,2.53E+01,5.97E+01,
RightLegBone,8.04E+00,2.75E+01,4.11E+01,3.64E+02,2.49E+01,1.33E+00,4.13E+00,4.72E+00,5.73E+00,9.99E+00,
RightLung,1.02E+01,3.51E+01,5.23E+01,4.65E+02,3.18E+01,1.70E+00,5.29E+00,6.03E+00,7.23E+00,1.27E+01,
RightOvary,0.00E+00,0.00E+00,0.00E+00,0.00E+00,0.00E+00,0.00E+00,0.00E+00,0.00E+00,0.00E+00,0.00E+00,
RightScapula,3.30E+01,1.89E+02,1.69E+02,3.66E+03,1.69E+02,6.41E+00,2.10E+01,2.89E+01,2.34E+01,4.85E+01,
RightTeste,1.91E+01,9.79E+01,9.78E+01,1.78E+03,8.76E+01,3.59E+00,1.16E+01,1.53E+01,1.35E+01,2.69E+01,
Skull,3.23E+01,2.15E+02,1.66E+02,4.29E+03,1.91E+02,6.47E+00,2.16E+01,3.18E+01,2.36E+01,5.06E+01,
SmallIntestine,2.82E+00,6.89E+00,1.44E+01,5.75E+01,6.31E+00,4.21E-01,1.25E+00,1.30E+00,2.16E+00,3.36E+00,
Spleen,6.25E+00,1.93E+01,3.20E+01,2.25E+02,1.75E+01,1.01E+00,3.08E+00,3.40E+00,4.51E+00,7.60E+00,
Stomach,6.03E+00,1.92E+01,3.08E+01,2.35E+02,1.74E+01,9.79E-01,3.01E+00,3.36E+00,4.35E+00,7.40E+00,
Thymus,7.56E+00,2.47E+01,3.87E+01,3.06E+02,2.23E+01,1.24E+00,3.83E+00,4.29E+00,5.37E+00,9.26E+00,
Thyroid,8.35E+00,2.66E+01,4.27E+01,3.19E+02,2.41E+01,1.36E+00,4.19E+00,4.66E+00,5.93E+00,1.02E+01,
Trunk,1.84E+01,1.79E+02,9.57E+01,3.55E+03,1.58E+02,3.95E+00,1.39E+01,2.45E+01,1.43E+01,3.40E+01,
UpperLargeIntestine,3.12E+00,7.96E+00,1.60E+01,7.11E+01,7.27E+00,4.73E-01,1.41E+00,1.48E+00,2.36E+00,3.73E+00,
UpperSpine,1.03E+01,3.53E+01,5.28E+01,4.65E+02,3.20E+01,1.72E+00,5.33E+00,6.07E+00,7.29E+00,1.28E+01,
UrinaryBladder,6.13E+00,2.01E+01,3.13E+01,2.57E+02,1.82E+01,1.00E+00,3.09E+00,3.49E+00,4.41E+00,7.58E+00,
Uterus,0.00E+00,0.00E+00,0.00E+00,0.00E+00,0.00E+00,0.00E+00,0.00E+00,0.00E+00,0.00E+00,0.00E+00,
,,,,,,,,,,,
,,,,,,,,,,,
\end{filecontents*}
\DTLloaddb{spe_events_2}{spe_events_10_2.csv}\label{tab:7}
\noindent\adjustbox{max width=\textwidth,max totalheight=\textheight-6ex}{
    \begin{tabular}{|c|c|c|c|c|c|c|c|c|c|c|}
    \hline
    \DTLforeach{spe_events_2}
    {\organ=organ,\eone=e1,\etwo=e2,\ethree=e3,\efour=e4,\efive=e5, \esix=e6, \eseven=e7, \eeight=e8, \enine=e9, \eten=e10}
    {%\DTLiffirstrow{}{\\}%
    \organ & \eone & \etwo & \ethree & \efour & \efive & \esix & \eseven & \eeight & \enine & \eten
    \DTLiflastrow{}{\DTLiffirstrow{\\ \hline}{\\}}
    }
    \\ \hline
    \end{tabular}\\
}
\captionof{table}{Radiation dose deposition in different organs for SEP events}

\begin{filecontents*}{spe_events_10_3.csv}
organ,e1,e2,e3,e4,e5,e6,e7,e8,e9,e10,
Organ,Event 21, Event 22, Event 23, Event 24, Event 25, Event 26, Event 27, Event 28, Event 29, Event 30, 
Brain,5.85E+00,8.28E+01,5.81E+00,3.46E+01,6.22E+00,1.87E+01,8.62E+00,4.64E+00,9.56E+01,2.35E+00,
Head,1.08E+01,3.31E+02,2.80E+01,1.63E+02,1.66E+01,5.86E+01,1.18E+01,5.36E+00,2.17E+02,2.59E+00,
Heart,1.24E+00,1.06E+01,9.24E-01,5.01E+00,1.26E+00,3.62E+00,1.90E+00,1.15E+00,1.69E+01,8.33E-01,
LeftAdrenal,2.73E+00,3.02E+01,2.35E+00,1.33E+01,2.82E+00,8.27E+00,4.13E+00,2.34E+00,4.13E+01,1.40E+00,
LeftArmBone,1.17E+01,2.32E+02,1.51E+01,9.70E+01,1.34E+01,4.21E+01,1.57E+01,7.90E+00,2.05E+02,3.68E+00,
LeftBreast,0.00E+00,0.00E+00,0.00E+00,0.00E+00,0.00E+00,0.00E+00,0.00E+00,0.00E+00,0.00E+00,0.00E+00,
LeftClavicle,7.24E+00,1.15E+02,7.67E+00,4.68E+01,7.79E+00,2.37E+01,1.04E+01,5.43E+00,1.21E+02,2.65E+00,
LeftKidney,3.40E+00,4.19E+01,3.11E+00,1.79E+01,3.55E+00,1.05E+01,5.11E+00,2.82E+00,5.36E+01,1.54E+00,
LeftLeg,1.40E+01,3.69E+02,2.84E+01,1.72E+02,1.92E+01,6.50E+01,1.67E+01,7.96E+00,2.67E+02,3.79E+00,
LeftLegBone,2.35E+00,2.71E+01,2.08E+00,1.19E+01,2.45E+00,7.24E+00,3.56E+00,2.02E+00,3.60E+01,1.19E+00,
LeftLung,3.29E+00,3.89E+01,2.94E+00,1.68E+01,3.43E+00,1.01E+01,4.99E+00,2.78E+00,5.13E+01,1.54E+00,
LeftOvary,0.00E+00,0.00E+00,0.00E+00,0.00E+00,0.00E+00,0.00E+00,0.00E+00,0.00E+00,0.00E+00,0.00E+00,
LeftScapula,1.23E+01,2.52E+02,1.58E+01,1.03E+02,1.40E+01,4.41E+01,1.63E+01,8.09E+00,2.17E+02,3.71E+00,
LeftTeste,9.84E+00,1.80E+02,1.17E+01,7.39E+01,1.09E+01,3.40E+01,1.36E+01,6.90E+00,1.70E+02,3.25E+00,
LowerLargeIntestine,8.89E-01,7.12E+00,6.36E-01,3.42E+00,8.96E-01,2.56E+00,1.35E+00,8.27E-01,1.17E+01,6.36E-01,
MaleGenitalia,2.03E+01,6.05E+02,4.73E+01,2.88E+02,2.96E+01,1.03E+02,2.25E+01,1.02E+01,4.01E+02,4.81E+00,
MiddleLowerSpine,1.65E+00,1.74E+01,1.37E+00,7.72E+00,1.70E+00,4.96E+00,2.49E+00,1.43E+00,2.43E+01,9.14E-01,
Pancreas,5.76E-01,3.72E+00,3.65E-01,1.91E+00,5.70E-01,1.60E+00,8.50E-01,5.49E-01,6.71E+00,4.99E-01,
Pelvis,2.14E+00,2.63E+01,1.94E+00,1.13E+01,2.24E+00,6.63E+00,3.18E+00,1.78E+00,3.30E+01,1.06E+00,
RibCage,1.30E+01,2.75E+02,1.75E+01,1.14E+02,1.51E+01,4.78E+01,1.71E+01,8.48E+00,2.32E+02,3.91E+00,
RightAdrenal,4.12E+00,5.03E+01,3.77E+00,2.17E+01,4.30E+00,1.28E+01,6.23E+00,3.43E+00,6.54E+01,1.83E+00,
RightArmBone,1.17E+01,2.33E+02,1.51E+01,9.72E+01,1.34E+01,4.23E+01,1.58E+01,7.97E+00,2.07E+02,3.73E+00,
RightBreast,0.00E+00,0.00E+00,0.00E+00,0.00E+00,0.00E+00,0.00E+00,0.00E+00,0.00E+00,0.00E+00,0.00E+00,
RightClavicle,4.13E+00,5.67E+01,4.01E+00,2.37E+01,4.37E+00,1.31E+01,6.08E+00,3.30E+00,6.65E+01,1.76E+00,
RightKidney,3.12E+00,3.78E+01,2.83E+00,1.63E+01,3.25E+00,9.64E+00,4.70E+00,2.61E+00,4.90E+01,1.44E+00,
RightLeg,1.40E+01,3.69E+02,2.84E+01,1.72E+02,1.92E+01,6.49E+01,1.67E+01,7.94E+00,2.67E+02,3.78E+00,
RightLegBone,2.39E+00,2.78E+01,2.13E+00,1.22E+01,2.50E+00,7.37E+00,3.62E+00,2.04E+00,3.68E+01,1.19E+00,
RightLung,3.04E+00,3.54E+01,2.70E+00,1.54E+01,3.17E+00,9.34E+00,4.62E+00,2.59E+00,4.72E+01,1.46E+00,
RightOvary,0.00E+00,0.00E+00,0.00E+00,0.00E+00,0.00E+00,0.00E+00,0.00E+00,0.00E+00,0.00E+00,0.00E+00,
RightScapula,1.19E+01,2.39E+02,1.51E+01,9.79E+01,1.35E+01,4.25E+01,1.60E+01,7.99E+00,2.10E+02,3.68E+00,
RightTeste,6.61E+00,1.19E+02,7.77E+00,4.90E+01,7.33E+00,2.27E+01,9.13E+00,4.68E+00,1.13E+02,2.28E+00,
Skull,1.23E+01,2.74E+02,1.79E+01,1.17E+02,1.47E+01,4.72E+01,1.59E+01,7.83E+00,2.22E+02,3.68E+00,
SmallIntestine,7.97E-01,6.03E+00,5.51E-01,2.94E+00,7.99E-01,2.27E+00,1.20E+00,7.45E-01,1.01E+01,6.08E-01,
Spleen,1.81E+00,1.84E+01,1.49E+00,8.28E+00,1.86E+00,5.43E+00,2.77E+00,1.61E+00,2.67E+01,1.01E+00,
Stomach,1.77E+00,1.88E+01,1.48E+00,8.34E+00,1.83E+00,5.34E+00,2.69E+00,1.54E+00,2.63E+01,9.68E-01,
Thymus,2.21E+00,2.40E+01,1.89E+00,1.06E+01,2.29E+00,6.72E+00,3.39E+00,1.93E+00,3.37E+01,1.12E+00,
Thyroid,2.42E+00,2.55E+01,2.04E+00,1.14E+01,2.50E+00,7.32E+00,3.74E+00,2.13E+00,3.68E+01,1.24E+00,
Trunk,7.91E+00,2.13E+02,1.68E+01,1.01E+02,1.11E+01,3.78E+01,9.31E+00,4.43E+00,1.51E+02,2.19E+00,
UpperLargeIntestine,8.85E-01,7.06E+00,6.31E-01,3.39E+00,8.91E-01,2.55E+00,1.34E+00,8.21E-01,1.16E+01,6.43E-01,
UpperSpine,3.07E+00,3.54E+01,2.71E+00,1.54E+01,3.19E+00,9.41E+00,4.67E+00,2.62E+00,4.76E+01,1.47E+00,
UrinaryBladder,1.82E+00,2.01E+01,1.55E+00,8.81E+00,1.88E+00,5.52E+00,2.74E+00,1.56E+00,2.72E+01,9.80E-01,
Uterus,0.00E+00,0.00E+00,0.00E+00,0.00E+00,0.00E+00,0.00E+00,0.00E+00,0.00E+00,0.00E+00,0.00E+00,
,,,,,,,,,,,
,,,,,,,,,,,
\end{filecontents*}
\DTLloaddb{spe_events_3}{spe_events_10_3.csv}\label{tab:8}
\noindent\adjustbox{max width=\textwidth,max totalheight=\textheight-6ex}{
    \begin{tabular}{|c|c|c|c|c|c|c|c|c|c|c|}
    \hline
    \DTLforeach{spe_events_3}
    {\organ=organ,\eone=e1,\etwo=e2,\ethree=e3,\efour=e4,\efive=e5, \esix=e6, \eseven=e7, \eeight=e8, \enine=e9, \eten=e10}
    {%\DTLiffirstrow{}{\\}%
    \organ & \eone & \etwo & \ethree & \efour & \efive & \esix & \eseven & \eeight & \enine & \eten
    \DTLiflastrow{}{\DTLiffirstrow{\\ \hline}{\\}}
    }
    \\ \hline
    \end{tabular}\\
}
\captionof{table}{Radiation dose deposition in different organs for SEP events}

\begin{filecontents*}{spe_events_10_4.csv}
organ,e1,e2,e3,e4,e5,e6,e7,e8,e9,e10,
Organ,Event 31, Event 32, Event 33, Event 34, Event 35, Event 36, Event 37, Event 38, Event 39, Event 40, 
Brain,3.27E+02,4.56E+02,3.43E+02,2.31E+02,3.01E+00,1.96E+01,2.56E+01,1.41E+01,1.33E+01,9.23E+01,
Head,4.08E+02,1.38E+03,7.20E+02,3.88E+02,3.70E+00,2.87E+01,3.28E+01,1.49E+01,1.76E+01,2.30E+02,
Heart,7.49E+01,8.01E+01,5.75E+01,5.37E+01,7.52E-01,4.12E+00,6.01E+00,3.70E+00,2.90E+00,1.44E+01,
LeftAdrenal,1.59E+02,1.92E+02,1.46E+02,1.12E+02,1.52E+00,9.21E+00,1.26E+01,7.29E+00,6.36E+00,3.78E+01,
LeftArmBone,5.76E+02,1.09E+03,7.33E+02,4.34E+02,5.19E+00,3.70E+01,4.52E+01,2.32E+01,2.42E+01,2.13E+02,
LeftBreast,0.00E+00,0.00E+00,0.00E+00,0.00E+00,0.00E+00,0.00E+00,0.00E+00,0.00E+00,0.00E+00,0.00E+00,
LeftClavicle,3.88E+02,5.93E+02,4.36E+02,2.78E+02,3.53E+00,2.39E+01,3.04E+01,1.63E+01,1.61E+01,1.21E+02,
LeftKidney,1.95E+02,2.49E+02,1.91E+02,1.37E+02,1.83E+00,1.15E+01,1.54E+01,8.68E+00,7.90E+00,5.02E+01,
LeftLeg,5.95E+02,1.60E+03,9.11E+02,5.10E+02,5.37E+00,4.01E+01,4.73E+01,2.27E+01,2.53E+01,2.81E+02,
LeftLegBone,1.37E+02,1.69E+02,1.27E+02,9.70E+01,1.31E+00,7.93E+00,1.09E+01,6.29E+00,5.49E+00,3.29E+01,
LeftLung,1.92E+02,2.38E+02,1.82E+02,1.34E+02,1.80E+00,1.12E+01,1.51E+01,8.62E+00,7.71E+00,4.74E+01,
LeftOvary,0.00E+00,0.00E+00,0.00E+00,0.00E+00,0.00E+00,0.00E+00,0.00E+00,0.00E+00,0.00E+00,0.00E+00,
LeftScapula,5.95E+02,1.17E+03,7.78E+02,4.51E+02,5.33E+00,3.86E+01,4.67E+01,2.36E+01,2.51E+01,2.28E+02,
LeftTeste,5.01E+02,8.74E+02,6.10E+02,3.69E+02,4.52E+00,3.18E+01,3.92E+01,2.04E+01,2.09E+01,1.75E+02,
LowerLargeIntestine,5.34E+01,5.60E+01,3.93E+01,3.87E+01,5.43E-01,2.91E+00,4.30E+00,2.69E+00,2.05E+00,9.84E+00,
MaleGenitalia,7.81E+02,2.52E+03,1.35E+03,7.21E+02,7.02E+00,5.49E+01,6.25E+01,2.84E+01,3.38E+01,4.30E+02,
MiddleLowerSpine,9.66E+01,1.14E+02,8.49E+01,6.85E+01,9.35E-01,5.52E+00,7.67E+00,4.52E+00,3.83E+00,2.19E+01,
Pancreas,3.42E+01,3.39E+01,2.15E+01,2.56E+01,3.65E-01,1.82E+00,2.80E+00,1.82E+00,1.28E+00,5.43E+00,
Pelvis,1.22E+02,1.57E+02,1.17E+02,8.68E+01,1.16E+00,7.15E+00,9.64E+00,5.52E+00,4.90E+00,3.09E+01,
RibCage,6.24E+02,1.26E+03,8.28E+02,4.78E+02,5.60E+00,4.07E+01,4.90E+01,2.47E+01,2.64E+01,2.44E+02,
RightAdrenal,2.38E+02,3.02E+02,2.33E+02,1.66E+02,2.22E+00,1.40E+01,1.87E+01,1.05E+01,9.65E+00,6.11E+01,
RightArmBone,5.81E+02,1.10E+03,7.37E+02,4.38E+02,5.24E+00,3.72E+01,4.56E+01,2.34E+01,2.44E+01,2.14E+02,
RightBreast,0.00E+00,0.00E+00,0.00E+00,0.00E+00,0.00E+00,0.00E+00,0.00E+00,0.00E+00,0.00E+00,0.00E+00,
RightClavicle,2.31E+02,3.17E+02,2.37E+02,1.64E+02,2.15E+00,1.38E+01,1.81E+01,1.01E+01,9.40E+00,6.40E+01,
RightKidney,1.80E+02,2.28E+02,1.74E+02,1.26E+02,1.69E+00,1.06E+01,1.42E+01,8.05E+00,7.28E+00,4.57E+01,
RightLeg,5.93E+02,1.59E+03,9.10E+02,5.09E+02,5.36E+00,4.00E+01,4.72E+01,2.27E+01,2.53E+01,2.81E+02,
RightLegBone,1.40E+02,1.72E+02,1.30E+02,9.84E+01,1.33E+00,8.08E+00,1.10E+01,6.36E+00,5.59E+00,3.38E+01,
RightLung,1.78E+02,2.19E+02,1.67E+02,1.25E+02,1.68E+00,1.03E+01,1.40E+01,8.04E+00,7.14E+00,4.34E+01,
RightOvary,0.00E+00,0.00E+00,0.00E+00,0.00E+00,0.00E+00,0.00E+00,0.00E+00,0.00E+00,0.00E+00,0.00E+00,
RightScapula,5.86E+02,1.12E+03,7.55E+02,4.41E+02,5.25E+00,3.79E+01,4.60E+01,2.34E+01,2.47E+01,2.20E+02,
RightTeste,3.38E+02,5.81E+02,4.06E+02,2.49E+02,3.06E+00,2.13E+01,2.65E+01,1.39E+01,1.41E+01,1.16E+02,
Skull,5.76E+02,1.24E+03,7.84E+02,4.52E+02,5.19E+00,3.77E+01,4.54E+01,2.27E+01,2.44E+01,2.34E+02,
SmallIntestine,4.76E+01,4.92E+01,3.35E+01,3.49E+01,4.92E-01,2.58E+00,3.85E+00,2.44E+00,1.81E+00,8.43E+00,
Spleen,1.08E+02,1.24E+02,9.33E+01,7.60E+01,1.04E+00,6.12E+00,8.56E+00,5.06E+00,4.27E+00,2.38E+01,
Stomach,1.04E+02,1.23E+02,9.20E+01,7.37E+01,1.01E+00,5.96E+00,8.26E+00,4.86E+00,4.13E+00,2.37E+01,
Thymus,1.31E+02,1.55E+02,1.19E+02,9.17E+01,1.25E+00,7.53E+00,1.04E+01,6.01E+00,5.24E+00,3.05E+01,
Thyroid,1.45E+02,1.68E+02,1.30E+02,1.01E+02,1.38E+00,8.29E+00,1.14E+01,6.67E+00,5.78E+00,3.29E+01,
Trunk,3.30E+02,9.15E+02,5.12E+02,2.88E+02,3.00E+00,2.23E+01,2.63E+01,1.27E+01,1.40E+01,1.59E+02,
UpperLargeIntestine,5.29E+01,5.57E+01,3.88E+01,3.85E+01,5.40E-01,2.89E+00,4.27E+00,2.67E+00,2.03E+00,9.75E+00,
UpperSpine,1.80E+02,2.20E+02,1.69E+02,1.26E+02,1.70E+00,1.04E+01,1.41E+01,8.11E+00,7.22E+00,4.37E+01,
UrinaryBladder,1.06E+02,1.28E+02,9.55E+01,7.51E+01,1.02E+00,6.11E+00,8.39E+00,4.90E+00,4.22E+00,2.49E+01,
Uterus,0.00E+00,0.00E+00,0.00E+00,0.00E+00,0.00E+00,0.00E+00,0.00E+00,0.00E+00,0.00E+00,0.00E+00,
,,,,,,,,,,,
,,,,,,,,,,,
\end{filecontents*}
\DTLloaddb{spe_events_4}{spe_events_10_4.csv}\label{tab:9}
\noindent\adjustbox{max width=\textwidth,max totalheight=\textheight-6ex}{
    \begin{tabular}{|c|c|c|c|c|c|c|c|c|c|c|}
    \hline
    \DTLforeach{spe_events_4}
    {\organ=organ,\eone=e1,\etwo=e2,\ethree=e3,\efour=e4,\efive=e5, \esix=e6, \eseven=e7, \eeight=e8, \enine=e9, \eten=e10}
    {%\DTLiffirstrow{}{\\}%
    \organ & \eone & \etwo & \ethree & \efour & \efive & \esix & \eseven & \eeight & \enine & \eten
    \DTLiflastrow{}{\DTLiffirstrow{\\ \hline}{\\}}
    }
    \\ \hline
    \end{tabular}\\
}

\captionof{table}{Radiation dose deposition in different organs for SEP events}
\begin{filecontents*}{spe_events_10_5.csv}
organ,e1,e2,e3,e4,e5,e6,e7,e8,e9,e10,
Organ,Event 41, Event 42, Event 43, Event 44, Event 45, Event 46, Event 47, Event 48, Event 49, Event 50, 
Brain,6.85E+01,9.74E+00,8.86E+01,6.48E+01,7.60E+00,2.61E+00,3.62E+00,3.50E+02,7.98E+01,1.68E+01,
Head,1.50E+02,3.01E+01,1.96E+02,1.00E+02,1.18E+01,5.53E+00,8.03E+00,5.63E+02,1.08E+02,3.04E+01,
Heart,1.30E+01,1.70E+00,1.30E+01,1.21E+01,1.62E+00,4.83E-01,7.85E-01,6.44E+01,1.95E+01,3.47E+00,
LeftAdrenal,3.04E+01,4.14E+00,3.58E+01,2.91E+01,3.57E+00,1.15E+00,1.69E+00,1.56E+02,3.98E+01,7.75E+00,
LeftArmBone,1.44E+02,2.24E+01,1.99E+02,1.26E+02,1.43E+01,5.45E+00,7.35E+00,6.89E+02,1.41E+02,3.30E+01,
LeftBreast,0.00E+00,0.00E+00,0.00E+00,0.00E+00,0.00E+00,0.00E+00,0.00E+00,0.00E+00,0.00E+00,0.00E+00,
LeftClavicle,8.59E+01,1.25E+01,1.16E+02,8.02E+01,9.22E+00,3.27E+00,4.46E+00,4.34E+02,9.44E+01,2.06E+01,
LeftKidney,3.89E+01,5.39E+00,4.82E+01,3.72E+01,4.45E+00,1.48E+00,2.10E+00,2.00E+02,4.81E+01,9.72E+00,
LeftLeg,1.85E+02,3.39E+01,2.49E+02,1.39E+02,1.61E+01,6.90E+00,9.72E+00,7.69E+02,1.52E+02,3.95E+01,
LeftLegBone,2.64E+01,3.63E+00,3.11E+01,2.51E+01,3.09E+00,9.99E-01,1.47E+00,1.35E+02,3.44E+01,6.72E+00,
LeftLung,3.74E+01,5.14E+00,4.54E+01,3.58E+01,4.34E+00,1.42E+00,2.04E+00,1.93E+02,4.74E+01,9.43E+00,
LeftOvary,0.00E+00,0.00E+00,0.00E+00,0.00E+00,0.00E+00,0.00E+00,0.00E+00,0.00E+00,0.00E+00,0.00E+00,
LeftScapula,1.51E+02,2.36E+01,2.14E+02,1.33E+02,1.49E+01,5.74E+00,7.68E+00,7.25E+02,1.46E+02,3.45E+01,
LeftTeste,1.19E+02,1.81E+01,1.65E+02,1.08E+02,1.22E+01,4.54E+00,6.11E+00,5.86E+02,1.22E+02,2.78E+01,
LowerLargeIntestine,9.04E+00,1.18E+00,8.67E+00,8.41E+00,1.15E+00,3.36E-01,5.61E-01,4.45E+01,1.41E+01,2.45E+00,
MaleGenitalia,2.76E+02,5.34E+01,3.72E+02,1.92E+02,2.23E+01,1.02E+01,1.45E+01,1.08E+03,2.04E+02,5.67E+01,
MiddleLowerSpine,1.80E+01,2.44E+00,2.05E+01,1.71E+01,2.15E+00,6.80E-01,1.03E+00,9.15E+01,2.44E+01,4.65E+00,
Pancreas,5.41E+00,6.90E-01,4.32E+00,4.86E+00,7.18E-01,1.98E-01,3.65E-01,2.53E+01,9.42E+00,1.53E+00,
Pelvis,2.41E+01,3.34E+00,2.92E+01,2.28E+01,2.77E+00,9.11E-01,1.33E+00,1.22E+02,3.05E+01,6.06E+00,
RibCage,1.61E+02,2.56E+01,2.28E+02,1.40E+02,1.58E+01,6.12E+00,8.22E+00,7.65E+02,1.53E+02,3.66E+01,
RightAdrenal,4.74E+01,6.55E+00,5.90E+01,4.56E+01,5.43E+00,1.80E+00,2.55E+00,2.45E+02,5.84E+01,1.18E+01,
RightArmBone,1.44E+02,2.25E+01,2.00E+02,1.27E+02,1.44E+01,5.48E+00,7.40E+00,6.93E+02,1.42E+02,3.32E+01,
RightBreast,0.00E+00,0.00E+00,0.00E+00,0.00E+00,0.00E+00,0.00E+00,0.00E+00,0.00E+00,0.00E+00,0.00E+00,
RightClavicle,4.78E+01,6.77E+00,6.11E+01,4.52E+01,5.35E+00,1.82E+00,2.55E+00,2.44E+02,5.68E+01,1.18E+01,
RightKidney,3.56E+01,4.92E+00,4.38E+01,3.41E+01,4.10E+00,1.35E+00,1.93E+00,1.83E+02,4.44E+01,8.93E+00,
RightLeg,1.85E+02,3.38E+01,2.48E+02,1.38E+02,1.60E+01,6.89E+00,9.71E+00,7.67E+02,1.51E+02,3.94E+01,
RightLegBone,2.70E+01,3.71E+00,3.21E+01,2.57E+01,3.15E+00,1.02E+00,1.49E+00,1.38E+02,3.48E+01,6.84E+00,
RightLung,3.45E+01,4.73E+00,4.15E+01,3.31E+01,4.02E+00,1.31E+00,1.89E+00,1.78E+02,4.41E+01,8.72E+00,
RightOvary,0.00E+00,0.00E+00,0.00E+00,0.00E+00,0.00E+00,0.00E+00,0.00E+00,0.00E+00,0.00E+00,0.00E+00,
RightScapula,1.47E+02,2.28E+01,2.07E+02,1.30E+02,1.46E+01,5.57E+00,7.46E+00,7.08E+02,1.43E+02,3.36E+01,
RightTeste,7.96E+01,1.20E+01,1.09E+02,7.21E+01,8.22E+00,3.03E+00,4.11E+00,3.91E+02,8.26E+01,1.87E+01,
Skull,1.54E+02,2.52E+01,2.15E+02,1.30E+02,1.47E+01,5.83E+00,7.91E+00,7.11E+02,1.43E+02,3.46E+01,
SmallIntestine,7.90E+00,1.02E+00,7.22E+00,7.28E+00,1.02E+00,2.92E-01,5.04E-01,3.83E+01,1.27E+01,2.17E+00,
Spleen,1.99E+01,2.67E+00,2.23E+01,1.90E+01,2.39E+00,7.50E-01,1.13E+00,1.01E+02,2.72E+01,5.15E+00,
Stomach,1.95E+01,2.64E+00,2.23E+01,1.85E+01,2.32E+00,7.35E-01,1.10E+00,9.91E+01,2.63E+01,5.02E+00,
Thymus,2.48E+01,3.36E+00,2.91E+01,2.38E+01,2.94E+00,9.40E-01,1.38E+00,1.28E+02,3.27E+01,6.34E+00,
Thyroid,2.71E+01,3.65E+00,3.15E+01,2.61E+01,3.23E+00,1.03E+00,1.51E+00,1.40E+02,3.61E+01,6.96E+00,
Trunk,1.05E+02,1.96E+01,1.39E+02,7.67E+01,8.99E+00,3.90E+00,5.57E+00,4.26E+02,8.52E+01,2.22E+01,
UpperLargeIntestine,8.96E+00,1.17E+00,8.55E+00,8.32E+00,1.14E+00,3.32E-01,5.58E-01,4.40E+01,1.40E+01,2.43E+00,
UpperSpine,3.48E+01,4.76E+00,4.19E+01,3.34E+01,4.06E+00,1.32E+00,1.91E+00,1.79E+02,4.44E+01,8.80E+00,
UrinaryBladder,2.01E+01,2.74E+00,2.34E+01,1.91E+01,2.37E+00,7.59E-01,1.13E+00,1.02E+02,2.67E+01,5.15E+00,
Uterus,0.00E+00,0.00E+00,0.00E+00,0.00E+00,0.00E+00,0.00E+00,0.00E+00,0.00E+00,0.00E+00,0.00E+00,
,,,,,,,,,,,
,,,,,,,,,,,
\end{filecontents*}
\DTLloaddb{spe_events_5}{spe_events_10_5.csv}\label{tab:10}
\noindent\adjustbox{max width=\textwidth,max totalheight=\textheight-6ex}{
    \begin{tabular}{|c|c|c|c|c|c|c|c|c|c|c|}
    \hline
    \DTLforeach{spe_events_5}
    {\organ=organ,\eone=e1,\etwo=e2,\ethree=e3,\efour=e4,\efive=e5, \esix=e6, \eseven=e7, \eeight=e8, \enine=e9, \eten=e10}
    {%\DTLiffirstrow{}{\\}%
    \organ & \eone & \etwo & \ethree & \efour & \efive & \esix & \eseven & \eeight & \enine & \eten
    \DTLiflastrow{}{\DTLiffirstrow{\\ \hline}{\\}}
    }
    \\ \hline
    \end{tabular}\\
}

\captionof{table}{Radiation dose deposition in different organs for SEP events}
\begin{filecontents*}{spe_events_10_6.csv}
organ,e1,e2,e3,e4,e5,e6,e7,e8,e9,e10,
Organ,Event 51, Event 52, Event 53, Event 54, Event 55, Event 56, Event 57, Event 58, Event 59, Event 60, 
Brain,7.23E+01,2.45E+01,1.49E+01,1.84E+02,1.42E+02,3.36E+01,7.00E+01,2.04E+02,7.03E+01,1.14E+01,
Head,1.47E+02,5.34E+01,3.75E+01,3.58E+02,3.17E+02,1.41E+02,2.85E+02,2.36E+02,1.04E+02,1.70E+01,
Heart,1.09E+01,4.37E+00,2.63E+00,2.79E+01,2.58E+01,5.15E+00,9.47E+00,5.21E+01,1.50E+01,2.57E+00,
LeftAdrenal,2.97E+01,1.06E+01,6.40E+00,7.58E+01,6.20E+01,1.34E+01,2.68E+01,1.04E+02,3.31E+01,5.50E+00,
LeftArmBone,1.58E+02,5.20E+01,3.27E+01,3.98E+02,3.01E+02,8.55E+01,1.84E+02,3.46E+02,1.31E+02,2.11E+01,
LeftBreast,0.00E+00,0.00E+00,0.00E+00,0.00E+00,0.00E+00,0.00E+00,0.00E+00,0.00E+00,0.00E+00,0.00E+00,
LeftClavicle,9.35E+01,3.09E+01,1.90E+01,2.38E+02,1.79E+02,4.44E+01,9.42E+01,2.38E+02,8.51E+01,1.38E+01,
LeftKidney,3.97E+01,1.38E+01,8.31E+00,1.01E+02,7.99E+01,1.80E+01,3.67E+01,1.25E+02,4.13E+01,6.77E+00,
LeftLeg,1.90E+02,6.64E+01,4.47E+01,4.70E+02,3.90E+02,1.48E+02,3.07E+02,3.50E+02,1.44E+02,2.33E+01,
LeftLegBone,2.57E+01,9.25E+00,5.59E+00,6.56E+01,5.39E+01,1.19E+01,2.37E+01,8.99E+01,2.86E+01,4.75E+00,
LeftLung,3.75E+01,1.32E+01,7.95E+00,9.56E+01,7.67E+01,1.70E+01,3.43E+01,1.23E+02,4.02E+01,6.63E+00,
LeftOvary,0.00E+00,0.00E+00,0.00E+00,0.00E+00,0.00E+00,0.00E+00,0.00E+00,0.00E+00,0.00E+00,0.00E+00,
LeftScapula,1.69E+02,5.49E+01,3.46E+01,4.27E+02,3.18E+02,9.03E+01,1.96E+02,3.53E+02,1.36E+02,2.19E+01,
LeftTeste,1.32E+02,4.32E+01,2.68E+01,3.34E+02,2.50E+02,6.70E+01,1.44E+02,3.02E+02,1.12E+02,1.81E+01,
LowerLargeIntestine,7.34E+00,3.02E+00,1.82E+00,1.87E+01,1.79E+01,3.50E+00,6.31E+00,3.77E+01,1.06E+01,1.83E+00,
MaleGenitalia,2.80E+02,9.89E+01,6.82E+01,6.86E+02,5.85E+02,2.43E+02,5.03E+02,4.49E+02,1.98E+02,3.20E+01,
MiddleLowerSpine,1.71E+01,6.25E+00,3.77E+00,4.35E+01,3.66E+01,7.82E+00,1.53E+01,6.43E+01,1.99E+01,3.33E+00,
Pancreas,3.72E+00,1.73E+00,1.05E+00,9.49E+00,1.05E+01,1.93E+00,3.17E+00,2.54E+01,6.61E+00,1.17E+00,
Pelvis,2.40E+01,8.47E+00,5.13E+00,6.12E+01,4.93E+01,1.11E+01,2.25E+01,7.92E+01,2.57E+01,4.24E+00,
RibCage,1.79E+02,5.86E+01,3.71E+01,4.52E+02,3.40E+02,9.91E+01,2.14E+02,3.71E+02,1.44E+02,2.31E+01,
RightAdrenal,4.86E+01,1.68E+01,1.01E+01,1.24E+02,9.75E+01,2.18E+01,4.45E+01,1.52E+02,5.03E+01,8.25E+00,
RightArmBone,1.59E+02,5.23E+01,3.29E+01,4.01E+02,3.03E+02,8.58E+01,1.84E+02,3.49E+02,1.32E+02,2.12E+01,
RightBreast,0.00E+00,0.00E+00,0.00E+00,0.00E+00,0.00E+00,0.00E+00,0.00E+00,0.00E+00,0.00E+00,0.00E+00,
RightClavicle,4.99E+01,1.70E+01,1.04E+01,1.27E+02,9.88E+01,2.32E+01,4.80E+01,1.46E+02,4.94E+01,8.08E+00,
RightKidney,3.61E+01,1.26E+01,7.59E+00,9.21E+01,7.31E+01,1.63E+01,3.32E+01,1.16E+02,3.80E+01,6.24E+00,
RightLeg,1.90E+02,6.63E+01,4.46E+01,4.70E+02,3.90E+02,1.48E+02,3.07E+02,3.49E+02,1.44E+02,2.33E+01,
RightLegBone,2.65E+01,9.47E+00,5.71E+00,6.76E+01,5.52E+01,1.22E+01,2.44E+01,9.10E+01,2.92E+01,4.83E+00,
RightLung,3.44E+01,1.21E+01,7.32E+00,8.76E+01,7.07E+01,1.55E+01,3.13E+01,1.15E+02,3.72E+01,6.15E+00,
RightOvary,0.00E+00,0.00E+00,0.00E+00,0.00E+00,0.00E+00,0.00E+00,0.00E+00,0.00E+00,0.00E+00,0.00E+00,
RightScapula,1.64E+02,5.33E+01,3.34E+01,4.14E+02,3.08E+02,8.64E+01,1.87E+02,3.49E+02,1.34E+02,2.15E+01,
RightTeste,8.74E+01,2.88E+01,1.79E+01,2.21E+02,1.67E+02,4.45E+01,9.52E+01,2.05E+02,7.55E+01,1.22E+01,
Skull,1.68E+02,5.58E+01,3.58E+01,4.22E+02,3.25E+02,9.99E+01,2.14E+02,3.43E+02,1.34E+02,2.16E+01,
SmallIntestine,6.14E+00,2.61E+00,1.57E+00,1.57E+01,1.55E+01,3.00E+00,5.28E+00,3.42E+01,9.38E+00,1.63E+00,
Spleen,1.86E+01,6.89E+00,4.14E+00,4.75E+01,4.03E+01,8.47E+00,1.65E+01,7.19E+01,2.22E+01,3.71E+00,
Stomach,1.85E+01,6.77E+00,4.08E+00,4.72E+01,3.96E+01,8.46E+00,1.66E+01,6.91E+01,2.15E+01,3.59E+00,
Thymus,2.42E+01,8.69E+00,5.22E+00,6.16E+01,5.06E+01,1.09E+01,2.16E+01,8.58E+01,2.72E+01,4.52E+00,
Thyroid,2.62E+01,9.48E+00,5.68E+00,6.69E+01,5.53E+01,1.17E+01,2.31E+01,9.49E+01,3.00E+01,4.98E+00,
Trunk,1.06E+02,3.75E+01,2.55E+01,2.61E+02,2.21E+02,8.68E+01,1.79E+02,1.95E+02,8.03E+01,1.30E+01,
UpperLargeIntestine,7.25E+00,2.98E+00,1.80E+00,1.85E+01,1.77E+01,3.47E+00,6.24E+00,3.75E+01,1.05E+01,1.82E+00,
UpperSpine,3.47E+01,1.22E+01,7.37E+00,8.84E+01,7.12E+01,1.56E+01,3.14E+01,1.16E+02,3.76E+01,6.20E+00,
UrinaryBladder,1.94E+01,7.00E+00,4.22E+00,4.94E+01,4.09E+01,8.88E+00,1.76E+01,6.99E+01,2.20E+01,3.66E+00,
Uterus,0.00E+00,0.00E+00,0.00E+00,0.00E+00,0.00E+00,0.00E+00,0.00E+00,0.00E+00,0.00E+00,0.00E+00,
,,,,,,,,,,,
,,,,,,,,,,,
\end{filecontents*}
\DTLloaddb{spe_events_6}{spe_events_10_6.csv}\label{tab:11}
\noindent\adjustbox{max width=\textwidth,max totalheight=\textheight-6ex}{
    \begin{tabular}{|c|c|c|c|c|c|c|c|c|c|c|}
    \hline
    \DTLforeach{spe_events_6}
    {\organ=organ,\eone=e1,\etwo=e2,\ethree=e3,\efour=e4,\efive=e5, \esix=e6, \eseven=e7, \eeight=e8, \enine=e9, \eten=e10}
    {%\DTLiffirstrow{}{\\}%
    \organ & \eone & \etwo & \ethree & \efour & \efive & \esix & \eseven & \eeight & \enine & \eten
    \DTLiflastrow{}{\DTLiffirstrow{\\ \hline}{\\}}
    }
    \\ \hline
    \end{tabular}\\
}
\captionof{table}{Radiation dose deposition in different organs for SEP events}

\begin{acknowledgements}

\noindent This work was supported by the New York University Abu Dhabi (NYUAD) Institute Research Grant G1502 and the ASPIRE Award for Research Excellence (AARE) Grant S1560 by the Advanced Technology Research Council (ATRC). We thank the Young Scientist Program of the Blue Marble Space Institute of Science for enabling this collaboration. Computations were carried out on High-Performance Computing (HPC) resources at NYUAD. KH gratefully acknowledges the support of the DFG priority program SPP 1992 “Exploring the Diversity of Extrasolar Planets (\textit{HE 8392/1-1})”. DA and KH also like to thank ISSI and the supported International Team 464 (ETERNAL).
\end{acknowledgements}

 \section*{Conflict of interest}
 The authors declare that they have no conflict of interest.

\bibliographystyle{spbasic}
\small{
\bibliography{mybibfile}
}
\end{document}